%% file: hdrhep.tex
\documentclass[11pt]{scrbook}                                                   

\usepackage{graphicx}
\usepackage{amsmath}
\usepackage{hhline}
\usepackage{amssymb}
\usepackage{times}
\usepackage{cite}
\usepackage{array}
\usepackage{multirow}
\usepackage{float}
\usepackage{latexsym}
\usepackage{epsfig}
\usepackage{rotating}
\usepackage{graphics}
\usepackage{mathrsfs}
\usepackage{setspace}
\usepackage{subfigure}

\usepackage{color}
\definecolor{rltred}{rgb}{0.75,0,0}
\definecolor{rltgreen}{rgb}{0,0.5,0}
\definecolor{rltblue}{rgb}{0,0,0.75}

\newif\ifpdf
\ifx\pdfoutput\undefined
    \pdffalse          % we are not running PDFLaTeX
\else
    \pdfoutput=1       % we are running PDFLaTeX
    \pdftrue
\fi

\ifpdf
\usepackage{thumbpdf}
\usepackage[pdftex,
        colorlinks=true,
        urlcolor=rltblue,       % \href{...}{...} external (URL)
        filecolor=rltgreen,     % \href{...} local file
        linkcolor=rltred,       % \ref{...} and \pageref{...}
        pdftitle={Example File for pdflatex},
        pdfauthor={H1},        pdfsubject={Template file},
        pdfkeywords={High-Energy Physics, Particle Physics},
        pdfpagemode=None,
        bookmarksopen=true]{hyperref}
 
\fi

\newlength{\dinwidth}
\newlength{\dinmargin}
\input{notations}

\begin{document}
\input{feynman}

\rm \normalsize

\thispagestyle{empty}
\begin{flushleft}
DESY-2004-06\\
CPPM-H-2003-01\\
January 2004
\end{flushleft}

\vspace{2.5cm}

\begin{center}
\begin{Large}

{\bf \Huge Isolated Lepton Production at Colliders }

\end{Large}
\vspace{3.5cm}

{\bf \Large  Cristinel Diaconu\footnote{Habilitation \`a diriger des recherches, December 15, 2003 
}}\\
Centre de Physique des Particules de Marseille\\
Case 902 , 163 Avenue de Luminy 
13288 Marseille cedex 09,
France\\
diaconu@cppm.in2p3.fr

\end{center}
\begin{center}

%\begin{itemize}
%\item[                                        ] {\bf Jury:}
%\item[                                        ] Prof. Elie Aslanides
%\item[                                        ] Prof. Jean-Jacques Aubert
%\item[                                        ] Prof. Philippe Bloch
%\item[                                        ] Prof. Jean-Fran\c{c}ois Grivaz
%\item[              ] Prof. Sylvain Tisserant
%\item[              ] Prof. Jos Vermaseren
%\end{itemize}
\end{center}

\vspace{2.5 cm}
The production of isolated leptons with high transverse momentum in high energy $ee$, $pp$ or $ep$ collisions is reviewed. The leptons are produced either through boson splitting or by boson-boson collision and yield experimentally simple and spectacular topologies which can be exploited to validate the \sm or to search for new phenomena.

\vspace{2.5cm}

% Table des matieres
\tableofcontents

\mainmatter

\include{intro}
\include{leptons}

\include{bsm}

\include{conclusion}

\bibliographystyle{utcaps.bst}
\bibliography{hdrhep}

\backmatter

\end{document}

%% file: notations.tex
\newcommand{\sm}{Standard Model }
\newcommand{\pbi}{\,\text{pb}^{-1}}   
\newcommand{\eVdist}{\kern-0.06667em} 
\newcommand{\gev}{{\,\text{Ge}\eVdist\text{V\/}}}

\def\ifmath#1{\relax\ifmmode #1\else $#1$\fi}%
\def\leqsim{\mathbin{\;\raise1pt\hbox{$<$}\kern-8pt\lower3pt\hbox{$\sim$}\;}}
\def\geqsim{\mathbin{\;\raise1pt\hbox{$>$}\kern-8pt\lower3pt\hbox{$\sim$}\;}}
\def\GeV{\ifmmode \hbox{\rm Ge\kern -0.1em V}\else
                  \hbox{\mathrm{Ge\kern -0.1em V}}\fi}%
\def\MeV{\ifmmode \hbox{\rm Me\kern -0.1em V}\else
                  \hbox{\mathrm{Me\kern -0.1em V}}\fi}%
\def\keV{\ifmmode \hbox{\rm ke\kern -0.1em V}\else
                  \hbox{\mathrm{ke\kern -0.1em V}}\fi}%
\def\eV{\ifmmode \hbox{\rm e\kern -0.1em V}\else
                 \hbox{\mathrm{e\kern -0.1em V}}\fi}%
\let\gev=\GeV

\newcommand{\CoM}        {centre-of-mass~}

\newcommand {\MW}      {m_{\mathrm{W}}}

\newcommand {\thw}        {\theta_{\mathrm{W}}}

\def\ra{\rightarrow}

\newcommand{\gz}{\mbox{$g_1^{\mathrm{Z}}$}}

\newcommand{\kg}{\mbox{$\kappa_\gamma$}}
\newcommand{\kz}{\mbox{$\kappa_{\mathrm{Z}}$}}
\renewcommand{\lg}{\mbox{$\lambda_\gamma$}}
\newcommand{\lz}{\mbox{$\lambda_{\mathrm{Z}}$}}

\newcommand{\twsq}{\ensuremath{\tan^2\thw}}

%% file: feynman.tex
%                     FEYNMAN(34).TEX 
%  CALLING ROUTINE FOR DRAWING FEYNMAN DIAGRAMS IN LATEX.
%  DOCUMENTATION IN "FEYNMAN - A LaTeX Routine for Generating Feynman Diagrams"
%  Cavendish-HEP 88/11  (Cavendish Labs, Cambridge, UK).
%  See also: Levine, M.J.S., A LaTeX Graphics Routine for Drawing Feynman
%  Diagrams, Cavendish - HEP 89/4.
%  USES THE FOLLOWING TEX FILES:
%   GLUONSETUP(31), PHOTONSETUP(28), FERMIONSETUP(7), SCALARSETUP(9)
%   VERTEX(25), GLUONLINKS, LOOPS(1) 
%
%   THIS PROGRAM PACKAGE NOT TO BE ALTERED WITHOUT THE EXPRESS WRITTEN
%   PERMISSION OF THE AUTHOR.
%
%**************************************************************************
%
%                           SAMPLE USAGE
%  
%  \documentstyle[12pt]{article}
%  \begin{document}
%  \input feynman
%  \textheight 800pt \textwidth 450pt
%  \begin{picture}(10000,18000)
%  \drawline\gluon[\S\REG](0,16000)[8]
%  \drawline\fermion[\SW\REG](\gluonbackx,\gluonbacky)[2000]
%  \drawline\fermion[\SE\REG](\gluonbackx,\gluonbacky)[2000]
%  \end{picture}
%  \end{document}
%
%**************************************************************************
%
\message{FEYNMAN:  For generating Feynman Diagrams in LaTex}
\message{Mark 1.0 Last Altered by MJSL 2/89}
\textheight 650pt \textwidth 400pt  % Page size set.
\setlength{\unitlength}{0.01pt}
\gdef\Feynmanlength{\setlength{\unitlength}{0.01pt}}  % Say \Feynmanlength
\gdef\unlock{\catcode`\@=11}
%  Allows use of "@" in macro names, like PLAIN.TEX does.
\gdef\lock{\catcode`\@=12}%  Change @'s back to their normal category code.
\global\newcount\LINETYPE                     
\global\newcount\LINEDIRECTION
\global\newcount\LINECONFIGURATION
\newcommand{\LTYPE}{\LINETYPE}
\newcommand{\LDIR}{\LINEDIRECTION}
\newcommand{\LCONFIG}{\LINECONFIGURATION}
%DEFAULTS:  Horizontal fermion.   
\global\LINETYPE=1  \global\LINEDIRECTION=0  \global\LINECONFIGURATION=0
%  The parametric code names.  Don't change these.
\global\newcount\fermion    \fermion=1
\global\newcount\scalar     \scalar=2
\global\newcount\photon     \photon=3
\global\newcount\gluon      \gluon=4
\global\newcount\especial   \especial=5
\gdef\N{0}  \gdef\NE{1}  \gdef\E{2}   \gdef\SE{3}
\gdef\S{4}  \gdef\SW{5}  \gdef\W{6}   \gdef\NW{7}
\global\newcount\REG            \global\REG=0
\global\newcount\FLIPPED        \global\FLIPPED=1
\global\newcount\CURLY          \global\CURLY=2
\global\newcount\FLIPPEDCURLY   \global\FLIPPEDCURLY=3
\global\newcount\FLAT           \global\FLAT=4
\global\newcount\FLIPPEDFLAT    \global\FLIPPEDFLAT=5
\global\newcount\CENTRAL        \global\CENTRAL=6
\global\newcount\FLIPPEDCENTRAL \global\FLIPPEDCENTRAL=7
\gdef\LONGPHOTON{6}             \gdef\FLIPPEDLONG{7}
\global\newcount\SQUASHEDGLUON  \global\SQUASHEDGLUON=8
\gdef\SQUASHED{\SQUASHEDGLUON}
%\global\newcount\FLIPPEDSQUASHEDGLUON  \FLIPPEDSQUASHEDGLUON=9
%
%%%%%%%%%%%%%%%%%%%%%%%%%%%%%%%%%%%%%%%%%%%%%%%%%%%%%%%%%%%%%%%%%%%%%%%%%%%%%%%
% SOME COUNTERS AND DEFINITIONS FOR POSITIONS AND LENGTHS OF LINES & FEATURES %
%%%%%%%%%%%%%%%%%%%%%%%%%%%%%%%%%%%%%%%%%%%%%%%%%%%%%%%%%%%%%%%%%%%%%%%%%%%%%%%
\newcount\adjx \adjx=0
\newcount\adjy \adjy=0
\global\newdimen\BIGPHOTONS     \BIGPHOTONS=0pt  %  DEFAULT:  10 & 11-PT PHOTONS
\gdef\bigphotons{\global\BIGPHOTONS=12pt}%FOR 12-PT DOCS DRAWING E-W PHOTONS.
\global\newdimen\THICKPHOTONS     \THICKPHOTONS=0pt  %  FOR E-W PHOTONS 
\global\newdimen\THICKPHOTONSWITCH    \THICKPHOTONSWITCH=0pt
\gdef\THICKPHOTONTEST{
\THICKPHOTONSWITCH=0pt
\ifdim\THICKPHOTONS=0pt \relax
  \else \ifnum\LTYPE=3
           \ifnum\LDIR=2 \THICKPHOTONSWITCH=1pt \fi % THICK \E PHOTON
           \ifnum\LDIR=6 \THICKPHOTONSWITCH=1pt \fi % THICK \W PHOTON
        \fi
\fi
}  % end of THICKPHOTONTEST
\gdef\THICKLINES{\thicklines  \THICKPHOTONS=1pt}
\gdef\THINLINES{\thinlines  \THICKPHOTONS=0pt}
\global\newcount\phantomswitch   \global\phantomswitch=0
\global\newcount\stemlength   \global\stemlength=275   % Default STEM length.
\global\newcount\absstemlength        % A copy of STEM length.
\global\newcount\stemlengthx          % FOR STEMS on particle lines
\global\newcount\stemlengthy          % FOR STEMS on particle lines
\newdimen\FRONTSTEM  \FRONTSTEM=0pt   % FOR STEMS
\newdimen\BACKSTEM   \BACKSTEM=0pt    % FOR STEMS
\newdimen\EITHERSTEM \EITHERSTEM=0pt  % FOR STEMS
\gdef\frontstemmed{\FRONTSTEM=1pt}            % FOR STEMS
\gdef\backstemmed{\BACKSTEM=1pt}              % FOR STEMS
\gdef\stemmed{\FRONTSTEM=1pt  \BACKSTEM=1pt}    % FOR STEMS
\global\newcount\arrowlength                % FOR ARROWS
\global\newdimen\ATTIP   \global\ATTIP=0pt  % FOR ARROWS
\global\newdimen\ATBASE  \global\ATBASE=1pt % FOR ARROWS
\global\newcount\unitboxnumber  % SHOWS THE NUMBER OF `UNIT BOXES' IN LINE
\global\newcount\unitboxnumberpo  % One more than \unitboxnumber (in GLUONSETUP)
\global\newcount\particlelengthx  % THE X-LENGTH OF THE PARTICLE LINE
\gdef\plengthx{\particlelengthx}
\global\newcount\particlelengthy  % THE Y-LENGTH OF THE PARTICLE LINE
\gdef\plengthy{\particlelengthy}  
\global\newcount\boxlengthx  % THE X-LENGTH OF THE BOX:  abs(plengthx) usually
\global\newcount\boxlengthy  % THE y-LENGTH OF THE box:  abs(plengthy) usually
\global\newcount\particleadjustx  % Replaces \gluonadjustx, \scalaradjustx etc.
\global\newcount\particleadjusty  % Replaces \gluonadjusty, \scalaradjusty etc.
\global\newcount\particlelength   % The LENGTH of a particle line BOX (x)
\global\newcount\particlefrontx
\gdef\pfrontx{\particlefrontx}
\global\newcount\PFRONTx
\global\newcount\particlefronty
\gdef\pfronty{\particlefronty}
\global\newcount\PFRONTy
\global\newcount\particlebackx
\gdef\pbackx{\particlebackx}
\global\newcount\particlebacky
\gdef\pbacky{\particlebacky}
\global\newcount\particlemidx
\gdef\pmidx{\particlemidx}
\global\newcount\particlemidy
\gdef\pmidy{\particlemidy}
% SOME SPECIAL DEFS FOR \SCALARs:
\global\newcount\seglength  \global\newcount\gaplength
\global\gaplength=850  %default
\global\seglength=1416  % Length of each seg not including `ends' for attachment
% Now some storage locations for the user:
\global\newcount\Xone    \global\newcount\Yone    % user co-ords (\Xone,\Yone)
\global\newcount\Xtwo    \global\newcount\Ytwo    % user co-ords (\Xtwo,\Ytwo)
\global\newcount\Xthree  \global\newcount\Ythree  % user's (\Xthree,\Ythree)
\global\newcount\Xfour   \global\newcount\Yfour   % user co-ords (\Xfour,\Yfour)
\global\newcount\Xfive   \global\newcount\Yfive   % user co-ords (\Xfive,\Yfive)
\global\newcount\Xsix    \global\newcount\Ysix    % user co-ords (\Xsix,\Ysix)
\global\newcount\Xseven  \global\newcount\Yseven  % user's (\Xseven,\Yseven)
\global\newcount\Xeight  \global\newcount\Yeight  % user's (\Xeight,\Yeight)
%
%  SOME COUNTERS IDENTIFYING VARIOUS LINE PORTIONS AND DIMENSIONS:
%
\newsavebox{\lastline}  %  Default name for an unnamed particle line.
\global\newcount\numlineparts   % Num of pieces each unitbox of the line needs
\global\newcount\upperlineadjx  \upperlineadjx=0  %Default
\global\newcount\upperlineadjy  \upperlineadjy=0  %Default
\global\newcount\lowerlineadjx  \lowerlineadjx=0  %Default
\global\newcount\lowerlineadjy  \lowerlineadjy=0  %Default
\global\newcount\thirdlineadjx  \thirdlineadjx=0  %Default
\global\newcount\thirdlineadjy  \thirdlineadjy=0  %Default
\global\newcount\fourthlineadjx \fourthlineadjx=0  %Default
\global\newcount\fourthlineadjy \fourthlineadjy=0  %Default
\global\newcount\unitboxwidth   \unitboxwidth=1000%Default
\global\newcount\unitboxheight  \unitboxheight=0  %Default
\global\newcount\numupperunits  \numupperunits=8  %Default
\global\newcount\numlowerunits  \numlowerunits=8  %Default
\global\newcount\numthirdunits  \numthirdunits=8  %Default
\global\newcount\numfourthunits \numfourthunits=8  %Default
%  Some counters.  =0 until a line-type is drawn. Then=1.
\global\newcount\fermioncount   \global\fermioncount=0    
\global\newcount\scalarcount    \global\scalarcount=0    
\global\newcount\photoncount    \global\photoncount=0    
\global\newcount\gluoncount     \global\gluoncount=0    
\global\newcount\especialcount  \global\especialcount=0    
\global\newcount\vertexcount    \global\vertexcount=-1
%
%%%%%%%%%%%%%%%%%%%%%%%%%%%%%%%%%%%%%%%%%%%%%%%%%%%%%%%%%%%%%%
%     AUXILIARY ROUTINES FOR SETTING PARTICLE DIRECTIONS     %
%%%%%%%%%%%%%%%%%%%%%%%%%%%%%%%%%%%%%%%%%%%%%%%%%%%%%%%%%%%%%%
\global\newcount\XDIR
\global\newcount\YDIR
\gdef\SETDIR{  % SETS THE DIRECTIONS
\ifcase\LDIR 
     \global\XDIR=0  \global\YDIR=1   %\N  case.
\or  \global\XDIR=1  \global\YDIR=1   %\NE case.
\or  \global\XDIR=1  \global\YDIR=0   %\E  case.
\or  \global\XDIR=1  \global\YDIR=-1  %\SE case.
\or  \global\XDIR=0  \global\YDIR=-1  %\S  case.
\or  \global\XDIR=-1 \global\YDIR=-1  %\SW case.
\or  \global\XDIR=-1 \global\YDIR=0   %\W  case.
\or  \global\XDIR=-1 \global\YDIR=1   %\NW case.
\else\DIRECTERROR 
\fi}  % END OF \SETDIR
\gdef\moduloeight#1{
\ifnum#1>7 \global\advance #1 by -8 
\relax
\moduloeight#1 
\relax
\else \relax  
\fi}
\gdef\multroothalf#1{\global\multiply #1 by 7071 \global\divide #1 by 10000}
\gdef\negate#1{\global\multiply #1 by -1}
\gdef\double#1{\global\multiply #1 by 2}
\gdef\slanttest(#1,#2){ 
\ifodd\LDIR
\multiply #1 by 7071  \divide #1 by 10000
\multiply #2 by 7071  \divide #2 by 10000
\fi
}
\gdef\gslanttest(#1,#2){
\ifodd\LDIR
\multroothalf#1
\multroothalf#2
\fi
}
%
%%%%%%%%%%%%%%%%%%%%%%%%%%%%%%%%%%%%%%%%%%%%%%%%%%%%%%%%%%%%%%
% AUXILIARY ROUTINES FOR SETTING PARTICLE LENGTHS & POSTIONS %
%%%%%%%%%%%%%%%%%%%%%%%%%%%%%%%%%%%%%%%%%%%%%%%%%%%%%%%%%%%%%%
%
\gdef\setplength{ % calcs length of particle line
\global\particlelengthx=\unitboxwidth
\global\particlelengthy=\unitboxheight
\global\multiply \particlelengthx by \unitboxnumber
\global\multiply \particlelengthy by \unitboxnumber
\global\advance \particlelengthx by \particleadjustx
\global\advance \particlelengthy by \particleadjusty
}
\gdef\boxlengthdefault{  % DEFAULT FOR BOX SIZES IN \drawas
\global\boxlengthx=\plengthx
\global\boxlengthy=\plengthy
\ifnum\plengthx<0 \global\multiply\boxlengthx by -1 \fi
\ifnum\plengthy<0 \global\multiply\boxlengthy by -1 \fi
}
\gdef\rearcoords{  %  CALCULATES THE CO-ORDINATES OF THE BACK OF PARTICLE LINE
\global\particlebacky=\particlefronty 
\global\particlebackx=\particlefrontx 
\global\advance \particlebackx by \particlelengthx
\global\advance \particlebacky by \particlelengthy
}
\gdef\midcoords{  %  CALCULATES THE CO-ORDINATES OF THE MID OF PARTICLE LINE
\global\particlemidy=\particlefronty
\global\particlemidx=\particlefrontx
\global\stemlengthx=\particlelengthx  % Convenient variables not being used
\global\stemlengthy=\particlelengthy  
\global\divide\stemlengthx by 2
\global\divide\stemlengthy by 2
\global\advance \particlemidx by \stemlengthx
\global\advance \particlemidy by \stemlengthy
}
\gdef\setparticle{\setplength\rearcoords\midcoords\boxlengthdefault}  %sets line
\gdef\setcoords(#1,#2,#3)(#4,#5,#6)[#7,#8]{  
% Sets co-ords of first 3 line-parts of a line and the unitbox height and width
% Used by photons and gluons.
\global\upperlineadjx=#1
\global\lowerlineadjx=#2
\global\thirdlineadjx=#3
\global\upperlineadjy=#4
\global\lowerlineadjy=#5
\global\thirdlineadjy=#6
\global\unitboxwidth=#7
\global\unitboxheight=#8
}
%
%%%%%%%%%%%%%%%%%%%%%%%%%%%%%%%%%%%%%%%%%%%%%%%%%%%%%%%%%%%%%%%%%%%%%%%%%%%%%%%
%                                                                             %
%                     ROUTINES FOR DRAWING LINES                              %
%                                                                             %
%%%%%%%%%%%%%%%%%%%%%%%%%%%%%%%%%%%%%%%%%%%%%%%%%%%%%%%%%%%%%%%%%%%%%%%%%%%%%%%
%
% **************   ROUTINE FOR DRAWING STORED LINES AND PICTURES   *************
%
\gdef\drawoldpic#1(#2,#3){  % DRAWS PRE-SAVED PICTURE
\global\particlefrontx=#2
\global\particlefronty=#3
\rearcoords  
\midcoords
\put(#2,#3){\usebox{#1}}
}
\gdef\drawsavedline`#1' as #2[#3#4](#5,#6)[#7]{
\global\LINETYPE=#2
\global\LINEDIRECTION=#3
\global\LINECONFIGURATION=#4
\global\particlefrontx=#5
\global\particlefronty=#6
\global\unitboxnumber=#7  
% Formerly called \numhalfwiggles,\numdashes, \numloops, \fermionlength
% #1 is saved linename;   #2 is \LINETYPE;    #3 is \LINEDIRECTION
% #4 is \LINECONFIGURATION (#5,#6)=(x,y) co-ords;  #7 is linelength (eg#wiggles)
\selectcase
\rearcoords% moved from before selectcase.
\midcoords
\ifnum\phantomswitch=0 \drawas{#1}\fi
% if \phantomswitch=1 then just set the line up and don't draw it.
}

\gdef\startphantom{\phantomswitch=1} % BEGIN PHANTOM MODE.
\gdef\stopphantom{\phantomswitch=0}  % END PHANTOM MODE.
% USE AS: 
% \startphantom...\drawline\gluon[...]...\drawvertex\photon...\stopphantom

\gdef\drawas#1{
\global\savebox{#1}(\boxlengthx,\boxlengthy){
\setlength{\unitlength}{0.01pt}
\begin{picture}(\boxlengthx,\boxlengthy)
\multiput(\upperlineadjx,\upperlineadjy)(\unitboxwidth,\unitboxheight)
{\numupperunits}{\upperunitbox}
\ifnum\numlineparts > 1  %  If the line needs 2 parts per unit or more
\multiput(\lowerlineadjx,\lowerlineadjy)(\unitboxwidth,\unitboxheight)
{\numlowerunits}{\lowerunitbox}  
\fi
\ifnum\numlineparts > 2  %  If the line needs 3 parts per unit or more
\multiput(\thirdlineadjx,\thirdlineadjy)(\unitboxwidth,\unitboxheight)
{\numthirdunits}{\thirdunitbox}  
\fi
\ifnum\numlineparts > 3  %  If the line needs 4 parts per unit or more
\multiput(\fourthlineadjx,\fourthlineadjy)(\unitboxwidth,\unitboxheight)
{\numfourthunits}{\lowerunitbox}  
\fi
\end{picture} }
% CHECK STEMS
\global\PFRONTx=\pfrontx  \global\PFRONTy=\pfronty   %save this value
\SETFRONTSTEM
% Now take into account the possibility of THICK E-W photons (drawn twice)
\THICKPHOTONTEST
\ifdim\THICKPHOTONSWITCH=1pt\global\advance\PFRONTy by 20  \fi
\put(\PFRONTx,\PFRONTy) {\usebox{#1}}   %\pfrontX,Y=\particlefrontx,y
%\put(\particlefrontx,\particlefronty) {\usebox{#1}}
\ifdim\THICKPHOTONSWITCH=1pt
\global\advance\PFRONTy by -40
\put(\PFRONTx,\PFRONTy) {\usebox{#1}}   % The second \E or \W photon ->thicker
\global\advance \PFRONTy by 20  %re-adjust:  advanced by -20 in total above.
\fi  %End of \ifdim\THICKPHOTONSWITCH=1
\SETBACKSTEM
\seglength=1416   \gaplength=850   % Re-set \SCALR defaults.
}
%
% *********   ROUTINES FOR STORING LINES  *******
%

\gdef\drawandsaveline`#1' as #2[#3#4](#5,#6)[#7]{
% #1 is saved linename;   #2 is \LINETYPE;    #3 is \LINEDIRECTION
% #4 is \LINECONFIGURATION (#5,#6)=(x,y) co-ords;  #7 is linelength (eg#wiggles)
\global\newsavebox{#1}
\drawsavedline`#1' as #2[#3#4](#5,#6)[#7]
}

\gdef\drawline#1[#2#3](#4,#5)[#6]{   % Draw line but don't name it.
\drawsavedline`\lastline' as #1[#2#3](#4,#5)[#6]}

\gdef\saveas#1{  %  For saving a line after the fact.
\global\newsavebox#1
\drawas#1}
%
%
%%%%%%%%%%%%%%%%%%%%%%%%%%%%%%%%%%%%%%%%%%%%%%%%%%%%%%%%%%%%%%%%%%%%%%%%
%                                                                      %
%                           C A S E S                                  %
%                           ---------                                  % 
%                                                                      %       
%%%%%%%%%%%%%%%%%%%%%%%%%%%%%%%%%%%%%%%%%%%%%%%%%%%%%%%%%%%%%%%%%%%%%%%%
%
% ERROR MESSAGES FOR INCORRECT CASE SPECIFICATION:
\gdef\TYPEERROR{\message{*** ERROR IN PARTICLE TYPE SELECTION ***}
\message{+++ Try with line type \fermion,\scalar,\photon,\gluon 
(see manual) +++}\SETERR}
\gdef\DIRECTERROR{\SETERR\message{*** ERROR IN PARTICLE DIRECTION SELECTION ***}
\message{+++ Try again with direction N, NE, E, SE  etc. or see manual +++}}
\gdef\UNIMPERROR{\message{*** ERROR IN PARTICLE OPTIONS SELECTION ***}
\message{
+++ The requested options combination has not yet been implemented +++}\SETERR}
\gdef\SETERR{\gdef\upperunitbox{{\tiny Error}}  % PRINTS `error' in diagram.
\gdef\lowerunitbox{\relax}
\gdef\thirdunitbox{\relax}
}
\gdef\neglengthcheck{\ifnum\unitboxnumber < 1 
\message{   *** ERROR:  PARTICLE OF NEGATIVE OR ZERO LENGTH REQUESTED. ***   }
\message{   ***         TAKING ABSOLUTE VALUE. ***   }\negate\unitboxnumber \fi}
%%%%%%%%%%%%%%%%%%%%%%%%%%%%%%%%%%%%%%%%%%%%%%%%%%%%%%%%%%%%%%%%%%%%%%%%%%%%%%%%
\gdef\selectcase{  
\neglengthcheck   %  check for particles of negative length.
% select PARTICLE alignment:
\SETDIR  
%  Select particle type
\ifcase\LINETYPE
\TYPEERROR  % \LINETYPE=0 case.
\or \selectfermion  % \LINETYPE=1 case.
\or \selectscalar   % \LINETYPE=2 case.
\or \selectphoton   % \LINETYPE=3 case.
\or \selectgluon    % \LINETYPE=4 case.
\or \selectespecial % \LINETYPE=5 case.
\else \TYPEERROR \fi  }
%%%%%%%%%%% (1) FERMIONS %%%%%%%%%%% 
\gdef\selectfermion{
% Input fermion-setup stuff ONLY IF HAVE NOT DONE SO YET.
% This avoids having to process a fermion if none are drawn.
\ifnum\fermioncount=0 \input fermionsetup \fi   
%                  CONTAINS fermion DEFINITIONS.
\global\advance\fermioncount by 1  % Counts number of fermions drawn. 
\ALLfermion   
}
%%%%%%%%%%% (2) SCALARS %%%%%%%%%%% 
\gdef\selectscalar{
% Input scalar-setup stuff ONLY IF HAVE NOT DONE SO YET.
% This avoids having to process a scalar if none are drawn.
\ifnum\scalarcount=0 \input scalarsetup \fi   
%                 CONTAINS scalar DEFINITIONS.
\global\advance\scalarcount by 1  % Counts number of scalars drawn. 
\ALLscalar
}
%%%%%%%%%%% (3) PHOTONS %%%%%%%%%%% 
\gdef\selectphoton{   % RECURSIVELY RE-DEFINED IN PHOTONSETUP(23+).TEX.
% Input photon-setup stuff ONLY IF HAVE NOT DONE SO YET.
% This avoids having to process a photon if none are drawn.
\ifnum\photoncount=0 \input photonsetup  \fi   
\selectphoton
%CONTAINS PHOTON DEFINITIONS. 
}
%%%%%%%%%%% (4) GLUONS %%%%%%%%%%% 
\gdef\selectgluon{   % RECURSIVELY RE-DEFINED IN GLUONSETUP(25+).TEX.
% Input gluon-setup stuff ONLY IF HAVE NOT DONE SO YET.
% This avoids having to process a gluon if none are drawn.
\ifnum\gluoncount=0 \input gluonsetup  \fi
\selectgluon
%                  CONTAINS gluon DEFINITIONS.
}
%%%%%%%%%%% (5) SPECIAL - USER DEFINED %%%%%%%%%%% 
\gdef\selectespecial{\UNIMPERROR}
%
%%%%%%%%%%%%%%%%%%%%%%%%%%%%%%%%%%%%%%%%%%%%%%%%%%%%%%%%%%%%%%%%%%%%%%%%%%%%%%%
%                                                                             %
%                   ROUTINES FOR DRAWING VERTICES                             %
%                                                                             %
%%%%%%%%%%%%%%%%%%%%%%%%%%%%%%%%%%%%%%%%%%%%%%%%%%%%%%%%%%%%%%%%%%%%%%%%%%%%%%%
%
% Input vertex-setup stuff ONLY IF HAVE NOT DONE SO YET.                      %
% This avoids having to process a vertex if none are drawn. 
\gdef\checkvertex{ %immediately re-defines \drawvertex,\vertexcap,\linkvertex...
\ifnum\vertexcount=-1   \input vertex  \fi}
% RECURSIVE DEFINITIONS:
\gdef\drawvertex#1[#2#3](#4,#5)[#6]{\checkvertex\drawvertex#1[#2#3](#4,#5)[#6]}
\gdef\vertexcap#1{\checkvertex\vertexcap#1}
\gdef\vertexcaps{\checkvertex\vertexcaps}
\gdef\vertexlink#1{\checkvertex\vertexlink#1}
\gdef\vertexlinks{\checkvertex\vertexlinks}
\gdef\stemvertex#1{\checkvertex\stemvertex#1}
\gdef\stemvertices{\checkvertex\stemvertices}
\gdef\flipvertex{\checkvertex\flipvertex}
%
%%%%%%%%%%%%%%%%%%%%%%%%%%%%%%%%%%%%%%%%%%%%%%%%%%%%%%%%%%%%%%%%%%%%%%%%%%%%%%%
%                                                                             %
%                   ROUTINES FOR DRAWING ARROWS                               %
%                                                                             %
%%%%%%%%%%%%%%%%%%%%%%%%%%%%%%%%%%%%%%%%%%%%%%%%%%%%%%%%%%%%%%%%%%%%%%%%%%%%%%%
%
% SYNTAX:  \drawarrow[\NW\ATBASE](\pmidx,\pmidy)  etc.
\global\arrowlength=349  % Length of arrow
\gdef\drawarrow[#1#2](#3,#4){
\global\LDIR=#1
\SETDIR
\global\boxlengthx=#3  % Just a convenient variable name.  No significance.
\global\boxlengthy=#4  % The Arrow co-ordinates.
\ifdim#2=1pt  % CASE \ATBASE WHERE THE CO-ORDS ARE AT THE ARROWS BASE.
   %   #2 IS either \ATTIP or \ATBASE...Depending where it is to be positioned.
\adjx=\arrowlength      \adjy=\arrowlength
\multiply\adjx by \XDIR \multiply\adjy by \YDIR  % Set in \SETDIR
\slanttest(\adjx,\adjy)
\global\advance\boxlengthx by \adjx    \global\advance\boxlengthy by \adjy
\fi
\ifnum\phantomswitch=0\put(\boxlengthx,\boxlengthy){\vector(\XDIR,\YDIR){0}}\fi
}  % END OF \drawarrow.
%
%%%%%%%%%%%%%%%%%%%%%%%%%%%%%%%%%%%%%%%%%%%%%%%%%%%%%%%%%%%%%%%%%%%%%%%%%%%%%%%
%                                                                             %
%                     ROUTINES FOR DRAWING STEMS                              %
%                                                                             %
%%%%%%%%%%%%%%%%%%%%%%%%%%%%%%%%%%%%%%%%%%%%%%%%%%%%%%%%%%%%%%%%%%%%%%%%%%%%%%%
%
\gdef\SETFRONTSTEM{
\EITHERSTEM=\FRONTSTEM   \advance\EITHERSTEM by \BACKSTEM
\ifdim\EITHERSTEM>0pt
\global\stemlengthx=\stemlength   \global\stemlengthy=\stemlength   
\global\absstemlength=\stemlength   
\SETDIR
\gslanttest(\stemlengthx,\stemlengthy)
\gslanttest(\absstemlength,\REG)  % the \REG is to use up the parameter space.
\ifnum\XDIR=0 \stemlengthx=0 \fi
\ifnum\YDIR=0 \stemlengthy=0 \fi
\global\multiply\stemlengthx by \XDIR
\global\multiply\stemlengthy by \YDIR
\ifdim\FRONTSTEM=1pt 
\ifnum\phantomswitch=0
          \put(\pfrontx,\pfronty){\line(\XDIR,\YDIR){\absstemlength}}\fi
\global\advance\plengthx by \stemlengthx
\global\advance\plengthy by \stemlengthy
\global\advance\PFRONTx by \stemlengthx   
\global\advance\PFRONTy by \stemlengthy
\global\advance\pmidx by \stemlengthx
\global\advance\pmidy by \stemlengthy
\global\advance\pbackx by \stemlengthx
\global\advance\pbacky by \stemlengthy
% FOR STEMMED PHOTONS AND GLUONS, \photonfront,back(x,y) are for the
% photon proper (no stem) while \pbackx,y include the stems.
\ifnum\LTYPE=3
\global\photonfrontx=\PFRONTx  \global\photonfronty=\PFRONTy
\global\photonbackx=\pbackx    \global\photonbacky=\pbacky
\fi  % END LTYPE
\ifnum\LTYPE=4
\global\gluonfrontx=\PFRONTx  \global\gluonfronty=\PFRONTy
\global\gluonbackx=\pbackx    \global\gluonbacky=\pbacky
\fi  % END LTYPE
\fi  % END FRONTSTEM
\fi  % END EITHERSTEM
}    % end of \SETFRONTSTEM
\gdef\SETBACKSTEM{
\ifdim\BACKSTEM=1pt 
\ifnum\phantomswitch=0
       \put(\pbackx,\pbacky){\line(\XDIR,\YDIR){\absstemlength}}\fi
\global\advance\plengthx by \stemlengthx
\global\advance\plengthy by \stemlengthy
\global\advance\pbackx by \stemlengthx
\global\advance\pbacky by \stemlengthy
\fi  % END BACKSTEM
\global\stemlength=275  \FRONTSTEM=0pt  \BACKSTEM=0pt % Reset default switches.
}    % END OF \SETBACKSTEM 
%%%%%%%%%%%%%%%%%%%%%%%%%%%%%%%%%%%%%%%%%%%%%%%%%%%%%%%%%%%%%%%%%%%%%%%%%%%%%
%                              LOOPS                                        %
%%%%%%%%%%%%%%%%%%%%%%%%%%%%%%%%%%%%%%%%%%%%%%%%%%%%%%%%%%%%%%%%%%%%%%%%%%%%%
\gdef\drawloop#1[#2#3](#4,#5){  %RECURSIVE.  
\input loops  % contains loops definitions
\drawloop#1[#2#3](#4,#5)}
%%%%%%%%%%%%%%%%%%%%%%%%%%%%%%%%%%%%%%%%%%%%%%%%%%%%%%%%%%%%%%%%%%%%%%%%%%%%%
\Feynmanlength  % Set length scale to centipoints.

%% file: fermionsetup.tex
%                        FERMIONSETUP(7).TEX
%  CALLED BY FEYNMAN(34).TEX.
% USED FOR GENERATING FERMION LINES IN FEYNMAN DIAGRAMS IN LATEX.
\global\newcount\fermionlength  %  THE TOTAL FERMION LINE LENGTH.
\global\newcount\fermionlengthx
\global\newcount\fermionlengthy
\global\newcount\fermionfrontx  %}(x,y) co-ord of left of fermion
\global\newcount\fermionfronty  %}
\global\newcount\fermionbackx
\global\newcount\fermionbacky
%%%%%%%%%%%%%%%%%%%%%%%%%%%%%%%%%%%%%%%%%%%%%%%%%%%%%%%%%%%%%%%%%%%%%%%%%%%
\gdef\ALLfermion{  % READ IN FROM FEYNMAN \selectfermion
\global\fermionfrontx=\particlefrontx \global\fermionfronty=\particlefronty
% Error messages for overly-long lines.  See FEYNMAN for negative-lengths.
\ifnum\unitboxnumber > 50000
\message{   *** WARNING *** Fermion of length
\the\unitboxnumber\space requested ***   }
\ifnum\unitboxnumber > 80000
\message{   *** Reducing fermion length to 30000 (max 80000) ***   }
\global\unitboxnumber=30000 \fi \fi  % end of length error
\global\fermionlength=\unitboxnumber % The TOTAL line length
\global\particleadjustx=0   \global\particleadjusty=0 %Default
\global\numlineparts = 1    \global\numupperunits=1
\global\upperlineadjx=-200  \global\upperlineadjy=0
\global\fermionlengthx=\fermionlength    \global\fermionlengthy=\fermionlength
\gslanttest(\fermionlengthx,\fermionlengthy)  % See FEYNMAN22.TEX (FOR \XDIR).
\global\multiply\fermionlengthx by \XDIR  %  In keeping with photons and gluons.
\global\multiply\fermionlengthy by \YDIR  %  In keeping with photons and gluons.
\global\unitboxheight=\fermionlengthy   \global\unitboxwidth=\fermionlengthx   
\global\advance \fermionlengthx by \particleadjustx
\global\advance \fermionlengthy by \particleadjusty
\global\particlelengthx=\fermionlengthx
\global\particlelengthy=\fermionlengthy  
\boxlengthdefault    \rearcoords    \midcoords
\global\fermionbackx=\particlebackx     \global\fermionbacky=\particlebacky
\ifcase\LINECONFIGURATION  %\REG case
\ifnum\XDIR=0 
\gdef\upperunitbox{\line(\XDIR,\YDIR){\boxlengthy}} %\N or \S
\else
\gdef\upperunitbox{\line(\XDIR,\YDIR){\boxlengthx}}
\fi
\else \UNIMPERROR
\fi
}

%% file: photonsetup.tex
%                            PHOTONSETUP(28).TEX
% CALLED BY FEYNMAN(34).TEX.
% USED FOR GENERATING PHOTON LINES IN FEYNMAN DIAGRAMS IN LATEX.
\newcount\numwiggles    \newcount\numwigglespo
\global\newcount\photonlengthx
\global\newcount\photonlengthy
\global\newcount\photonfrontx  %}(x,y) co-ord of left of photon
\global\newcount\photonfronty  %}
\global\newcount\photonbackx
\global\newcount\photonbacky
\newcount\halfwigglelength
\global\font\Twelverom=cmr12
\global\font\Tenrom=cmr10
\gdef\Lbr{{\Twelverom(}}   \gdef\Rbr{{\Twelverom)}}
\gdef\SLbr{{\Tenrom(}}     \gdef\SRbr{{\Tenrom)}}
%  Want \smile,\frown to always be 12-point but won't work!
\gdef\Smile{{\large$\smile$}}  % Default for 10 and 11-point documents.
\gdef\Frown{{\large$\frown$}}  % Default for 10 and 11-point documents.
\ifdim\BIGPHOTONS>0pt  \gdef\Smile{$\smile$} \gdef\Frown{$\frown$} \fi
%  For use with 12-point documents only.  Invoked by saying \bigphotons.
%
%%%%%%%%%%%%%%%%%%%%%%%%%%%%%%%%%%%%%%%%%%%%%%%%%%%%%%%%%%%%%%%%%%%%%%%%%%%
\gdef\selectphoton{   % RECURSIVELY RE-DEFINED HERE.  Define in FEYNMAN.
\global\advance\photoncount by 1  % Counts number of photons drawn. 
\global\photonfrontx=\particlefrontx   % READ IN FROM FEYNMAN \selectphoton
\global\photonfronty=\particlefronty   % READ IN FROM FEYNMAN \selectphoton
% Error messages for overly-long lines.  See FEYNMAN for negative-lengths.
\ifnum\unitboxnumber > 50
\message{   *** WARNING *** Photon with 
\the\unitboxnumber\space half-wiggles requested ***   }
\ifnum\unitboxnumber > 150
\message{   *** Reducing photon length to 10 half-wiggles (max 150) ***   }
\ifnum\unitboxnumber > 1000
\message{   *** Probable Cause:  Photon selected instead of Fermion ***   }
\fi \global\unitboxnumber=10 \fi \fi  % end of length error
\numwiggles=\unitboxnumber
\divide\numwiggles by 2
\global\unitboxnumberpo=\numwiggles % here \unitboxnumberpo is an unused counter
\global\multiply \unitboxnumberpo by -1
\numwigglespo=\unitboxnumber
\advance\numwigglespo by \unitboxnumberpo %\numwigglespo is one greater than 
\global\numlineparts = 2  % DEFAULT                %\numwiggles in this case.
\global\numupperunits=\numwigglespo  % DEFAULT
\global\numlowerunits=\numwiggles  % DEFAULT
\particleadjustx=0  %DEFAULT
\particleadjusty=0  %DEFAULT
% select photon alignment:
\ifcase\LINEDIRECTION
     \Nphoton    %\LINEDIRECTION=0 (NORTH) CASE
\or  \NEphoton   % 1 case
\or  \Ephoton    % 2 case...horizontal photon.
\or  \SEphoton   % .
\or  \Sphoton    % .
\or  \SWphoton   % .
\or  \Wphoton    % .
\or  \NWphoton   % 7 case
\else\DIRECTERROR \fi
\setplength
\global\divide\plengthx by 2  \global\divide\plengthy by 2
\rearcoords  \boxlengthdefault   \midcoords
\global\photonbackx=\pbackx  %PHOTONSETUP26
\global\photonbacky=\pbacky  %PHOTONSETUP26
\global\photonlengthx=\plengthx  %PHOTONSETUP26
\global\photonlengthy=\plengthy  %PHOTONSETUP26
}
%%%%%%%%%%%%%%%%%%%%%%%%%%%%%%%%%%%%%%%%%%%%%%%%%%%%%%%%%%%%%%%%%%%%%%%%%%%
\gdef\SETUNITBOX(#1)[#2][#3]{ % For slanted photons only.
\gdef\upperunitbox{\oval(#1,#1)[#2]}
\gdef\lowerunitbox{\oval(#1,#1)[#3]}
}
%%%%%%%%%%%%%%%%%%%%%%%%%%%%%%%%%%%%%%%%%%%%%%%%%%%%%%%%%%%%%%%%%%%%%%%%%%%
\gdef\Nphoton{  % VERTICAL PHOTONS
\ifcase\LINECONFIGURATION  %\REG case
\setcoords(-490,-250,0)(260,1250,0)[0,2000]
\gdef\upperunitbox{\SLbr}   \gdef\lowerunitbox{\SRbr}
\particleadjusty=10
\or % \FLIPPED case
\setcoords(-271,-501,0)(250,1250,0)[0,2000]   
\gdef\upperunitbox{\SRbr}   \gdef\lowerunitbox{\SLbr}
\or %\CURLY case (a bit shorter).
\particleadjusty=0
\setcoords(-501,-351,0)(300,1400,0)[0,2200]
\gdef\upperunitbox{\Lbr}   \gdef\lowerunitbox{\Rbr}
\or %\FLIPPEDCURLY case.
\setcoords(-353,-499,0)(300,1400,0)[0,2200]
\gdef\upperunitbox{\Rbr}   \gdef\lowerunitbox{\Lbr}
\or % \FLAT case.  Flatter and shorter than \CURLY.
\setcoords(-481,-371,0)(280,1300,0)[0,2000]
\gdef\upperunitbox{\Lbr}   \gdef\lowerunitbox{\Rbr}
\particleadjusty=150
\ifnum\numwiggles=\number\numwigglespo \particleadjustx=-50 \fi
\or %\FLIPPEDFLAT case.  \LINECONFIGURATION=5.
\setcoords(-321,-391,0)(280,1300,0)[0,2000]
\gdef\upperunitbox{\Rbr}   \gdef\lowerunitbox{\Lbr}
\particleadjusty=150
\ifnum\numwiggles=\number\numwigglespo \particleadjustx=80 \fi
\or % \LONGPHOTON
\setcoords(-490,-260,0)(300,1500,0)[0,2400]
\gdef\upperunitbox{\Lbr}   \gdef\lowerunitbox{\Rbr}
\or % \FLIPPEDLONGPHOTON
\setcoords(-301,-531,0)(300,1500,0)[0,2400]
\gdef\upperunitbox{\Rbr}   \gdef\lowerunitbox{\Lbr}
\else \UNIMPERROR
\fi
}
%%%%%%%%%%%%%%%%%%%%%%%%%%%%%%%%%%%%%%%%%%%%%%%%%%%%%%%%%%%%%%%%%%%%%%%%%%%
\gdef\NEphoton{    % NE   SLANTED PHOTONS:  RE-ORDERED IN PHOTONSETUP27
\ifcase\LINECONFIGURATION  %\REG case
\setcoords(425,425,0)(1250,0,0)[1250,1250]       \SETUNITBOX(1250)[br][tl]  
\ifnum\numwigglespo > \number \numwiggles \particleadjustx=15 \fi
\or % \FLIPPED case
\setcoords(1050,-200,0)(625,625,0)[1250,1250]    \SETUNITBOX(1250)[tl][br]
\ifnum\numwigglespo > \number \numwiggles \particleadjustx=25 \fi
\or % \CURLY case.
\setcoords(500,500,0)(1400,0,0)[1400,1400]       \SETUNITBOX(1400)[br][tl]
\or % \FLIPPEDCURLY case
\setcoords(1200,-200,0)(700,700,0)[1400,1400]    \SETUNITBOX(1400)[tl][br]  
\or % \FLAT case
\setcoords(400,400,0)(1200,0,0)[1200,1200]       \SETUNITBOX(1200)[br][tl]  
\or % \FLIPPEDFLAT case
\setcoords(1000,-200,0)(600,600,0)[1200,1200]    \SETUNITBOX(1200)[tl][br]
\else \UNIMPERROR
\fi
\numupperunits=\numwiggles   \numlowerunits=\numwigglespo
}
%%%%%%%%%%%%%%%%%%%%%%%%%%%%%%%%%%%%%%%%%%%%%%%%%%%%%%%%%%%%%%%%%%%%%%%%%%%
\gdef\Ephoton{    %  EASTWARD  HORIZONTAL PHOTONS
\ifcase\LINECONFIGURATION  % REG case
\setcoords(-285,715,0)(-150,-400,0)[2005,0]
\gdef\upperunitbox{\Frown}   \gdef\lowerunitbox{\Smile}
\or  % \FLIPPED case
\setcoords(-285,715,0)(-420,-170,0)[2005,0]
\gdef\upperunitbox{\Smile}   \gdef\lowerunitbox{\Frown}
\else \UNIMPERROR
\fi
\particleadjustx=-15 % Lengths are in centipoints.
}
%%%%%%%%%%%%%%%%%%%%%%%%%%%%%%%%%%%%%%%%%%%%%%%%%%%%%%%%%%%%%%%%%%%%%%%%%%%
\gdef\SEphoton{   % SE   SLANTED PHOTONS:  RE-ORDERED IN PHOTONSETUP27
\ifcase\LINECONFIGURATION  %\REG case
\setcoords(-200,1050,0)(-625,-625,0)[1250,-1250] \SETUNITBOX(1250)[tr][bl]
\ifnum\numwigglespo > \number \numwiggles \particleadjustx=25 \fi
\or % \FLIPPED case
\setcoords(425,425,0)(0,-1250,0)[1250,-1250]     \SETUNITBOX(1250)[bl][tr]
\ifnum\numwigglespo > \number \numwiggles \particleadjustx=15 \fi
\or % \CURLY case.
\setcoords(-200,1200,0)(-700,-700,0)[1400,-1400] \SETUNITBOX(1400)[tr][bl]  
\or % \FLIPPEDCURLY case
\setcoords(500,500,0)(0,-1400,0)[1400,-1400]     \SETUNITBOX(1400)[bl][tr]  
\or % \FLAT case
\setcoords(-200,1000,0)(-600,-600,0)[1200,-1200] \SETUNITBOX(1200)[tr][bl]
\particleadjustx=-20
\or % \FLIPPEDFLAT case
\setcoords(420,420,0)(0,-1200,0)[1200,-1200]     \SETUNITBOX(1200)[bl][tr]
\particleadjustx=40
\else \UNIMPERROR
\fi
}
%%%%%%%%%%%%%%%%%%%%%%%%%%%%%%%%%%%%%%%%%%%%%%%%%%%%%%%%%%%%%%%%%%%%%%%%%%%
\gdef\Sphoton{  % DOWN, DOWN VERTICAL PHOTONS
\ifcase\LINECONFIGURATION  %\REG case
\setcoords(-252,-490,0)(-740,-1740,0)[0,-2000]
\gdef\upperunitbox{\SRbr}   \gdef\lowerunitbox{\SLbr}
\or % \FLIPPED case
\setcoords(-490,-260,0)(-740,-1740,0)[0,-2002]
\gdef\upperunitbox{\SLbr}   \gdef\lowerunitbox{\SRbr}
\or %\CURLY case (a bit shorter).
\setcoords(-299,-449,0)(-870,-1970,0)[0,-2200]
\gdef\upperunitbox{\Rbr}    \gdef\lowerunitbox{\Lbr}
\particleadjusty=-95
\or %\FLIPPEDCURLY case.
\setcoords(-517,-371,0)(-900,-2000,0)[0,-2200]
\gdef\upperunitbox{\Lbr}    \gdef\lowerunitbox{\Rbr}
\particleadjusty=-165
\or % \FLAT case.  Flatter and shorter than \CURLY.  \LINECONFIGURATION=4.
\setcoords(-299,-409,0)(-885,-1905,0)[0,-2000]
\gdef\upperunitbox{\Rbr}   \gdef\lowerunitbox{\Lbr}
\particleadjustx=50     \particleadjusty=-380
\ifodd\unitboxnumber\relax\else\particleadjustx=250 \particleadjusty=-400 \fi
\or %\FLIPPEDFLAT case.  \LINECONFIGURATION=5.
\setcoords(-519,-449,0)(-900,-1920,0)[0,-2000]
\gdef\upperunitbox{\Lbr}   \gdef\lowerunitbox{\Rbr}
\particleadjusty=-370
\ifodd\unitboxnumber\relax\else\particleadjustx=-240 \particleadjusty=-400 \fi
\or % \LONGPHOTON
\gdef\upperunitbox{\Rbr}   \gdef\lowerunitbox{\Lbr}
\setcoords(-325,-555,0)(-900,-2100,0)[0,-2400]
\particleadjusty=-40
\or % \FLIPPEDLONG
\setcoords(-505,-275,0)(-900,-2100,0)[0,-2400]
\gdef\upperunitbox{\Lbr}   \gdef\lowerunitbox{\Rbr}
\particleadjusty=-30  % Lengths are in centipoints.
\else \UNIMPERROR
\fi
}
%%%%%%%%%%%%%%%%%%%%%%%%%%%%%%%%%%%%%%%%%%%%%%%%%%%%%%%%%%%%%%%%%%%%%%%%%%%
\gdef\SWphoton{  % SW SLANTED PHOTONS:  RE-ORDERED IN PHOTONSETUP27
\ifcase\LINECONFIGURATION  %\REG case
\setcoords(-825,-825,0)(0,-1250,0)[-1250,-1250]     \SETUNITBOX(1250)[br][tl]  
\or % \FLIPPED case
\setcoords(-175,-1425,0)(-625,-625,0)[-1250,-1250]  \SETUNITBOX(1250)[tl][br]  
\or % \CURLY case.
\setcoords(-900,-900,0)(0,-1410,0)[-1400,-1400]     \SETUNITBOX(1400)[br][tl]  
\or % \FLIPPEDCURLY case
\setcoords(-200,-1600,0)(-700,-700,0)[-1400,-1400]  \SETUNITBOX(1400)[tl][br]  
\or % \FLAT case
\setcoords(-800,-800,0)(0,-1200,0)[-1200,-1200]     \SETUNITBOX(1200)[br][tl]  
\or % \FLIPPEDFLAT case
\setcoords(-200,-1400,0)(-600,-600,0)[-1200,-1200]  \SETUNITBOX(1200)[tl][br]  
\else \UNIMPERROR
\fi
}
%%%%%%%%%%%%%%%%%%%%%%%%%%%%%%%%%%%%%%%%%%%%%%%%%%%%%%%%%%%%%%%%%%%%%%%%%%%
\gdef\Wphoton{
\ifcase\LINECONFIGURATION %\REG case
\setcoords(-2245,-1245,0)(-150,-400,0)[-2005,0]
\gdef\upperunitbox{\Frown}   \gdef\lowerunitbox{\Smile}
\or % \FLIPPED case
\setcoords(-2245,-1245,0)(-400,-150,0)[-2005,0]
\gdef\upperunitbox{\Smile}   \gdef\lowerunitbox{\Frown}
\else \UNIMPERROR
\fi
\particleadjustx=57 % Lengths are in centipoints.
\ifnum\numwigglespo=\number\numwiggles \particleadjustx=0  \fi
\numlowerunits=\numwigglespo   \numupperunits=\numwiggles
}
%%%%%%%%%%%%%%%%%%%%%%%%%%%%%%%%%%%%%%%%%%%%%%%%%%%%%%%%%%%%%%%%%%%%%%%%%%%
\gdef\NWphoton{  % NW   SLANTED PHOTONS:  RE-ORDERED IN PHOTONSETUP27
\ifcase\LINECONFIGURATION  %\REG case
\setcoords(-200,-1425,0)(625,625,0)[-1250,1250]   \SETUNITBOX(1250)[bl][tr]
\or % \FLIPPED case
\setcoords(-825,-825,0)(0,1250,0)[-1250,1250]     \SETUNITBOX(1250)[tr][bl]
\ifnum\numwigglespo > \number \numwiggles \particleadjusty=-15 \fi
\or % \CURLY case.
\setcoords(-200,-1600,0)(700,700,0)[-1400,1400]   \SETUNITBOX(1400)[bl][tr]
\or % \FLIPPEDCURLY case
\setcoords(-900,-900,0)(0,1400,0)[-1400,1400]     \SETUNITBOX(1400)[tr][bl]
\or % \FLAT case.
\setcoords(-200,-1400,0)(600,600,0)[-1200,1200]   \SETUNITBOX(1200)[bl][tr]  
\or % \FLIPPEDFLAT case
\setcoords(-800,-800,0)(0,1200,0)[-1200,1200]     \SETUNITBOX(1200)[tr][bl]  
\else \UNIMPERROR
\fi
}

%% file: intro.tex
\chapter{Leptons from Collisions}
 
\section{Definition}
\par
The major goal of the high energy particle colliders is to investigate the structure of the matter at very short distances. The signature of the processes that access the shortest distances is the large transverse momenta of the produced particles in the final state. In addition, the creation of new particles, of different type form the colliding particles, provides an useful tool for investigating the nature of the high energy phenomena.
\par
During the collision process, the beam particles can act as composed systems. The main interaction, also called ``hard scattering'', takes place actually between two constituents or ``partons''. This is obvious for the case of the proton, for which most phenomena occuring in collisions can be explained by the proton's quarks or gluons. Partially this is also the case for the electron or even for the photon, as will be explained below.
\par
In particle collisions, the primary (or ``hard'') processes can be classified in several cases.
\begin{itemize}
\item[{\bf A}] ``elastic'' scattering of two elementary partons   $ab\rightarrow ab$ (for instance the Bhabha scattering   $e^+e^- \rightarrow e^+e^-$ or the deep inelastic scattering $eq\rightarrow eq$).
\item[{\bf B}]  ``quasi-elastic hard scattering'' where the two initial partons are also in the final state, but new particles are created $ab\rightarrow ab cd$. This is the case for instance in the four fermion production at LEP if the electron-positron pair is found in the final state: $e^+e^- \ra e^+e^- f\bar{f}$.
\item[{\bf C}] ``inelastic hard scattering'' where the initial state partons are transformed into new particles like for instance $ab\rightarrow a'b'$ (for instance the Drell--Yan process $q\bar{q} \ra \ell^+\ell^-$), with possibly multiplication of final state particles by the mechanism related to the previous point: $ab\rightarrow a'b' cd$ (for instance $e^+e^-\ra \mu^+\mu^- q\bar{q}$).
\end{itemize}

 In this paper we are going to review processes that yield new particles different from the colliding partons (processes of type {\bf B} and {\bf C}). In addition we focus on the production of leptons in the final state.
 These processes are defined by the condition:
\begin{center}
{\Large \bf
$\Delta N_L=N_L^f-N_L^i>0$,  }
\end{center}
where $N_L^i$ ($N_L^f$) is the number of charged or neutral leptons in the initial (final) state.
\par
The production of isolated leptons at high transverse momentum in high energetic collisions is particularly interesting due to the clean experimental signature. The production mechanisms in the Standard Model framework include weak boson decays or boson-boson fusion, as  will be described below. 
\newpage
\par
The high $P_T$ lepton production reveals important information related to the matter structure at short distances:
\begin{itemize}
\item gauge structure and coupling strengths of the Standard Model;
\item composed particles structure: proton structure and study of hadronic fluctuation of the photon;
\item search for physics beyond the Standard Model.
\end{itemize}
In this paper we summarize the lepton production in high energy collisions at present colliders. In the following, we will shortly present the production mechanisms and the experimental setups at  LEP, HERA and Tevatron. 
\section{Basic lepton production mechanisms in the \sm}
\label{sec:prod-lep}
In the \sm leptons couple to the photon and to the weak boson $W$ and $Z$ which in turn couple to initial state partons that can be of different nature than the produced leptons. It is important to note that leptons can only be produced in pairs in the \sm framework. The leptonic number is conserved and  $\Delta N_\ell$ can only be an even integer.
\par
   There are two production mechanisms that may produce a pair of leptons.
\par {\bf boson  conversion}\par
A pair of leptons can be issued from a real or virtual boson $\gamma^*,W^{(*)},Z^{(*)}$ or by the decays of an on-shell weak boson $W$ or $Z$. The lepton pair production at sufficient high energies is the natural signature for weak boson production and used in the experimental analyses of the weak bosons properties.  The lepton--boson couplings to charged or neutral bosons are described below~\cite{leader}. \\ The charged current coupling is given by: \\
\vspace{0.3cm}
\par
\begin{picture}(10000,8000)
\drawline\photon[\E\REG](8000,4000)[4]
\put(4000,\photonfronty){$W^{\pm}_{\mu}$}
\drawline\fermion[\NE\REG](\photonbackx,\photonbacky)[\photonlengthx]
\drawarrow[\NE\ATTIP](\pmidx,\pmidy)
\global\advance\fermionbackx by 500
\put(\fermionbackx,\fermionbacky){$\ell^-$ or $\nu_e$}
\drawline\fermion[\SE\REG](\photonbackx,\photonbacky)[\photonlengthx]
\drawarrow[\NW\ATTIP](\pmidx,\pmidy)
\global\advance\fermionbackx by 500
\put(\fermionbackx,\fermionbacky){$\bar{\nu}_\ell$ or $\ell^+$}
\global\advance\photonbackx by \plengthx
\global\advance\photonbackx by \plengthx
\global\advance\photonbackx by \plengthx
\put(\photonbackx,\photonbacky){$ie\gamma_{\mu}(1-\gamma_{5})\displaystyle \frac{1}
{2\sqrt{2}\sin \theta_{W}}$}
\end{picture}
\\
$\ell$ is $e$, $\mu$ or $\tau$. Due to the $1-\gamma_5$ factor, only left-handed leptons are created or annihilated at this vertex. There is no charged current interaction between leptons of different flavours, which correspond to no mixing in the leptonic sector, in contrast to the quark sector where the Cabibbo-Kobayashi-Maskawa matrix element is an extra factor in the coupling. This assumption was based on the fact that neutrinos were believed to have no mass.  The recent experimental evidence for non-zero neutrinos masses opens the possibility of some mixing also in the leptonic sector.
\par
The photon couples to the electrical charge of the lepton:
\\
\begin{picture}(10000,8000)
\drawline\photon[\E\REG](8000,4000)[4]
\put(4000,\photonfronty){$\gamma_{\mu}$}
\drawline\fermion[\NE\REG](\photonbackx,\photonbacky)[\photonlengthx]
\drawarrow[\NE\ATTIP](\pmidx,\pmidy)
\global\advance\fermionbackx by 500
\put(\fermionbackx,\fermionbacky){$f$}
\drawline\fermion[\SE\REG](\photonbackx,\photonbacky)[\photonlengthx]
\drawarrow[\NW\ATTIP](\pmidx,\pmidy)
\global\advance\fermionbackx by 500
\put(\fermionbackx,\fermionbacky){$\bar f$}
\global\advance\photonbackx by \plengthx
\global\advance\photonbackx by \plengthx
\global\advance\photonbackx by \plengthx
\put(\photonbackx,\photonbacky){$-ieQ_{f}\gamma_{\mu}$}
\end{picture}
\\
The weak interaction by neutral current is described by:
\\
\begin{picture}(10000,8000)
\drawline\photon[\E\REG](8000,4000)[4]
\put(4000,\photonfronty){$Z^{0}_{\mu}$}
\drawline\fermion[\NE\REG](\photonbackx,\photonbacky)[\photonlengthx]
\drawarrow[\NE\ATTIP](\pmidx,\pmidy)
\global\advance\fermionbackx by 500
\put(\fermionbackx,\fermionbacky){$f$}
\drawline\fermion[\SE\REG](\photonbackx,\photonbacky)[\photonlengthx]
\drawarrow[\NW\ATTIP](\pmidx,\pmidy)
\global\advance\fermionbackx by 500
\put(\fermionbackx,\fermionbacky){$\bar f$}
\global\advance\photonbackx by \plengthx
\global\advance\photonbackx by \plengthx
\global\advance\photonbackx by \plengthx
\put(\photonbackx,\photonbacky){$ie\gamma_{\mu}(v_{f}-a_{f}\gamma_{5})$}
\end{picture}
\\
where the coupling constants are : \\
$$
\begin{array}{ll}
v_{f} = \displaystyle \frac{I_{3}-2Q_{f}\sin^{2}\theta_{W}}{2\sin\theta_{W}\cos\theta_{W}},&
\;\; a_{f} = \displaystyle \frac{I_{3}}{2\sin\theta_{W}\cos\theta_{W}}
\end{array}
$$
\\
with $\theta_{W}$ the weak mixing angle, $Q_f$ fermion charge and  $I_3$
the third component of the weak isospin ($-1/2$ for charged leptons and $+1/2$ for the neutrinos). 
In the \sm the neutral currents do not mix flavours, an assumption based on present experimental non-observation of processes with flavour changing neutral currents (FCNC). The neutral current weak interactions trough a $Z$ involve a superposition of left-- and right--handed charged leptons. This leads to interesting helicity effects in the experimental observables like charge/spin asymmetries.
\par {\bf boson-boson fusion } \par
In this case two bosons merge to produce a pair of leptons as showed below. The most common mechanism in this case is the photon--photon collision. The photon fusion with a weak boson is not favoured within the \sm due to the large weak boson mass. 
\par
\begin{picture}(10000,8000)
\drawline\photon[\E\REG](8000,4000)[4]
\put(4000,\photonfronty){$\gamma(Z)$}
\drawline\fermion[\NE\REG](\photonbackx,\photonbacky)[\photonlengthx]
\drawarrow[\NE\ATTIP](\pmidx,\pmidy)
\global\advance\fermionbackx by 500
\put(\fermionbackx,\fermionbacky){$f$}
\drawline\fermion[\S\REG](\photonbackx,\photonbacky)[3000]
\global\advance\photonbackx by 10000
\drawarrow[\N\ATTIP](\pmidx,\pmidy)
\drawline\photon[\E\REG](8000,1000)[4]
\put(4000,\photonfronty){$\gamma(Z)$}
\drawline\fermion[\SE\REG](\photonbackx,\photonbacky)[3000]
\drawarrow[\NW\ATTIP](\pmidx,\pmidy)
\global\advance\fermionbackx by 500
\put(\fermionbackx,\fermionbacky){$\bar f$}
\drawline\photon[\E\REG](28000,4000)[4]
\put(24000,\photonfronty){$W^{\pm}_{\mu}$}
\drawline\fermion[\NE\REG](\photonbackx,\photonbacky)[\photonlengthx]
\drawarrow[\NE\ATTIP](\pmidx,\pmidy)
\global\advance\fermionbackx by 500
\put(\fermionbackx,\fermionbacky){$f_{1}$}
\drawline\fermion[\S\REG](\photonbackx,\photonbacky)[3000]
\global\advance\photonbackx by 10000
\drawarrow[\N\ATTIP](\pmidx,\pmidy)
\drawline\photon[\E\REG](28000,1000)[4]
\put(24000,\photonfronty){$\gamma(Z)$}
\drawline\fermion[\SE\REG](\photonbackx,\photonbacky)[3000]
\drawarrow[\NW\ATTIP](\pmidx,\pmidy)
\global\advance\fermionbackx by 500
\put(\fermionbackx,\fermionbacky){$\bar f_{2}$}
\end{picture}

\section{The colliders and the ingredients}

The present study covers the three highest energy colliders LEP ($e^+e^-$) HERA ($e^\pm p$) et Tevatron ($p\bar{p}$), the characteristics of which are summarized in table~\ref{tab:acc}. In total, LEP experiments accumulated approximately 3.5~fb$^{-1}$ for \CoM energy ranging from 89 to 209~GeV. In their first running period, HERA and Tevatron experiments only collected 0.25~fb$^{-1}$, respectively. The second phase, recently started both at DESY and Fermilab, should bring the available luminosities in $ep$ and $p\bar{p}$ collision modes close to the LEP value, for \CoM energies of 319~GeV (HERA) and 1.96~GeV (Tevatron).
\renewcommand{\arraystretch}{1.2}     
\begin{table}[htb]
\begin{center}
\begin{tabular}{lccc}
    & {\bf HERA} & {\bf LEP} & {\bf TEVATRON} \\
%    & \includegraphics[width=0.33\textwidth]{fig/hera.eps} &  
%    \includegraphics[width=0.33\textwidth]{fig/lep.eps} & 
%    \includegraphics[width=0.33\textwidth]{fig/tev.eps}   \\
Beams &$e^\pm-p$ & $e^+-e^-$ & $p-\bar{p}$\\
Expt. & H1  ZEUS & ADLO & CDF,D0 \\
$\cal L$/expt. & 120 pb$^{-1}$ & 800 pb$^{-1}$ & 150 pb$^{-1}$ \\
Future (goal) & HERA II (1~fb$^{-1}$)  &  & run 2 ($4\div 8$~fb$^{-1}$) \\
 \end{tabular}
\caption{ Summary of LEP, HERA et Tevatron characteristics and performances }
\label{tab:acc}
\end{center}
\end{table} 
\par
The highly energetic collisions of particle beams lead to a variety of primary hard scattering processes. Besides the direct interaction between the colliding particles, the initial state processes (radiation, parton substructure etc.) can lead to a picture where the intial beams produce  secondary beams of different particles that enter the true hard scattering. For instance, the direct electron-proton interaction at HERA may take place in elastic scattering. In Deep Inelastic Scattering (DIS), the proton is seen as a bag filled with quarks or gluons and HERA collides (in a ``second'' mode)  electrons with quarks or electrons with gluons. In the same way at LEP the flux of real photons from initial state radiation on one side can collide with the electron beam from the other side or with its associated photon beam and therefore LEP can function as $e-\gamma$ or $\gamma-\gamma$ collider. 
\par
This ``second'' mode functionning for the considered colliders is summarized in table~\ref{tab:partons}.
\begin{table} 
 \begin{center}
\begin{tabular}{lllll}
Collider & Beams & ``Second'' mode collisions & Comment \\ \hline
{\bf LEP}            & $e^+-e^-$ &  &  full use of the $\sqrt{s}$ \\
& & & dominates at high $P_T$ \\
         &  & $e^\pm-\gamma$ & \\
& & $\gamma-\gamma$ \\ \hline
{\bf HERA}           & $e^\pm-p$ &  & elastic/diffraction \\
          & & $e^\pm-q$ & DIS\\
& & $e^\pm-\gamma$ & Compton\\
& & $\gamma-q$ & \\ 
& & $\gamma-g$ & \\ 
& & $q-q$ & photo--production (from $e\ra \gamma\ra$hadrons)\\
& & $\gamma-\gamma$ & \\ \hline
{\bf TEVATRON}       & $p-\bar{p}$     & & elastic/diffraction\\
& & $q-q$ & main collision mode\\
& & $q-g$ & for high $P_T$ physics \\
& & $g-g$ & for high $P_T$ physics \\
\end{tabular} 
\label{tab:partons}
\caption{ Possible parton-parton interactions at colliders with electron and/or proton beams}
\end{center}
\end{table}

\begin{figure}[hhh]
  \begin{center}
    \includegraphics[width=0.7\textwidth]{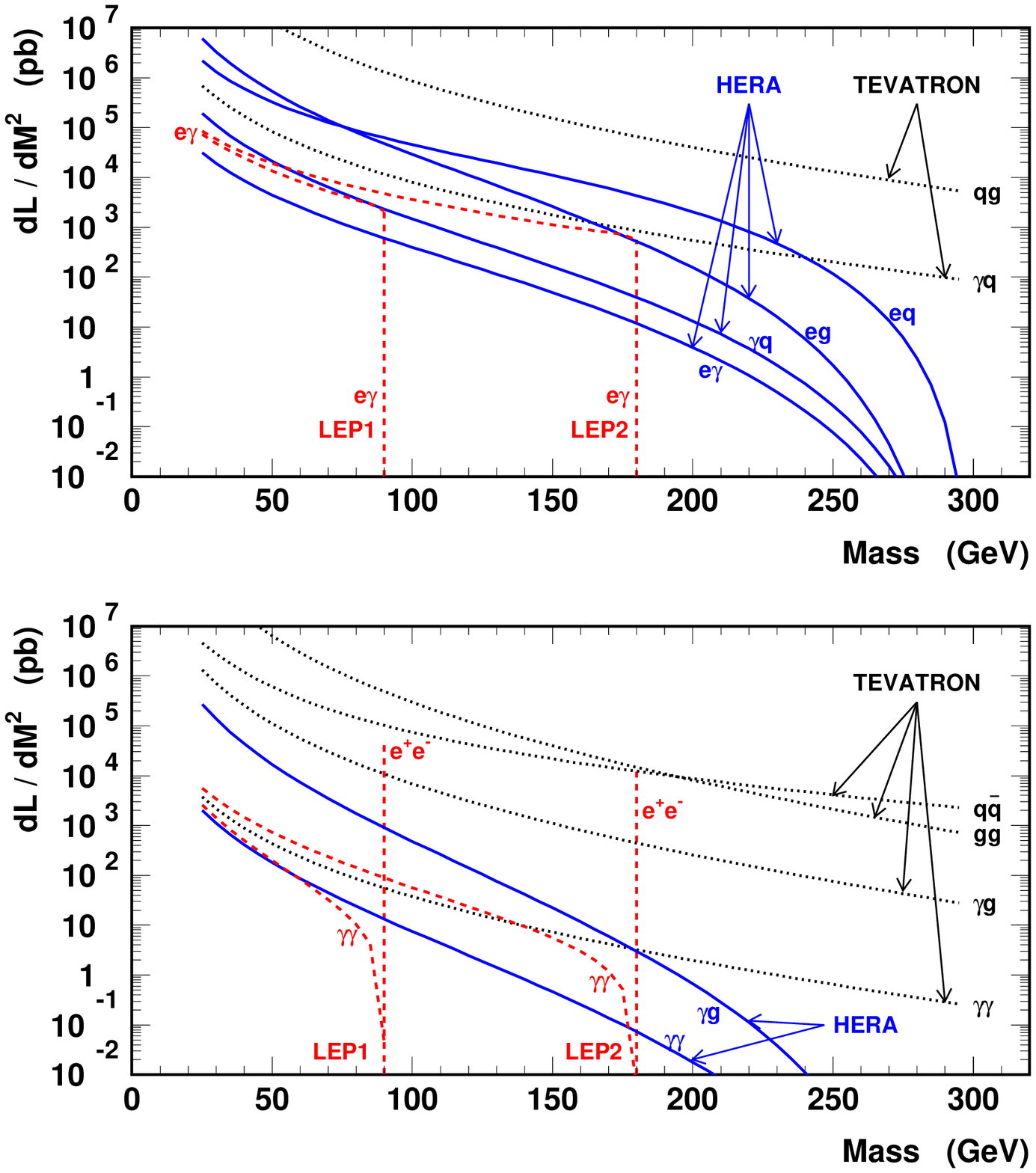}
    \caption{ Diferential partonic luminosities ($dL/dM^2$) at LEP HERA and Tevatron as a function of the \CoM energy  of the parton-parton collision. In this figure $\sqrt{s}(e^+e^-)_{LEP}=180$~GeV and  $\sqrt{s}(p\bar{p})_{Tevatron}=1.8$~TeV. } 
    \label{fig:partlum}
  \end{center}
\end{figure}

Altough this picture of the hard scattering is schematical and incomplete, it helps understanding the main features in the capabilities of the three colliders to test the \sm and to search for the new physics. The partonic luminosities~\cite{peter_partlum} are shown in figure~\ref{fig:partlum}. From that picture one can see for instance that HERA  collides electrons with quarks at highest energy and luminosity up to 300 GeV. LEP has the priority on $e-\gamma$ collisions up to its \CoM energy but HERA takes over at higher energies.

The physics potential of a collider with respect to a given process depends on the following three factors:
\begin{itemize}
\item[{\bf Luminosity}] This includes the partonic luminosity as discussed above but also the total integrated collision luminosity, which has to do rather with the existing tehnology for accelerating and colliding a given type of particles.
\item[{\bf Cross section}] of the hard scattering. The hard scattering cross section is convoluted with the partonic fluxes and with the experimental acceptance in order to obtain the observed cross section. The higher the cross section, the better the  sensitivity of a given collider to the searched process.  
\item[{\bf Background}] The observability of the searched process depends on the rate of irreducible \sm background processes.

\end{itemize}

The study of the processes that yield leptons ($\Delta N_L>0$) is of course dependent on the experimental luminosity. In addition, due to the partonic nature of the interaction at HERA and Tevatron, the effective \CoM energy is reduced and a part of the input energy is lost with the proton remnants. The phase space for particle (lepton) production at high $P_T$ is in this way reduced, in contrast to the LEP case, where most of the time the full available energy is ``consumed'' for particle production. The factor of ten in collision energy at Tevatron with respect to LEP helps to recover from that handicap, while HERA has only a modest advantage in \CoM energy versus LEP. Conversely, the small production rate within the \sm for a given  multi-lepton topology product may be useful in the frame of some searches for new physics, where the smaller background increases the sensitivity.
\par
The experimental backgrounds for lepton identification are small at LEP, while at HERA and Tevatron the hadronic processes or the radiative processes with photons in the final state have a larger cross section and represent often a challenge for the lepton identification.
\par
In the next chapter we summarize the main experimental analyses related to the lepton production at LEP, HERA and Tevatron. The possible models and experimental searches of the new physics using high energy lepton signatures are described in the last chapter.

%% file: leptons.tex
\chapter{Lepton Production in the \sm}
 
\section{Motivation }
 
The measurement of energetic and isolated leptons in the final state provides a useful probe of the underlying processes occuring in high energy collisions. 
This is due to the following facts:
\begin{itemize}
 \item The leptons are not sensitive to the strong interactions. The final state interactions are therefore small and well understood within the Standard Model. 
 \item The leptons couple to gauge bosons. Final states containing high energy isolated leptons may signal large boson masses or virtualities. The highly virtual photons resolve the matter to short distances with interesting insight into partonic structure. The leptonic decays of heavy electro-weak boson are gold plated channels for further electroweak sector tests. 
 \item In view of searches for non standard phenomena, events with leptons provide a clean signature. The production cross sections are significantly lower than the quark(jet) production in the same mass or $P_T$ range. 
 \item The charge is experimentally accessible for the charged leptons, in contrast with quarks.  A  high energy neutrinos in the final state signals an underlying charged current interaction or a real $W^\pm$ boson produced in the final state. This implies that a final state with high energetic leptons provides a better signature for specific subprocesses of the \sm or for signals beyond the \sm.

\end{itemize}
The \sm  predicts the relationship between the leptonic channels and the hadronic final states. Therefore, if the background from QCD-like processes is not too high, the analysis of leptonic final states is correlated to and complemented by the hadronic channels. 
\par
In this chapter the production of high $P_T$ isolated leptons is summarized for the experimental measurements done at LEP, Tevatron and HERA colliders. The production mechanisms within the \sm are briefly explained. Possible BSM interpretations of the final states with extra-leptons with high $P_T$ will be discussed in the next chapter.

\section{High $P_T$ lepton production at LEP}
In electron--positron collisions the annihilation process dominates the production rate of high $P_T$ leptons. The  $e^+e^-$ pair fuses into a massive boson ($\gamma^*$ or $Z^{(*)}$) that subsequently materialises into a lepton pair ($\Delta N_L$=0) or a quark--antiquark pair  ($\Delta N_L=-2$). 
\begin{figure}[hhh]
  \begin{center}
    \includegraphics[width=0.54\textwidth]{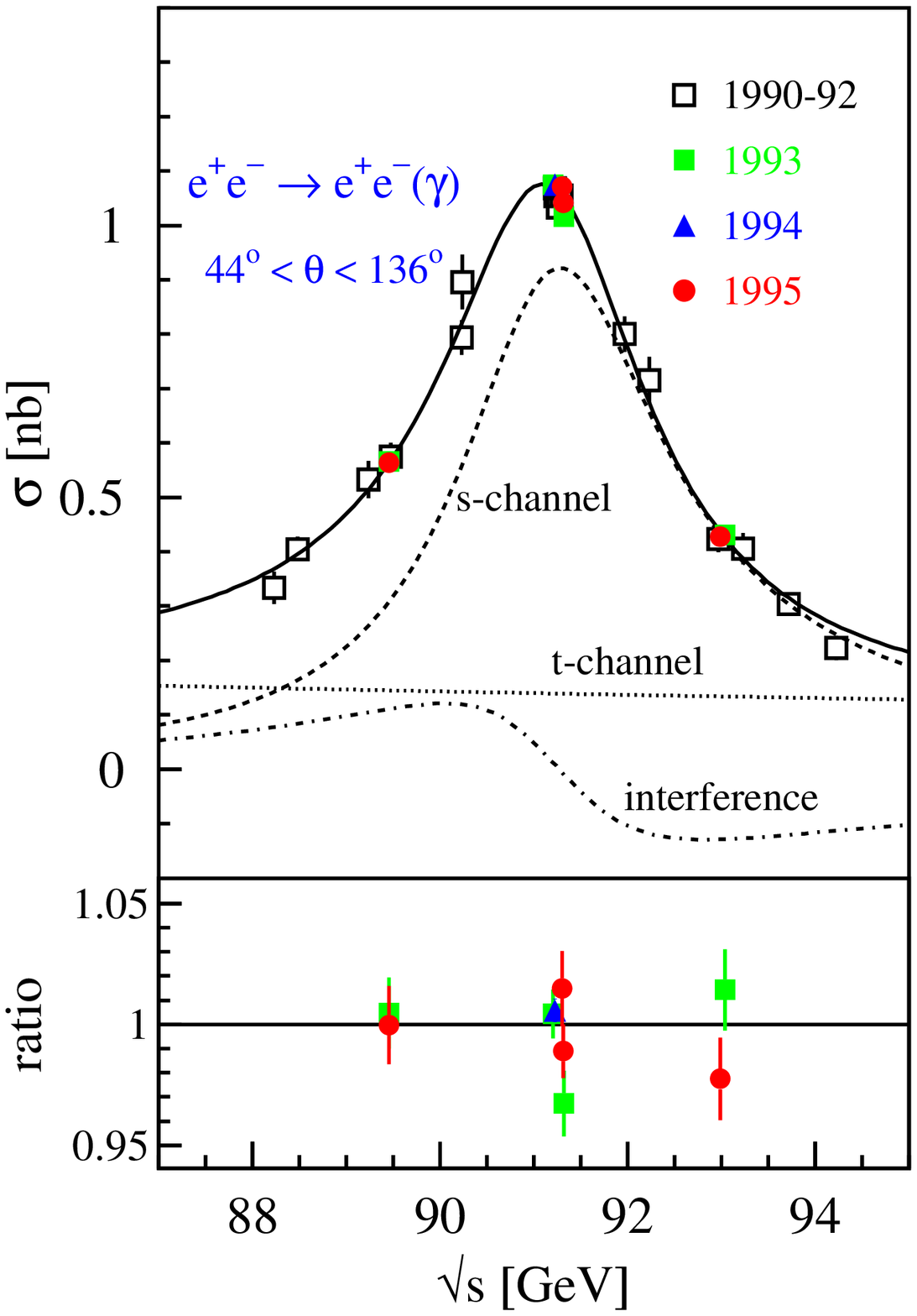}
    \includegraphics[width=0.56\textwidth]{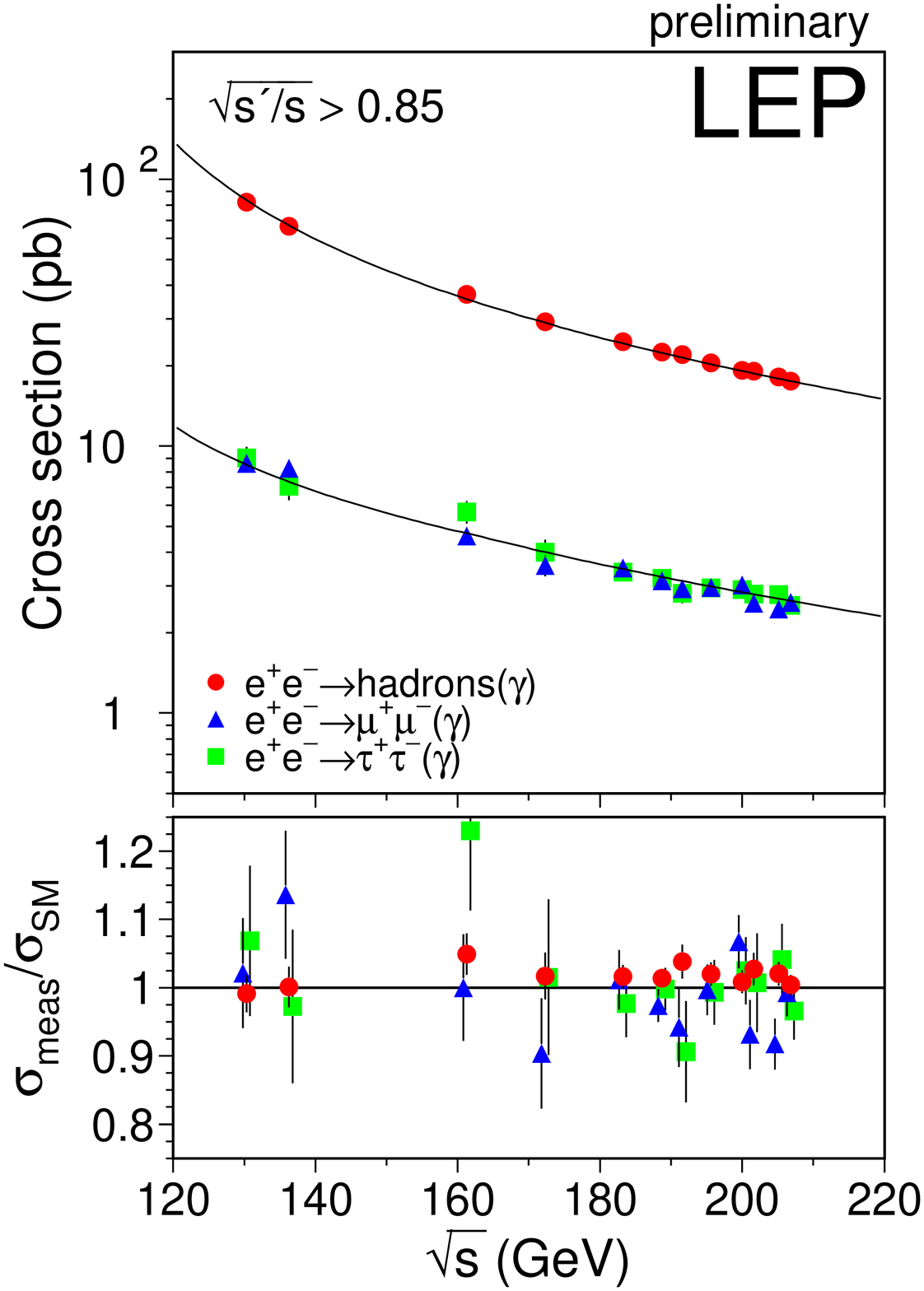}
    \caption{ 
Cross section for $e^+e^-\rightarrow f\bar{f}$ as a function of center of mass energy measured by LEP experiments around the $Z$ pole (left) and at higher LEP2 energies (right).
}
    \label{fig:2f_lep}
  \end{center}
\end{figure}
The fermion pair production at LEP is an important tool for measuring the electroweak parameters~\cite{ADLO:2002mc}. The measurement of the $Z$ boson mass and width, the hadronic pole cross section o-f $Z$ exchange, the ratios of hadron to lepton cross-sections for the three types of leptons and for the $b$ and the $c$ quarks together with the corresponding forward-backward asymmetries are used to extract the \sm parameters with high precision~\cite{Renton:2002wy}.  The cross section of fermion pair production as a function of the center of mass energy is shown in figure~\ref{fig:2f_lep} for the \CoM energy around the $Z$ mass (left) and in the higher \CoM energy regime of LEP2 (right).
\par
\begin{figure}[hhh]
  \begin{center}
    \includegraphics[width=0.49\textwidth]{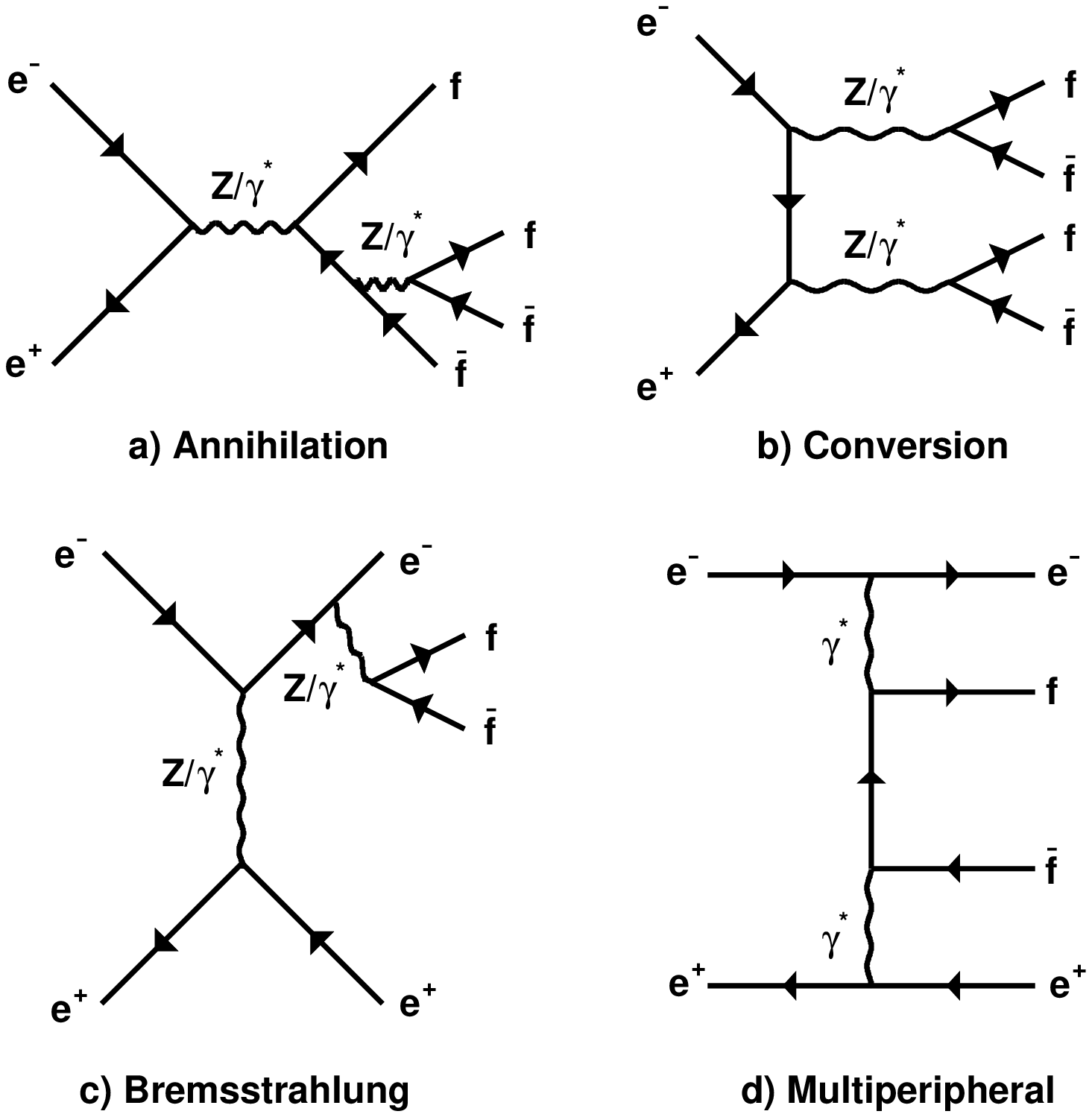}
    \includegraphics[width=0.49\textwidth]{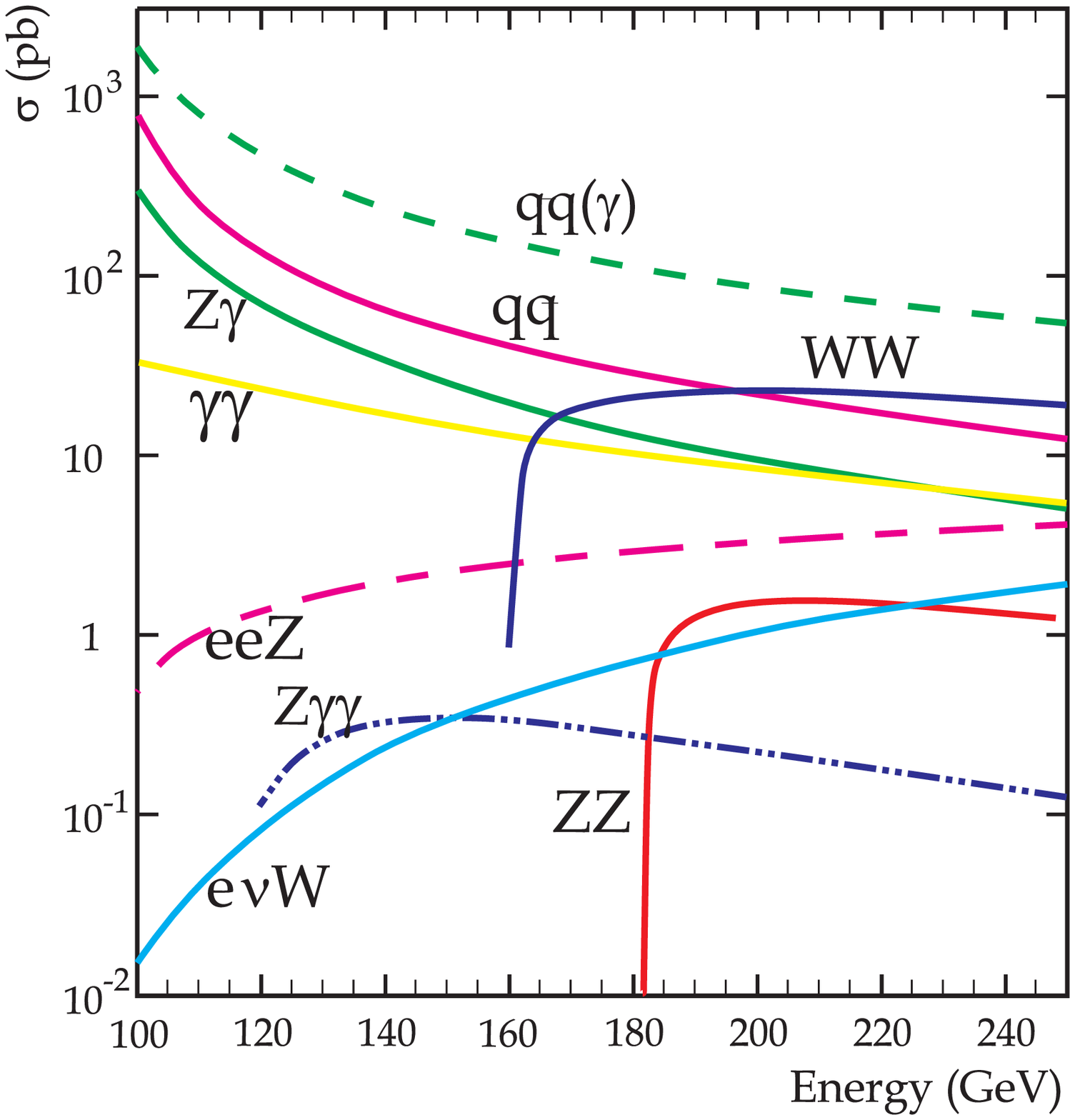}
    \caption{ 
Four fermions production mechanisms in $e^+e^-$ collisions for the neutral current production (left) and \sm fermion pair  and boson pair production cross section as a function of $\sqrt{s}$ at LEP2 energies (right).
}
    \label{fig:4f_all}
  \end{center}
\end{figure}

Within the Standard Model, extra lepton production ($\Delta N_L=2$) is part of a higher order process (${\cal O}(\alpha^4)$) leading to four fermions in the final state. As an example, the production mechanisms for four charged fermions are shown in figure~\ref{fig:4f_all} (left). 
The final state with electrons  is described by more diagrams, including  the $t$-channel scattering that is possible only in the $e^+e^-\ra e^+e^- ff$ processes.
At LEP2 energies,  the four lepton production can also proceed through  single or double resonant processes with one or two weak bosons ($W$ or $Z$) in the final state. The cross section of weak boson production at LEP2 energies is illustrated in figure~\ref{fig:4f_all} (right).
\par
In an electron--positron collider, the partons that enter the hard scattering leading to more than two leptons in the final state can be:
\begin{itemize}
\item $e^+e^-$ : up to 4 leptons at high $P_T$ in the final state. The extra pair of leptons is produced during the hard $e^+e^-$ scattering to four particles. The full $2\rightarrow 4$ matrix elements calculation is used for the theoretical calculation of this process. Resonant vector boson production is possible if enough \CoM energy is available. The final state is either fully detected or, in the case of neutrinos in the final state, the energy--momentum conservation may be used to constrain the kinematics.
\item $e\gamma$ :  one of the incident electrons escapes down the beampipe after producing the interacting photon.  Besides the classical QED Compton scattering  $e\gamma\rightarrow e\gamma$, multilepton final states can be produced through the so-called electroweak Compton mechanism  $e\gamma\rightarrow e\gamma^*$ or  $e\gamma\rightarrow eZ$, with subsequent boson conversion into a lepton pair. In this case three leptons of the final state are usually in the detector acceptance. The calculation relies on a convolution of photon flux with the electroweak Compton scattering matrix elements. The process is tagged either by the presence of an electron close to the beam pipe or by the missing energy aligned with the beam direction, due to the escaping electron. The charged current EW Compton process  $e\gamma \rightarrow \nu W$ is also possible.
The electroweak Compton scattering provides the main mechanism of single vector boson production at LEP.

\item $\gamma\gamma$ : two leptons in the final state. At high transverse momentum, this contribution is modest with respect to the $s$--channel di-lepton production. However, it can well be measured by using  ``tagged'' events, where at least one of the scattered  electrons is measured in a low angle  detector near the beam pipe.
\end{itemize}

The neutral current interactions lead to final states with charged leptons only in the final states in all cases except for the Z boson conversion into a neutrino pair. The charged current interactions, mediated in the \sm by W bosons, lead to final states containing neutrinos. 
\par
From the experimental point of view, the excellent hermeticity and tracking performance allow to detect charged leptons with transverse momenta as low as 1 GeV and to signal the presence of an undetected particle by the missing transverse energy down to a few GeV. Depending on the analysed final state, the energy--momentum conservation is used to improve the resolution or to separate the signal from the background.
\par
In the following,  main topologies with charged or neutral leptons in the final state are described. After a short insight into  a LEP1 study of events with four fermions, we will concentrate on LEP2 data obtained at highest $e^+e^-$ collision energies.

\subsection{Four fermion production at LEP1}
 Four fermion production at LEP I is based on final state topologies with two leptons plus two other fermions $\ell^+\ell^-f\bar{f}$. The ALEPH collaboration studied events in which, besides two charged leptons used as a tag, two other fermions are identified~\cite{Buskulic:1995gk}. Using tracking information, low multiplicity events, also called  $\ell^+\ell^-V$, are selected. Part of this sample is identified as four-lepton events.  The rest is attributed to the production of two hadrons $\pi^+\pi^-$ or $K^+K^-$ in addition to the lepton pair. The hadron pair may originate from  the decay of the resonance formed by the quark-antiquark pair produced in the final state.

\begin{figure}[hhh]
  \begin{center}
    \includegraphics[width=0.55\textwidth]{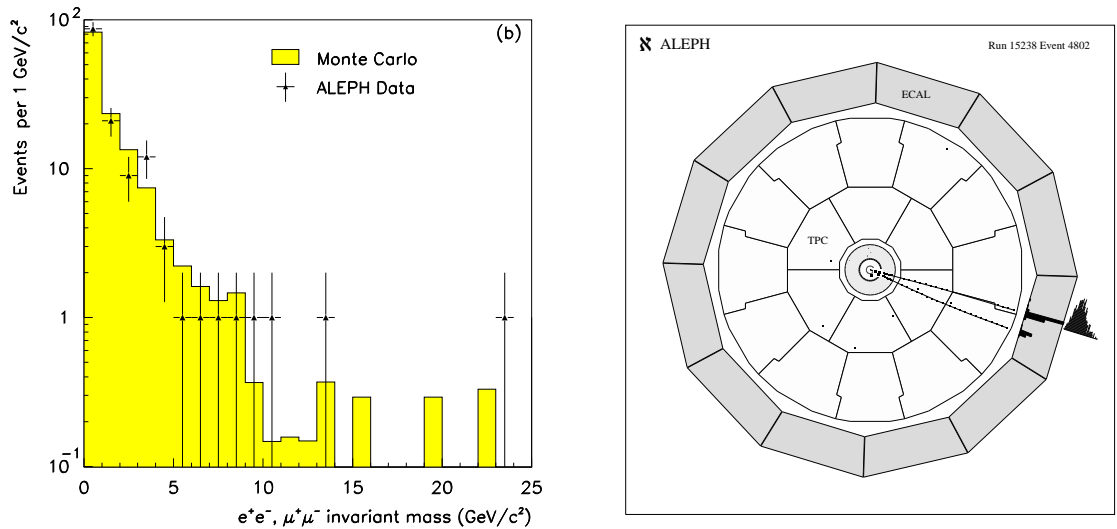}
    \includegraphics[width=0.55\textwidth]{fig/aleph_4f_acop.ps}
    \caption{ 
Minimal di-fermion invariant mass distribution for events with four fermions at LEP (left) and a  candidate for $e^+e^-\rightarrow \nu\bar{\nu} e^+e^-$ process (right).
}
    \label{fig:4f_picZ}
  \end{center}
\end{figure}

\par
This study revealed a good agreement between the data and the \sm prediction. The  minimimum di-fermion invariant mass is   shown in figure~\ref{fig:4f_picZ} (left). The selected events are concentrated at low values consistent with a production mechanism based mainly on $\gamma^*$ internal conversion.
  Besides events with four charged fermions, evidence for the production of charged leptons in events with neutrinos has also been found -- an event with two charged leptons associated with large missing energy is shown in figure~\ref{fig:4f_picZ} (right).

\subsection{Events with charged leptons at LEP2}
\par
\begin{figure}[hhh]
  \begin{center}
    \includegraphics[width=0.35\textwidth,height=7.5cm]{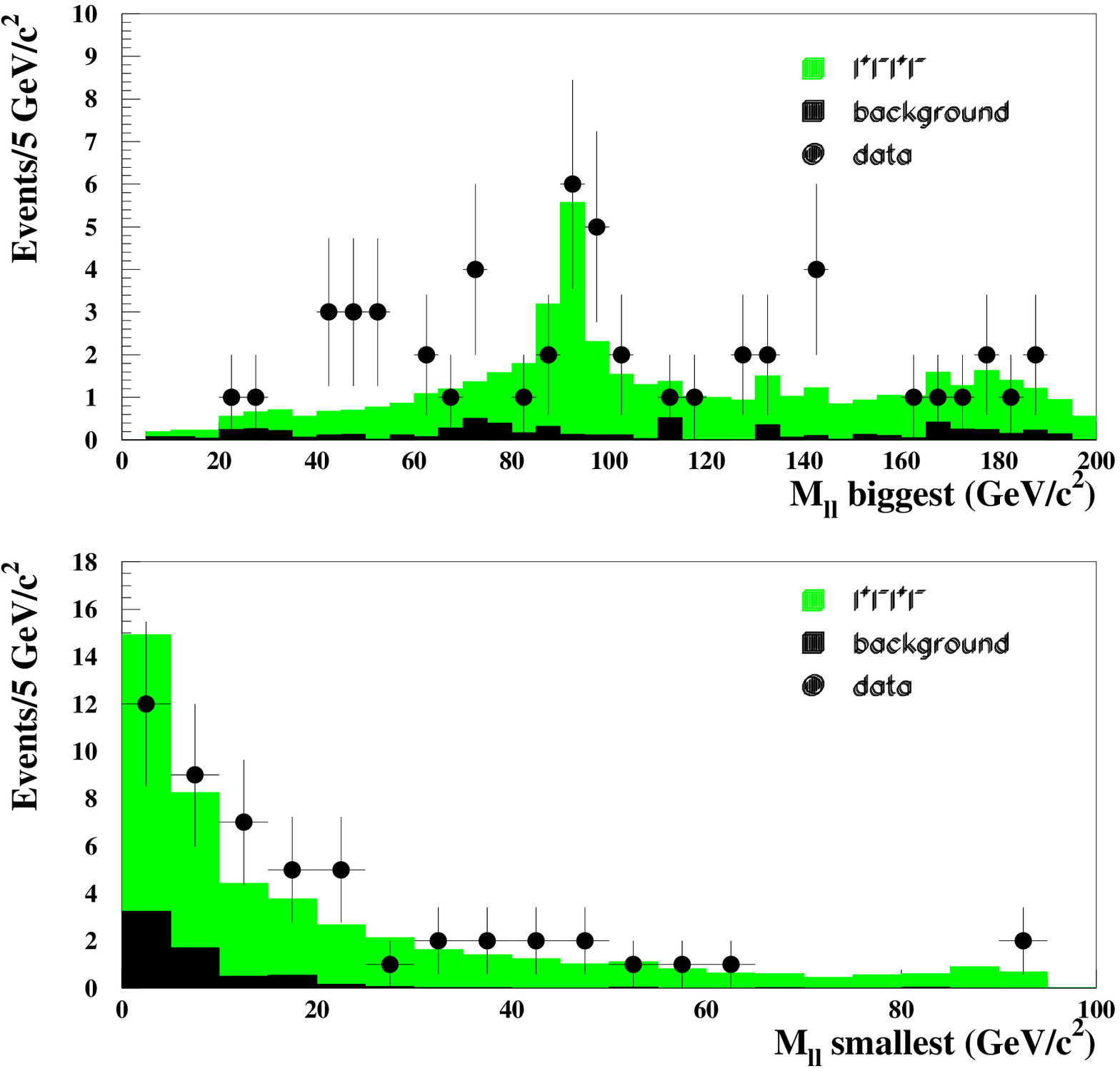}
    \includegraphics[width=0.38\textwidth,height=8cm]{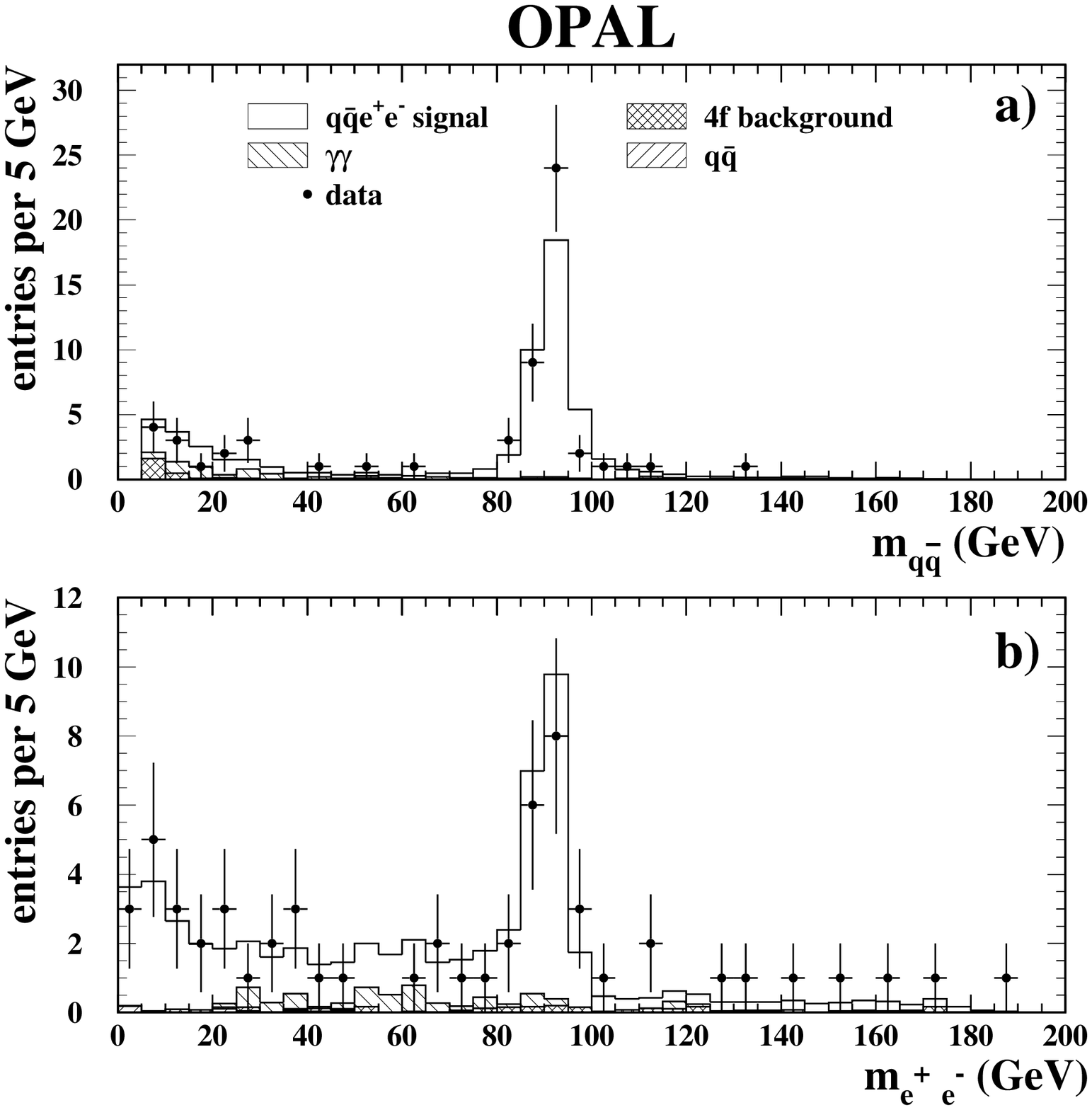}
    \includegraphics[width=0.38\textwidth,height=8cm]{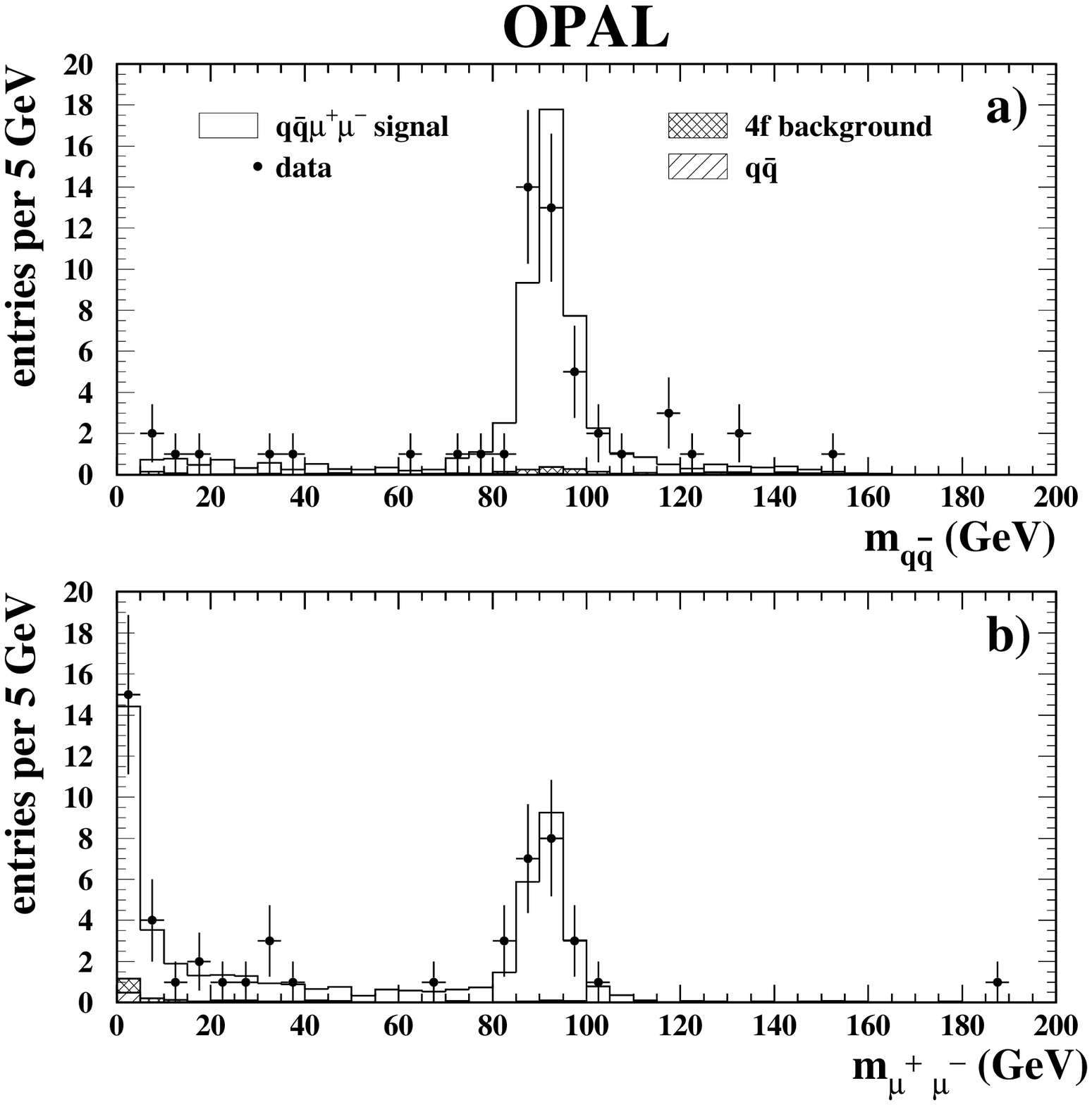}
    \caption{ 
Maximum and minimum di-lepton invariant mass distribution for four lepton events detected with a flavour independent method by DELPHI(left). Invariant mass of same flavour fermion pairs in  $e^+e^- q\bar{q}$ (center) and $\mu^+\mu^- q\bar{q}$ (right) channels(from OPAL analysis). 
}
    \label{fig:opaldelphi_4f}
  \end{center}
\end{figure}

 
The search for the process $e^+e^-\rightarrow 4\ell^\pm$ relies on the well understood lepton identification.  As an example, we consider  
the flavour independent analysis performed by the DELPHI collaboration~\cite{delphi_4f}. The analysis is part of a study of four fermion production through neutral current interactions at $\sqrt{s}=183\div 208$~GeV.  The distributions of the minimal and maximal invariant mass of the lepton pairs in the event are shown in figure~\ref{fig:opaldelphi_4f}(left). The $Z$ peak is observed in the maximal mass distribution.
\par
The four charged leptons final state is a rather rare process at LEP. For instance, in the flavour-blind analysis (including therefore electrons, muons or taus among the four leptons in the final state) only 16 events were selected in the data compared with a prediction of 14.6 from the \sm in the data sample corresponding to an integrated luminosity of 221~pb$^{-1}$ with $\sqrt{s}=204\div 209$~GeV.
\par
The total cross section of four lepton production was measured at highest LEP2  \CoM energies in the range from 183 to 208~GeV. The result in the highest energy bin is $\sigma_{204-208}=0.55_{-0.13}^{+0.15}\pm 0.04$~pb for the flavour blind measurement of leptons with $|\cos \theta| < 0.98$.
\par 
The contributions of different poles can be studied if the interference effects are neglected. However, the pure leptonic channel is not well suited for separating $ZZ$, $Z\gamma^*$ and $\gamma^*\gamma^*$ contributions due to the low statistics.  The production of a pair of quarks together with a pair of leptons is included in the analysis by using $e^+e^- q \bar{q}$, $\mu^+\mu^- q \bar{q}$ and $\nu\bar{\nu} q \bar{q}$ final states. The purely hadronic final state $q \bar{q}q \bar{q}$ (four jets) is hampered by the  background from QCD processes or $WW$ pair production and cannot be used for NC four fermion production studies.

\par
Figure~\ref{fig:opaldelphi_4f} shows the di-fermion invariant mass distributions obtained in a similar study done by the OPAL collaboration~\cite{Abbiendi:2003va}. A significat peak at $Z$ mass due to resonant production $Z\rightarrow f\bar{f}$ is visible for both channels and both fermion pairs. The low mass peak  due to internal conversion $\gamma^*\rightarrow f\bar{f}$ is only visible in $M_{\mu\mu}$ and $M_{ee}$, while it is supressed by the analysis cuts for the other distributions. No other significant peak, possibly due to a new boson, is found. 
\par
The cross sections corresponding to the semileptonic channels $eeqq$ and $\mu\mu qq$ for various production mechanisms are extracted by using a fit of the invariant mass spectra.  The contribution from non-resonant process $ \sigma_{\gamma^*\gamma^*}$, from semi-resonant process $ \sigma_{Z\gamma^*}$ and from double resonant$ \sigma_{ZZ}$ are measured. The $t$-channel contribution is measured in the $eeqq$ channel. The results are~\cite{delphi_4f}:\\
\begin{center}
\begin{tabular}{l}
$ \sigma_{Z\gamma^*}=0.129\pm0.020$~pb, \\
$ \sigma_{ZZ}=0.029\pm0.006$~pb, \\
$ \sigma_{\gamma^*\gamma^*}=0.017\pm0.008$~pb and\\
$ \sigma_{t-channel}=0.245\pm0.045$~pb (for electrons).
\end{tabular}
\end{center}

\begin{figure}[hhh]
  \begin{center}
 \includegraphics[width=0.45\textwidth]{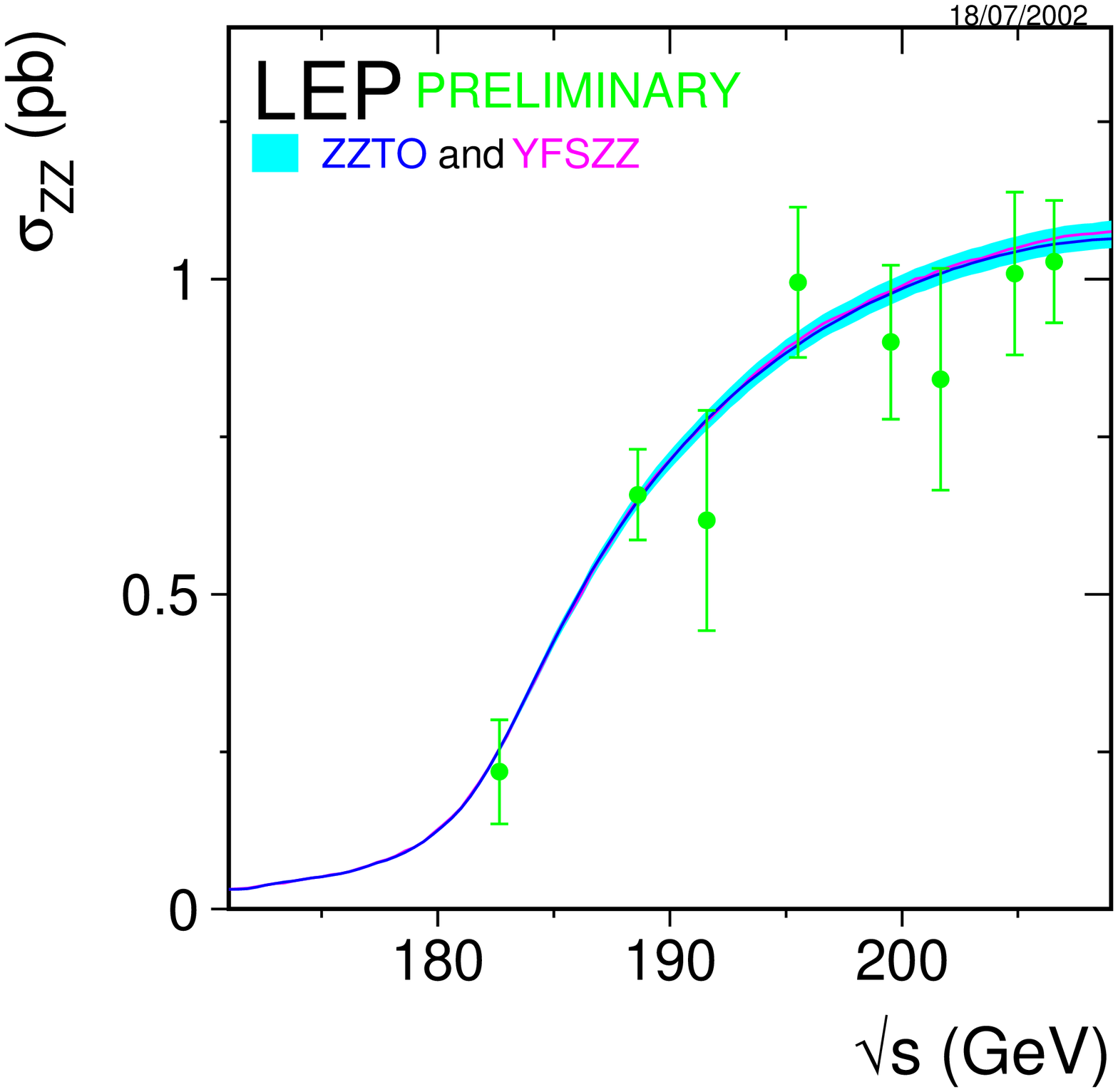}
 \includegraphics[width=0.45\textwidth]{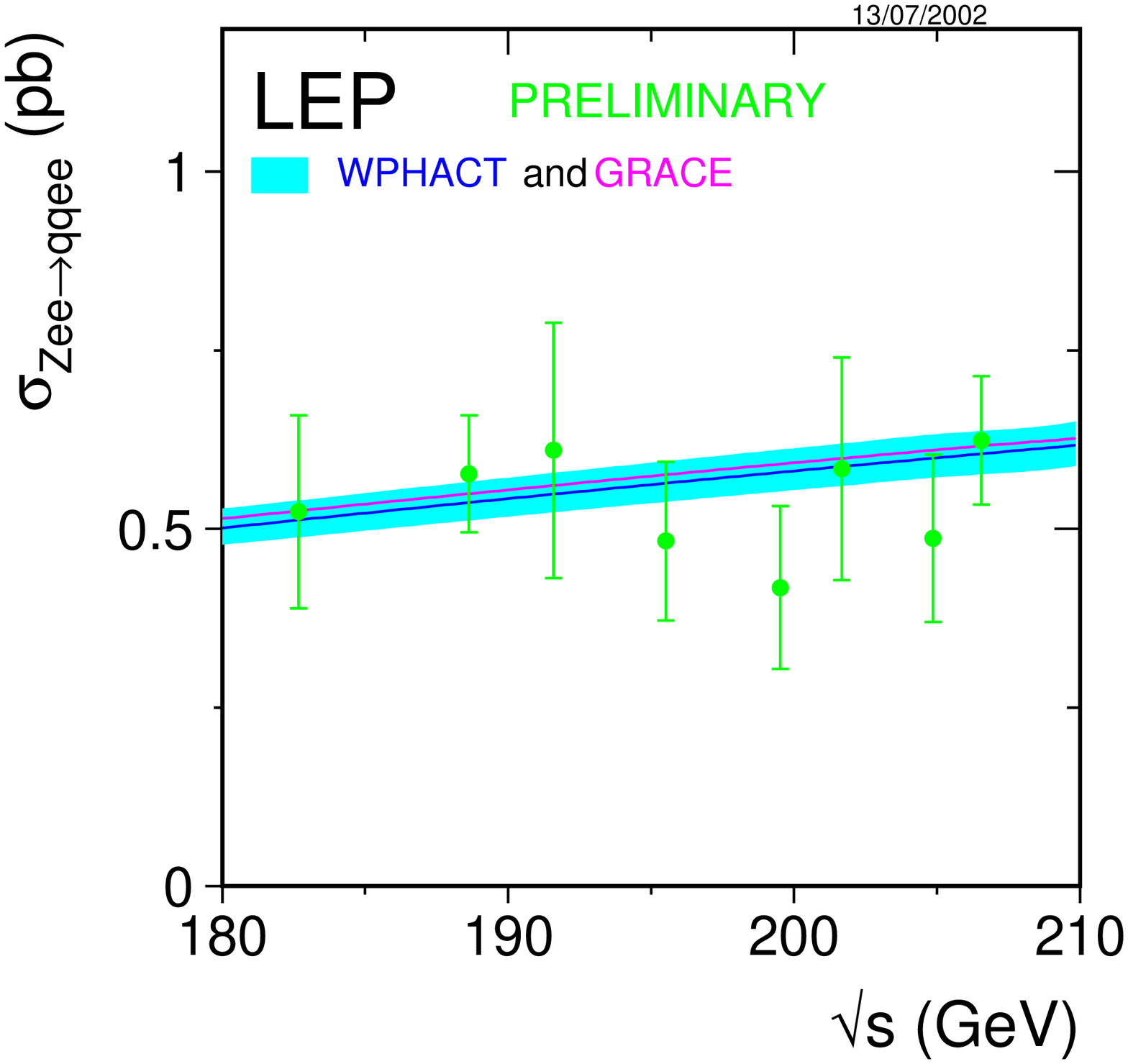}
 \includegraphics[width=0.45\textwidth]{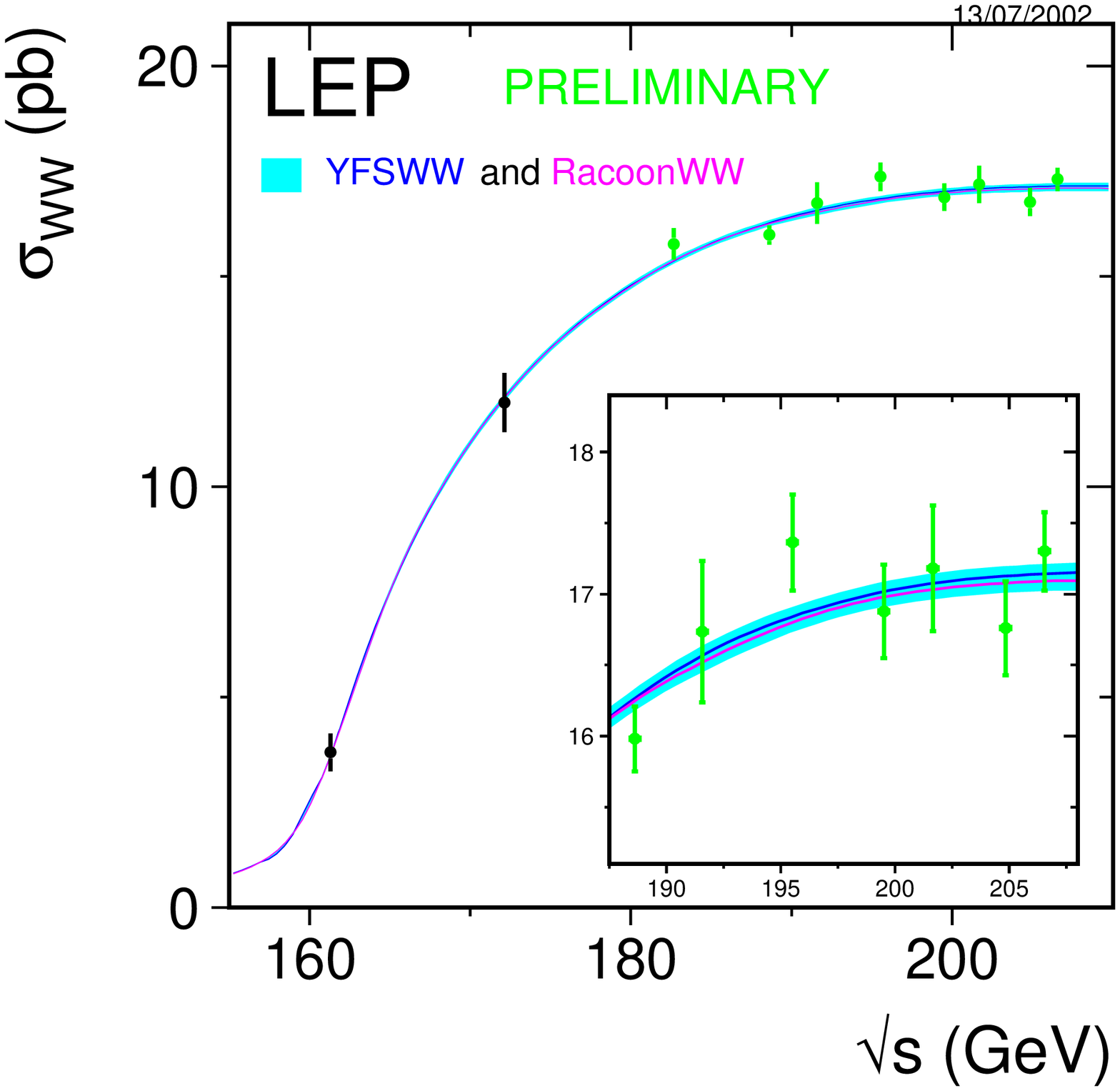}
 \includegraphics[width=0.45\textwidth]{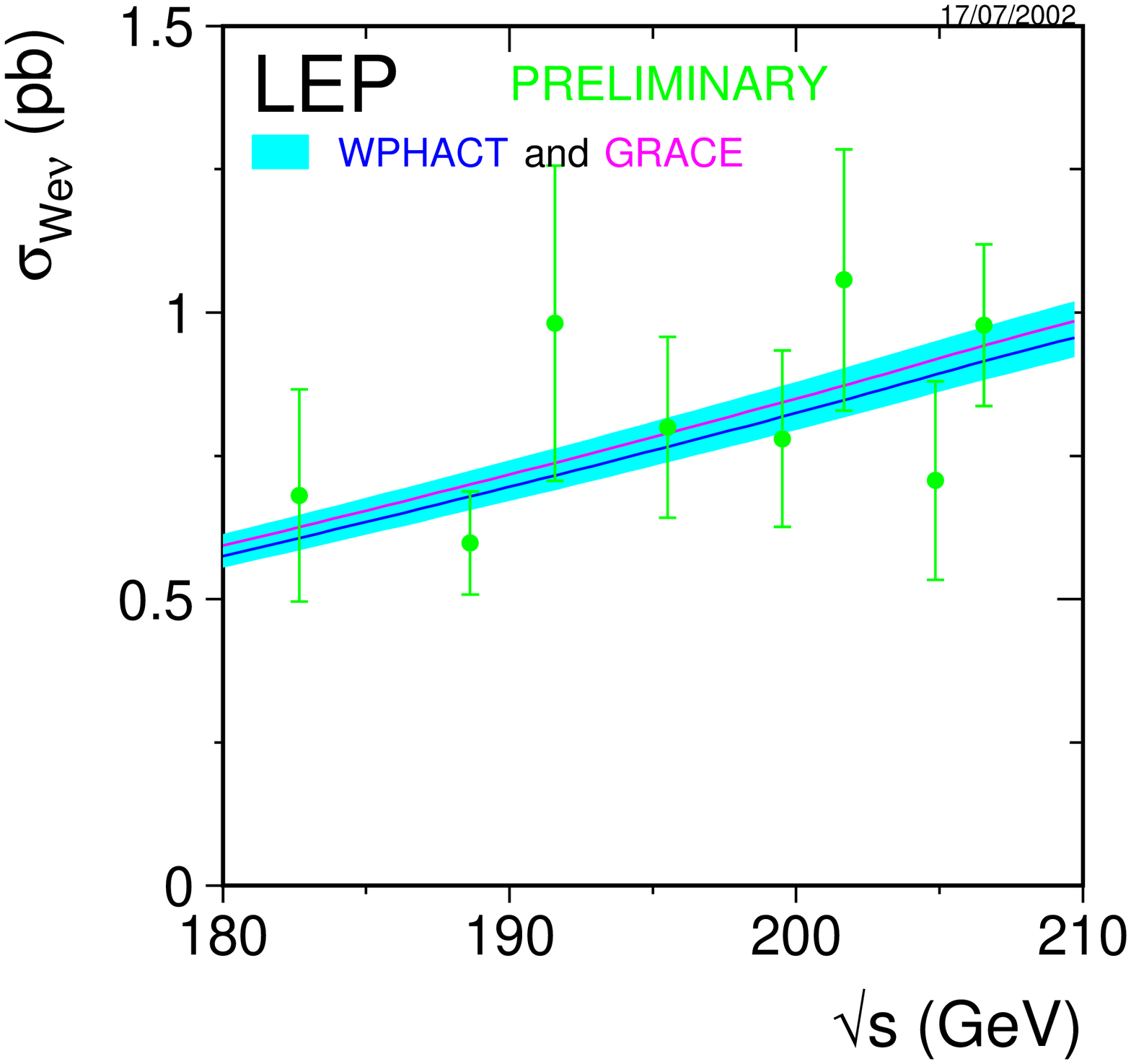}
    \caption{ 
The cross sections of weak boson(s) production at LEP as a function of $\sqrt{s}$. 
}
    \label{fig:lep_bos}
  \end{center}
\end{figure}

\par
The production of $Z$ bosons pairs is possible at LEP in the highest \CoM  energy range.  The cross section of $Z$ boson pair production has been measured by LEP collaborations and is presented in figure~\ref{fig:lep_bos}(up left). The raise around the kinematical threshold $\sqrt{s}\simeq 2M_Z$ is observed. At highest energy the $Z$ pair production cross section is around 1~pb. 

\subsubsection{Three leptons at high $P_T$ in the final state}

Events with three visible leptons are produced if one of the beam electrons is scattered at low angles and escapes down the beam pipe. In this configuration, the virtuality of the photon emitted by this low angle electron is close to zero and therefore the photon is quasi-real. The corresponding interaction mechanism is  the EW Compton scattering $e\gamma \rightarrow  e\gamma^*/Z$. This mechanism is responsible for single $Z$ boson production at LEP.
The final state consists of a low energy electron recoiling against a pair of leptons or jets. The missing momentum in the event should point along the beam direction, corresponding to the lost electron. 
\par
The contribution from $\gamma^*$ and $Z$ can be separated by using the invariant mass spectrum of the boson candidate decay $\gamma^*/Z \ra f\bar{f}$~\cite{Abbiendi:1998va,Abreu:2001fk}. Low invariant masses of the boson candidate, typically below 60~GeV, are dominated by the non-resonant contribution from $\gamma^*$, larger masses are due mainly to the single $Z$ production. The combined measurements of the single $Z$ production as a function of the LEP center of mass energy is presented in figure~\ref{fig:lep_bos}(up right). The cross section is defined for the hadronic decay channel of the $Z$ boson and slowly increase as a function of \CoM energy. At highest LEP energy $\sigma_{Zee\ra qqee}=0.6$~pb.

\subsubsection{Charged lepton pairs produced in $\gamma\gamma$ collisions}
At $e^+e^-$ colliders, leptons pairs can also be produced in $\gamma\gamma$ collisions. This process allows  for both \sm tests and searches for new phenomena coupled to the two photon collisions~\cite{Arteaga-Romero:1971wq,Parisi:1971su}. The fermion pair production mechanism can be understood from the diagram d) in figure~\ref{fig:4f_all}. The incident electrons are scattered at low angles and are most of the time lost down the beampipe. However, in order to separate this process from the $s$--channel annihilation,  the ``tagged'' events are analyzed for which one of the scattered electrons is detected in the sub-detectors situated close to the beam pipe or at low angle in the main calorimeter. 
\par
The production of muon pairs in $\gamma\gamma$ collisions allows to study QED up to the fourth order of $\alpha$. This process has been studied by DELPHI~\cite{Abreu:2000de} using LEP1 data. The selected sample contains single tag events only, in order to insure that one of the incoming photons is real, i.e. its virtuality $Q^2\simeq 0$. The virtuality of the other photon is calculated from the tag electron and is in the range $Q^2=2.5\div 750$~GeV$^2$. The muons with transverse momenta up to 20~GeV are observed in this sample. This analysis allows the measurement of the leptonic photon structure function $F_{2,\ell}^\gamma$. The angle between the di-muon plane and the plane formed by the four vector of the tag electron and the beam direction is sensitive to the helicity structure of the photon--photon interaction. The ratios of the helicity related structure functions $F_2^A$ and $F_2^B$ to $F_{2,\ell}^\gamma$ are measured from this azimuthal correlations and found in good agreement with the QED prediction.
\par
Using the same technique to tag the photon--photon collision by the low angle electrons, the $\gamma\gamma\rightarrow q\bar{q}$ process can be studied by events with hadrons in the final state. The hadron production in $\gamma\gamma$ collisions is a laboratory for QCD studies and the kinematical variables from the data are compared with the prediction from models or NLO-QCD calculation~\cite{aleph_gghad,Achard:2001kr,LEP:2002sz}. The modelling of the hadronic final state is of interest also for the production by this mechanism of high mass states decaying to hadrons. However, the effective \CoM energy is lower than from the direct $e^+e^-$ collisions and no anomaly has been reported for this type of events. The photon structure is studied in deep inelastic $e\gamma$ scattering~\cite{Nisius:1999cv}. 
 
\subsection{Events with charged leptons and missing energy at LEP2}
The production of events with high energy leptons and missing energy at LEP proceeds mainly through W bosons. The contribution of events in which the missing transverse momentum originate from $Z\rightarrow \nu\bar{\nu}$ is usually small and it will not be discussed here~\footnote{The leptons in events with a decay $Z\rightarrow \nu\bar{\nu}$ are of opposite sign and same flavour. This process can be used in principle used to measure the $Z$ production, but the improvement of the measurement is tiny due to the modest branching ratios involved. The signature in the electroweak scattering of a $Z\ra \nu\bar{\nu}$ decay would be an electron at low energy and nothing else in the detector.}. 
\subsubsection{Events with one prominent lepton and missing transverse energy}
In the \sm framework, events with one charged lepton and missing transverse energy are produced in $e^+e^-$ collisions through the production of single $W$ bosons. In this case, one of the incident electrons is scattered at low angles and escapes down the beampipe ($|\cos\theta_e|>0.95$). The basic interaction is a charged current electroweak Compton scattering $e\gamma \rightarrow \nu W$ with a subsequent leptonic decay of the $W$ boson. In addition to this mechanism, the production of single $W$ bosons involves the triple gauge coupling $\gamma WW$ ($\gamma+W \ra W$). The peripheral contribution from $\gamma-W$ fusion into a $\ell^\pm\nu$ pair via a $t$-channel neutrino exchange corresponds to the non-resonant continuum.
\par
The single $W$ production in $e\gamma$ collisions has been measured including the hadronic channel $W\rightarrow q\bar{q}$. The W identification is rather loose (cut on lepton energy $E>20$~GeV in the leptonic channels or a cut on di-jet mass $>45$~GeV in the hadronic channel). The measured cross section as a function of $\sqrt{s}$ and is shown in figure~\ref{fig:lep_bos}(down right).
The measured cross section slowly increase as a function of \CoM energy in agreement with the theoretical expectation. At highest LEP energy $\sigma_{Wev}=0.8$~pb.
\par
Events with one lepton and missing energy in events with at least two jets are produced from $W$ pair production processes when one $W$ boson decays leptonically and the other hadronically. 
 
\subsubsection{Events with two leptons and missing transverse energy}
This topology is mainly produced by the $W$ pair production process, with subsequest leptonic decay of both $W$ bosons. 
The rate of such events is sizeable at LEP. In the analysis performed by the OPAL collaboration~\cite{Abbiendi:2003ji}, 416 events are found at highest LEP energies ($\sqrt{s}=205\div 209$~GeV) for $423.3\pm2.5$ predicted. In total 1317 events were detected in the highest energy range  ($\sqrt{s}=183\div 209$~GeV).
\par
The cross section of the $W$ pair production has been measured including also the hadronic decays channels and the combined LEP result is presented in figure~\ref{fig:lep_bos}(down left). The increase of the cross section at the kinematical threshold $\sqrt{s}\simeq 2M_W$ is measured. Good agreement with the \sm prediction is observed. At highest LEP energy $\sigma_{WW}=17$~pb.

\subsection{Events with leptons and photons}

Due to the initial state radiation, lower energy beam electrons are produced accompanied by photons colinear with the beam direction. The hard scattering involves either the lower energy electrons (radiative $e^+e^-$ process) or the photon (Compton scattering). The radiative processes scan a large domain in the center of mass energy.
The Compton scattering $e\gamma \rightarrow e\gamma$ yields events with an electron and a photon at large polar angles and transverse momenta in the final state, while the non-colliding electron is either lost down the beam--pipe or detected close to it. A similar pattern is observed for the non-colliding photon in radiative $e^+e^-$ interactions. The effective center of mass energy $\sqrt{s'}$ is calculated from the detected particles. 
\par
The analysis done by the L3 collaboration~\cite{Acciarri:1998vr} uses a  data sample including  center of mass energies from 89 to 180 GeV. After the deconvolution of the photon flux, the Bhabha scattering cross section is measured for an effective cms energy $\sqrt{s'}=40\div 180$~GeV.  The Compton scattering cross section is measured for   $\sqrt{s'}=15\div 170$~GeV. Both Bhabha and Compton measurements are found in good agreement with the expectations. Limits on excited electron production and decay $e^*\rightarrow e\gamma$ are obtained.
\par
The final states of four leptons plus a photon has also been studied at LEP~\cite{ADLO:2002mc,delphi_wwg}. The cross section  is dominated by the ISR and therefore the topology of the observed events displays a photon detected close to the beam pipe. The production of $4\ell\gamma$ events where the photon is produced during the main hard scattering is below 1~fb at LEP and therefore undetectable. This final state is suited for the search of anomalous quartic gauge couplings. 
\par Events with one or more photons and missing energy can be produced at LEP through two processes: radiative
returns to the $Z$ resonance ($ \rm e^{+} e^{-} \rightarrow \gamma  Z$)  with Z$\rightarrow \nu \bar{\nu}$, and $t$-channel W
exchange with photon(s) radiated from the beam electrons or the $W$. The cross section has been measured~\cite{Barate:1998ue} $ \mathrm \sigma(e^{+}e^{-}\rightarrow   \nu \bar{\nu} \gamma (\gamma)) = 4.7 \pm 0.8 \pm 0.2$~pb for $\sqrt{s}=172$~GeV and found consistent with the Standard Model predictions  of $4.85 \pm 0.04$~pb.

\section{Lepton production in $p\bar{p}$ collisions at Tevatron}
The production of leptons is a precious tool for \sm tests and new physics searches in $p\bar{p}$ collisions. The hadronic environment implies difficult background conditions for the exclusive final states relying on multi--jet topologies. The enormous QCD jet production is still the main background for the leptonic channels due to the jet misidentification into leptons. This experimental effect is however under control and can be studied with data. Careful energy calibration is required in order to insure the measurement of events with missing energy in the final state.

\subsection{Events with charged leptons}
\begin{figure}[hhh]
  \begin{center}
 \includegraphics[width=0.47\textwidth]{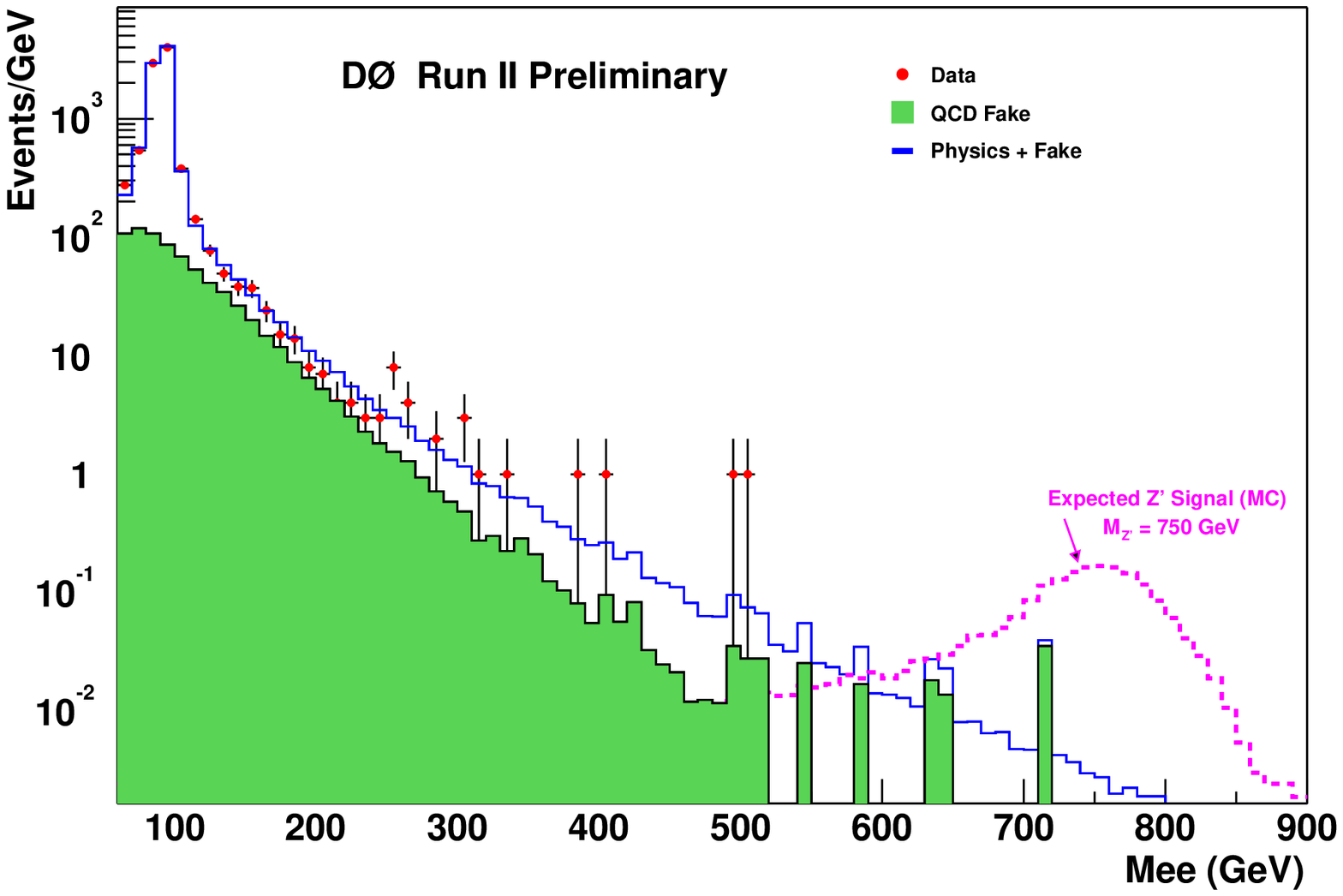}
 \includegraphics[width=0.55\textwidth]{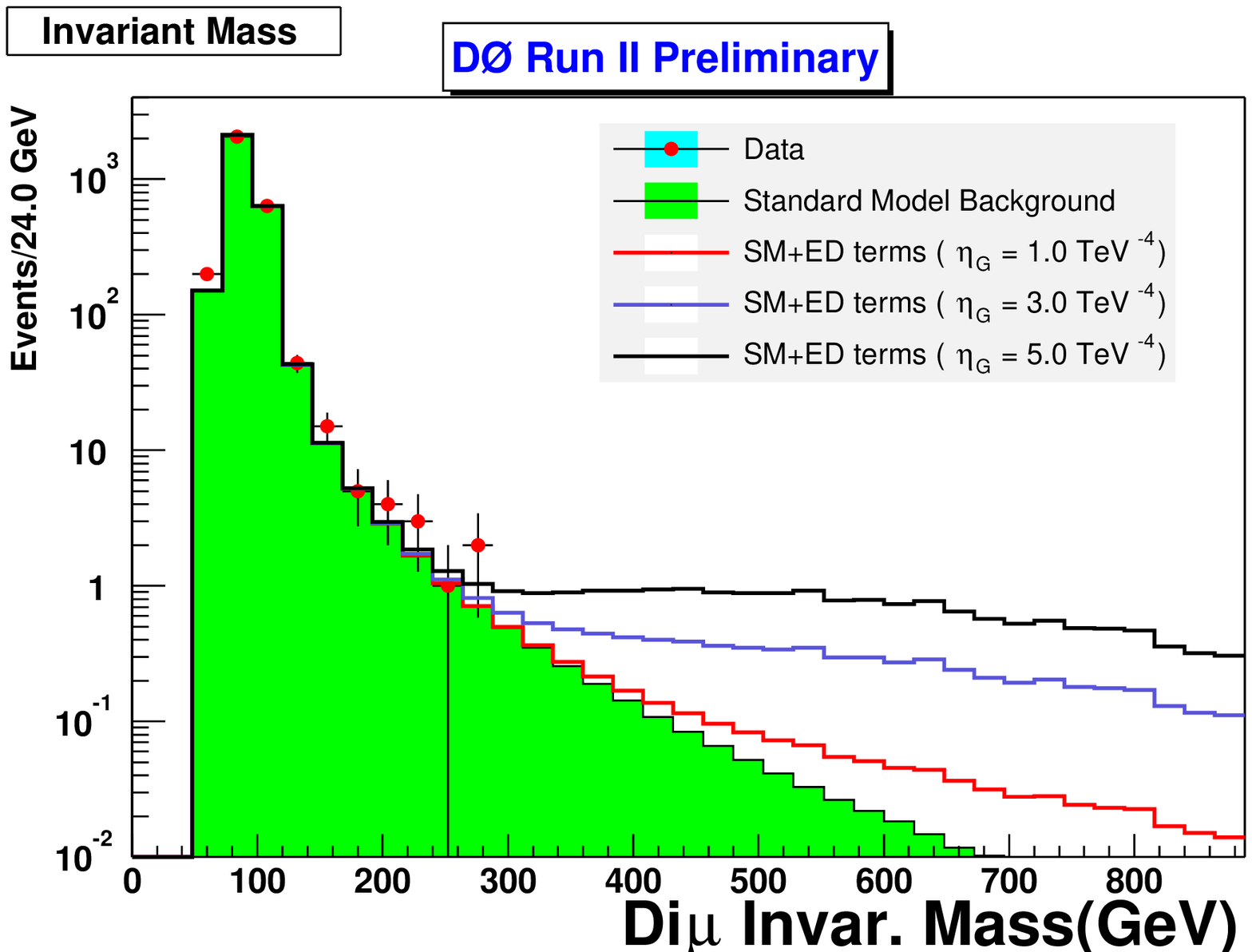}
    \caption{ 
The spectra of di-lepton invariant mass obtained in runI by the D0 experiment for electrons(left) and muons(right). 
}
    \label{fig:d0_dilep}
  \end{center}
\end{figure}
The production of two charged leptons at Tevatron proceeds mainly through the Drell-Yan mechanism~\cite{DY}. The neutral current annihilation $q\bar{q}\rightarrow \ell^+\ell^-$ proceeds mainly through the $\gamma^*$ or the $Z$ boson in the $s$--channel and gives rise to a pair of opposite charge leptons in the final state\footnote{The production of a pair of leptons by quark-antiquark annihilation, the Drell-Yan mechanism, can be used to extract information about the proton structure, in particular about the sea quark distribution~\cite{Towell:2001nh}}. 
\par
The Z production cross section at $\sqrt{s}=1.96$~TeV has been measured using the most recent data from Tevatron run~2 : $\sigma(p\bar{p}\ra Z + X \ra e^+e^- + X) = 267\pm 23$~pb.
 This  measurement is in agreement with the \sm prediction of $252.1\pm 8.8$~pb that takes into account the NNLO--QCD corrections~\cite{Martin:2002dr}.
\par
The events with high mass $\ell^+\ell^-$ pairs are potential candidates for the physics beyond the \sm. The search for such events  has been performed on run~1 data~\cite{Abe:1997gt} and repeated on the recent run 2 data. The invariant mass spectra for the di-electron and di-muon channels are shown in figure~\ref{fig:d0_dilep}. Events with di-lepton invariant masses up to 500~GeV have been measured. The forward backward charge asymmetry can be measured and used to extract information about possible non-standard contribution~\cite{Affolder:2001ha}.

\subsection{Events with charged leptons and missing energy}
\par
The search for events with at a lepton and missing transverse momentum has been performed on the run~1 data~\cite{Affolder:2001gr} and recently updated with the new run 2 data. The main \sm process that contribute to this final state is the production of W boson through a Drell-Yan annihilation $q\bar{q'}\ra W \ra \ell\nu$. The main quark flavour pairs that produce W's are $u\bar{d}\ra W^+$ and $d\bar{u}\ra W^-$. By measuring the lepton charge asymmetries in the $W$ production useful information is provided for the valence quark flavour composition of the proton~\cite{Abe:1998rv}
.
\par
\begin{figure}[hhh]
  \begin{center}
 \includegraphics[width=0.47\textwidth]{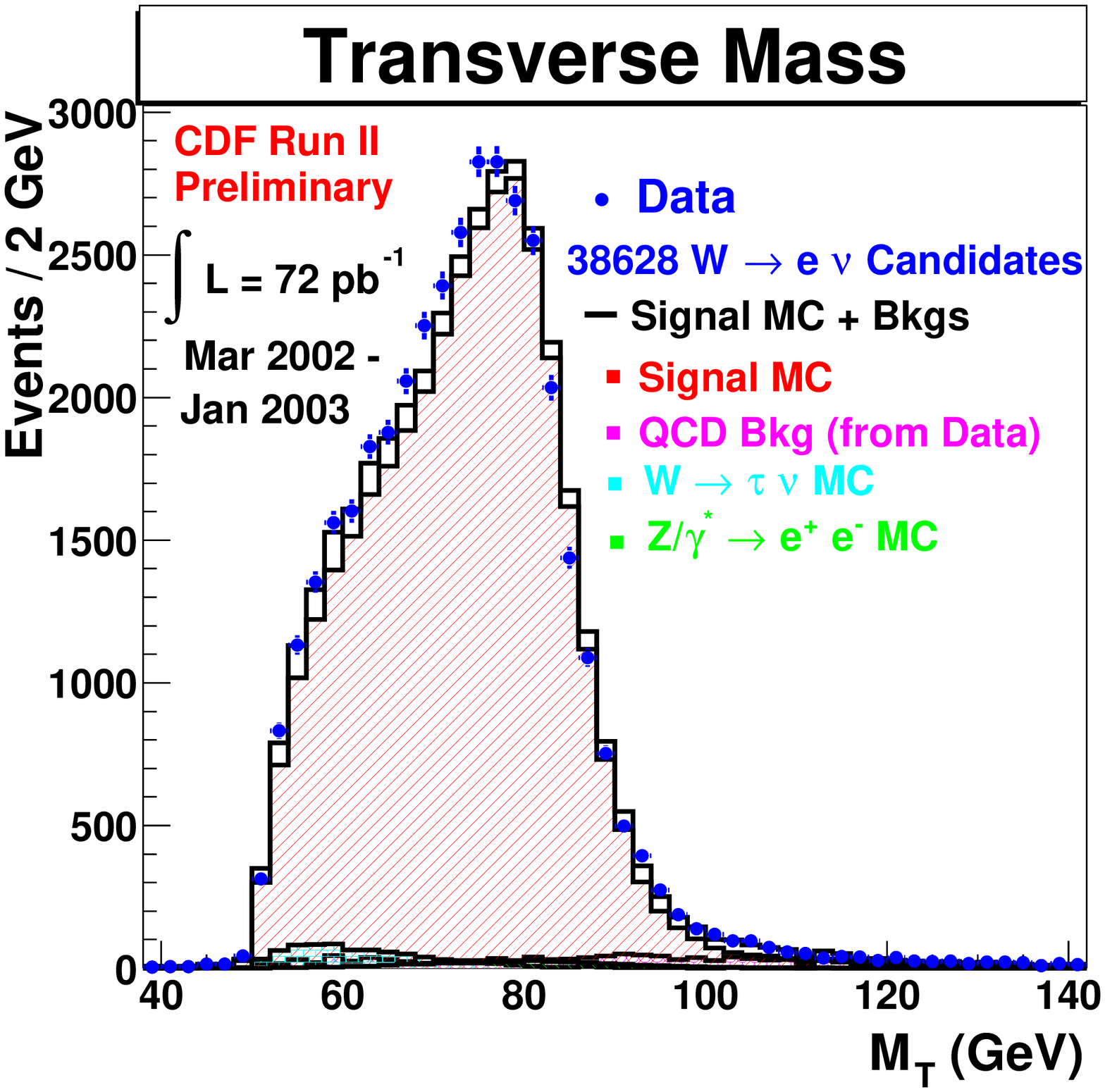}
 \includegraphics[width=0.63\textwidth]{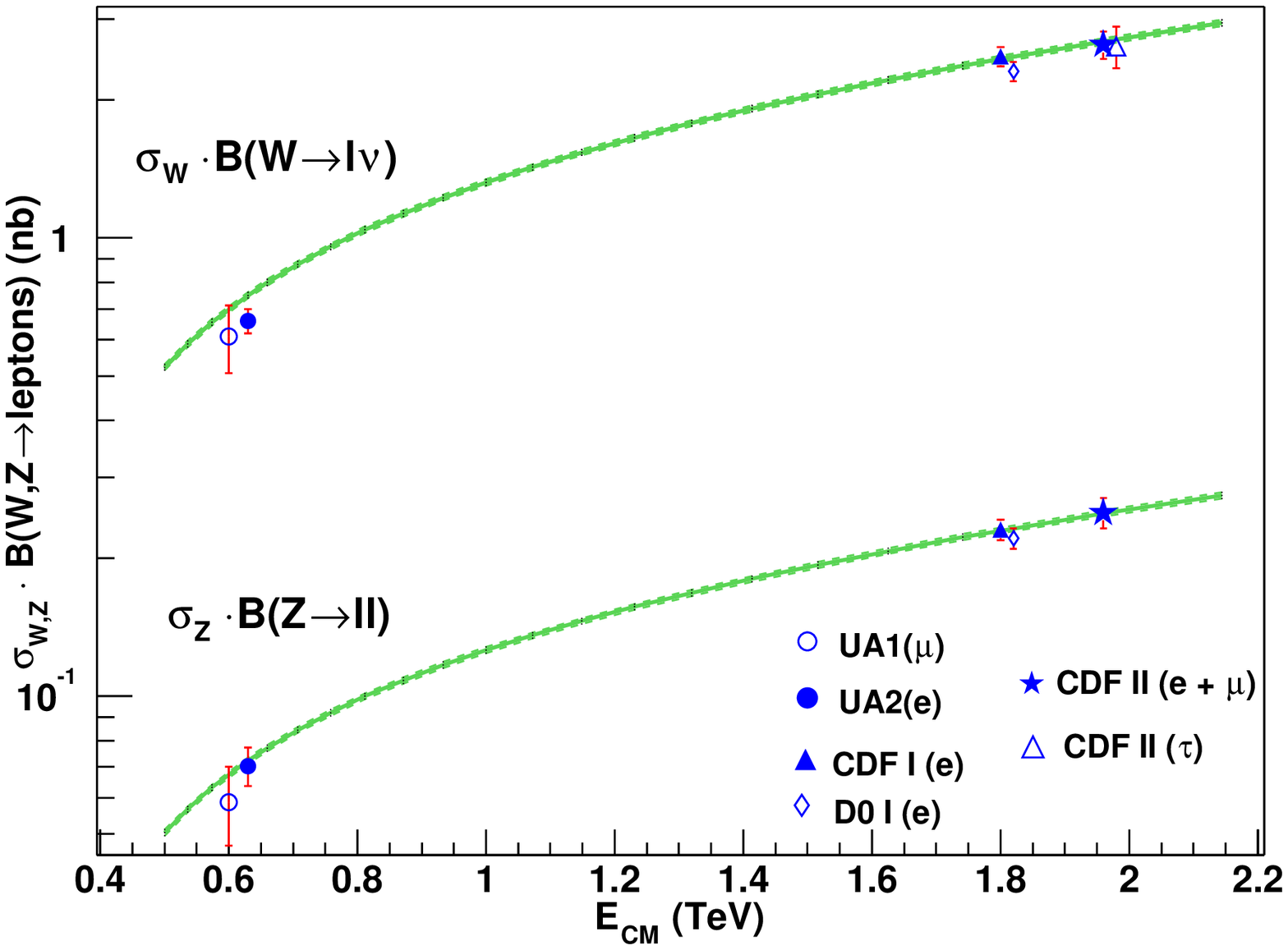}
    \caption{ 
Left: the transverse mass distribution of events with isolated leptons and missing transverse energy produced in $p\bar{p}$ collisions at Tevatron RUN II ($\sqrt{s}=1.96$~TeV). Right: the cross sections of weak boson(s) production in $p\bar{p}$ collisions as a function of $\sqrt{s}$. 
}
    \label{fig:tev_bos}
  \end{center}
\end{figure}

Events with isolated leptons (electrons or muons) at high transverse momentum $P_T^\ell>25$~GeV and significant transverse energy $P_T^{\mathrm miss}>25$~GeV are selected. The transverse mass is shown in figure~\ref{fig:tev_bos}(left) for an analysis done by the CDF collaboration on run 2 data~\cite{eperez:lp03}.
\par
The weak boson production is a powerful test of both the electroweak sector of the \sm and the proton stucture. The cross sections of the weak boson production in $p\bar{p}$ collisions is shown in figure ~\ref{fig:tev_bos}(right) as a function of $\sqrt{s}$.
The main systematical error is due to the luminosity measurement. The ratio of $W$ to $Z$ production cross sections is insensitive to this error and can be used to extract the $W$ boson width.
\begin{figure}[hhh]
  \begin{center}
 \includegraphics[width=0.9\textwidth]{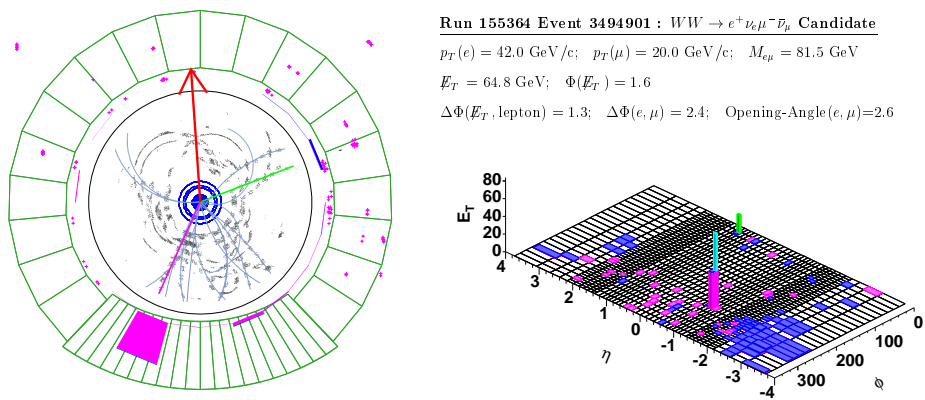}
    \caption{ 
A $WW$ candidate in $e\mu+P_T^{miss}$ channel at CDF.
}
    \label{fig:ww_cdf}
  \end{center}
\end{figure}

\par
The production of two charged leptons in events with missing transverse energy has also been studied at the Tevatron. The signal within the \sm is the $W$ pair production. This channel has a potential interest in triple gauge couplings ($\gamma WW$) and Higgs production at the Tevatron ($H\ra W^+W^-$).
 In a recent CDF analysis, five candidate events are found against an expected $WW$ signal of $6.89\pm 1.53$ events and a background of $2.34\pm 0.38$ events in a run~2 data sample with a luminosity of approximately ~126 pb$^{-1}$. The cross section is measured:
\begin{displaymath}
\sigma^{p\bar{p}{\rightarrow}WW}_{meas} = 5.1^{+5.4}_{-3.6}\;{\rm (stat)}\;\pm\;1.3\;{\rm (syst)}\;\pm\;0.3\;{\rm (lumi)}\;{\rm pb}\;\;
\end{displaymath}
 in agreement with the NLO-QCD calculation~\cite{Campbell:1999ah}:
\begin{displaymath}
\sigma^{p\bar{p}{\rightarrow}WW}_{theo:NLO} = 13.25 \pm 0.25 \;{\rm pb}\;.
\end{displaymath}
A display of a $WW$ candidate event found in run 2 data with an electron, a muon and missing transverse momentum in shown in figure~\ref{fig:ww_cdf}.
\par
The production of top quarks can also lead to events with leptons and missing energy in the final state, if one or both of the top quarks decay semileptonically $t\ra b W \ra b l \nu$. The top candidate events also have at least two jets in the final state. The leptonic channel profits from a small and well understood background and can be used for the measurement of the top mass by a fit to the kinematical distributions~\cite{Abbott:1998dn} of the selected events. In the present run 2 data  sample (CDF,${\cal L}=126\pbi$) at $\sqrt{s}=1.96$~TeV, a few top candidates have been identified in the semileptonic channel and the production cross section has been measured: 
\begin{displaymath}
\sigma^{p\bar{p}{\rightarrow}t\bar{t}} = 7.3 \pm 3.4 {\small \rm (stat.)} \;\pm\; 1.7{\small \rm (syst.)}{\rm pb}\;.
\end{displaymath}

The top quark can also be  single produced in association with a $b$ quark through an $s$-channel $W$ boson splitting or via and $W$-gluon fusion mechanism~\cite{Stelzer:1998ni}. The single top is therefore produced through electroweak mechanisms and the production rate is sensitive to the $V_{tb}$ CKM matrix element. The search for  single top production is difficult due to the high background from top pair production~\cite{Acosta:2001un,Abazov:2001ns}.  The predicted cross section is around 2~pb and no signal has been found at present. The existing limits of the single top  production cross section obtained from the data are typically around 20~pb.

\section{Lepton production at HERA}

In electron-proton collisions, events with high energetic leptons in the final state are mainly produced through the deep inelastic scattering process. In the interaction process, the exchanged virtual boson resolves the quark structure of the proton. In this case the proton can be seen as a flux of quarks. The hard $eq$ interaction may proceed through neutral currents $eq\rightarrow eq$ (NC) or charged currents $eq\rightarrow \nu q'$ (CC). In the case of a large virtuality ($Q^2$) of the exchanged boson, the scattered lepton has a large transverse momentum and leads to a prominent experimental signature: an isolated electron in the central detector for NC and significat missing transverse momentum in the case of CC, due to the undetected neutrino.

\begin{figure}[hhh]
  \begin{center}
    \includegraphics[width=0.8\textwidth]{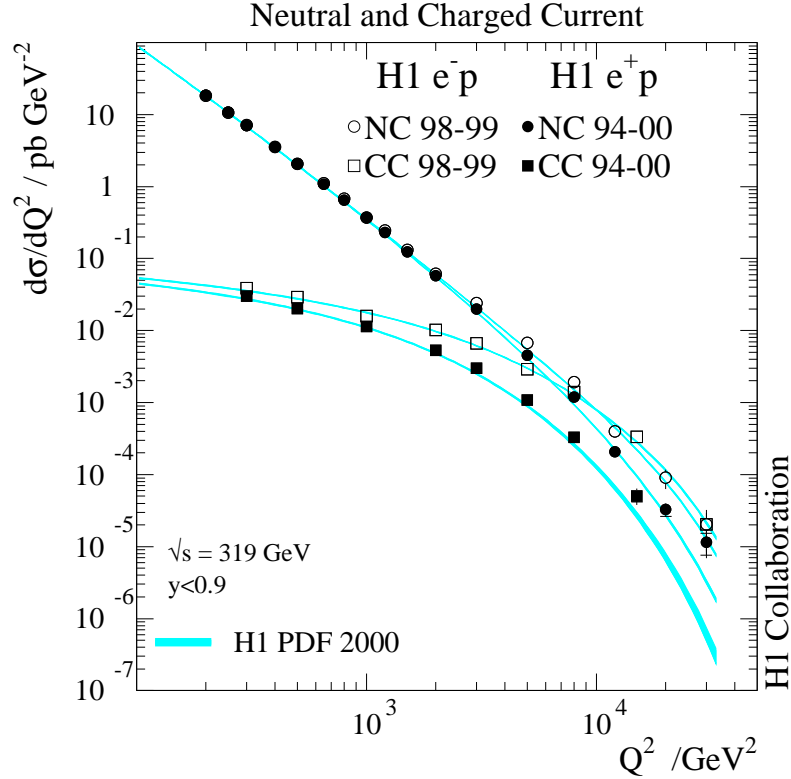}
    \caption{ 
The $Q^2$ dependences of the NC(circles) and CC(squares) cross sections $d\sigma/d Q^2$ shown for the combined 94--00 $e^+p$ (solid points) and 98--99  $e^-p$ (open points) data. The results are compared with the corresponding \sm expectations determined from the H1 PDF 2000 fit. 
}
    \label{fig:ccnc_q2}
  \end{center}
\end{figure}

\par
The CC interaction proceeds only through weak interactions ($t$-channel exchange of a $W$ boson) and is therefore in general much less intense than the NC process which has the important electromagnetic component ($\gamma^*$ exchange) together with the weak contribution ($Z$ exchange).
Only at very large virtualities $Q^2\simeq M_{W,Z}^2$ the CC and NC interactions have comparable rates. This can be seen in figure~\ref{fig:ccnc_q2} where the cross sections for the NC and CC interactions  are shown as a function of $Q^2$~\cite{Adloff:2003uh}. This figure beautyfully show the electroweak scale unification. At high $Q^2\simeq M_{W,Z}$, the parity violating couplings play a role and the differences between the cross sections measured in $e^+p$ and $e^-p$ collisions are observed also for NC.
\par
The production of more than one lepton ($\Delta N_L>0$) in $ep$ collisions is a higher order process with respect to CC/NC processes and appears therefore at a much lower rate. The production mechanisms involve both boson--boson fusion and boson conversion. The production mechanisms within the \sm framework together with the experimental observations are described below.

\subsection{Events with isolated leptons and missing transverse momentum}

\subsubsection{\sm mechanisms}

\begin{figure}[hhh]
  \begin{center}
    \includegraphics[width=0.85\textwidth]{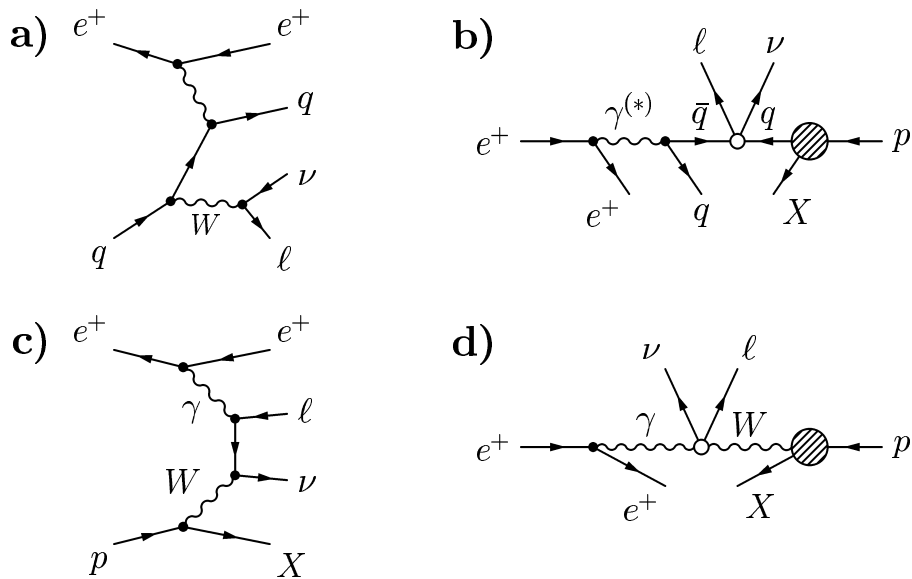}
    \caption{ 
The schematic view of the lepton-neutrino production in $ep$ collisions. The real $W$ boson production as radiation from the quark line (a) is mainly produced in photon-production process sketched in diagram (b). The non-resonant $\ell \nu$ pair can also be produced in a $t$-channel fermion exchange between two boson $\gamma-W$. This corresponds to a boson-boson fusion sketched in (d). A similar mechanism yields real $W$ bosons through triple boson couplings $WW\gamma$.
}
    \label{fig:whera-diag}
  \end{center}
\end{figure}

At parton level a lepton-neutrino pair can be produced via the
reactions
$ e^+ q \rightarrow e^+ q' \ell \nu $. The typical Feynman diagrams together with schematic view of the poles are presented in figure~\ref{fig:whera-diag}. 
 The dominant contribution is due to the real $W$ boson production and its subsequent leptonic decay, shown in figure~\ref{fig:whera-diag} (a and b). The boson--boson fusion, shown in figure~\ref{fig:whera-diag} (c and d), count for about 5\% and yields events with flat distribution of the $\ell\nu$ invariant mass. The CC-like reaction $ e^+ q \rightarrow \overline{\nu}  q' \ell \nu $ involves an extra  $W$ boson propagator connected to the initial electron in the diagrams and is therefore supressed. 

\par
The real $W$ production mainly proceeds through the coupling of the $W$ boson to the quark line, shown in figure~\ref{fig:whera-diag}(a). It is dominated by the photoproduction mechanism: the photon entering the reaction from the electron side (figure~\ref{fig:whera-diag}(b)) is close to its mass-shell ($Q^2\simeq 0$) and fluctuates to a $q\bar{q}$ pair. The $W$ boson is then produced in a   $q\bar{q'}$ fusion between a quark originating from the photon hadronic fluctuation and another (different flavour) quark from the proton. Interestingly enough, the production mechanism is of hadronic nature: the photon provides the hadronic flux in front of the incoming proton. Due to the $q\bar{q'}$ knock-on nature of the hard scattering, the transverse momenta of the proton or photon rests are typically small and therefore the transverse momentum of the hadronic final state is expected to be close to zero. 
\par
In case of large photon virtuality, the $W$ production is a DIS-like scattering with a $W$-strahlung from the quark line. In this case the scattered electron is visible in the detector together with the charged lepton issued from the $W$ decay. This is the case for 25\% of the simulated events in H1 detector.
In this configuration, the hadronic system $X$ can also acquire important transverse momentum.
\par
The production of a $\ell-\nu$ pair in boson--boson  ($\gamma-W$ or $Z-W$) collision is sketched in the diagrams c) and d) in figure~\ref{fig:whera-diag}. This non-resonant production contributes to less than 5\% to the total cross section.
The $W$ boson can also be produced by a triple boson coupling $WW\gamma$, by the $\gamma-W$ fusion. HERA sensitivity to an anomalous coupling will be discussed in the next chapter.
\par 
The first complete calculation of $W$ production in $ep$ collisions has been done in~\cite{Baur:1992pp}. An event generator (EPVEC) has been used to interface the calculation to the full detector simulation. The EPVEC generator includes the QCD parton shower simulation in the final state~\cite{Diaconu:1998kr}.
\par
  The SM prediction for $W$ production via
  $ep\rightarrow e W^\pm X$ is calculated by using a next to leading
  order (NLO) Quantum Chromodynamics (QCD) calculation
  \cite{Diener:2002if} in the framework of the EPVEC
  \cite{Baur:1992pp} event generator. Each event generated by EPVEC
  according to its default LO cross section is weighted by a factor
  dependent on the transverse momentum and rapidity of the $W$
  \cite{Diener:2003df}, such that the resulting cross section
  corresponds to the NLO calculation. 
\par 
  The NLO corrections are found to be of the order of $30\%$ at low
  $W$ transverse momentum (resolved photon interactions) and typically
  $10\%$ at high $W$ transverse momentum (direct photon interactions)
  \cite{Diener:2002if,Spira:1999ja,Nason:1999xs}.  The NLO calculation reduces the theory error
  to $15\%$ (from $30\%$ at leading order).
 \par
  The charged current process $ep\rightarrow \nu W^\pm X$ is
  calculated with EPVEC \cite{Baur:1992pp} and found to contribute
  less than $7\%$ of the predicted signal cross section.
 \par
  The total predicted $W$ production cross section amounts to $1.1$ pb
  for an electron--proton centre of mass energy of $\sqrt{s}=300$ GeV
  and $1.3$ pb for $\sqrt{s}=318$ GeV.
\par Another possible contribution to the signal in the electron channel only is the production of the Z boson with the subsequent decay to neutrinos: $ep \ra e Z X \ra e \nu\bar{\nu} X $. In this case the isolated lepton is the ``scattered'' electron. This process is also calculated in EPVEC. The main contribution at large transverse momenta is not from the photoproduction mechanism, for which the scattered electron tend to be rather backward\footnote{The origin of the coordinate system at HERA experiments is the nominal
$ep$ interaction point. The direction of the proton beam defines
the positive $z$--axis (forward direction).}, but the Cabibbo-Parisi process, induced by a fluctuation of a photon from the proton into an $e^+e^-$ pair, with the annihilation of one of those electron with the beam positron to produce the $Z$ boson\footnote{see also next section for the explanation of this mechanism}. The contribution of this process to the cross section is around 3\%.

\subsubsection{First observations}
The observation of events with isolated leptons and missing transverse momentum has been made first at HERA back in 1994. In a sample of CC events with a transverse momentum in the calorimeter $P_T^{calo}$ above 25~GeV, a spectacular event  with an isolated muon in the final state has been observed~\cite{Ahmed:1994mx}. The event was detected in a data sample corresponding to an integrated luminosity of 4~pb$^{-1}$. The sum of the expected contributions from the \sm processes was estimated to be 0.03 events.
\begin{figure}[hhh]
\centering
\includegraphics[width=0.45\textwidth]{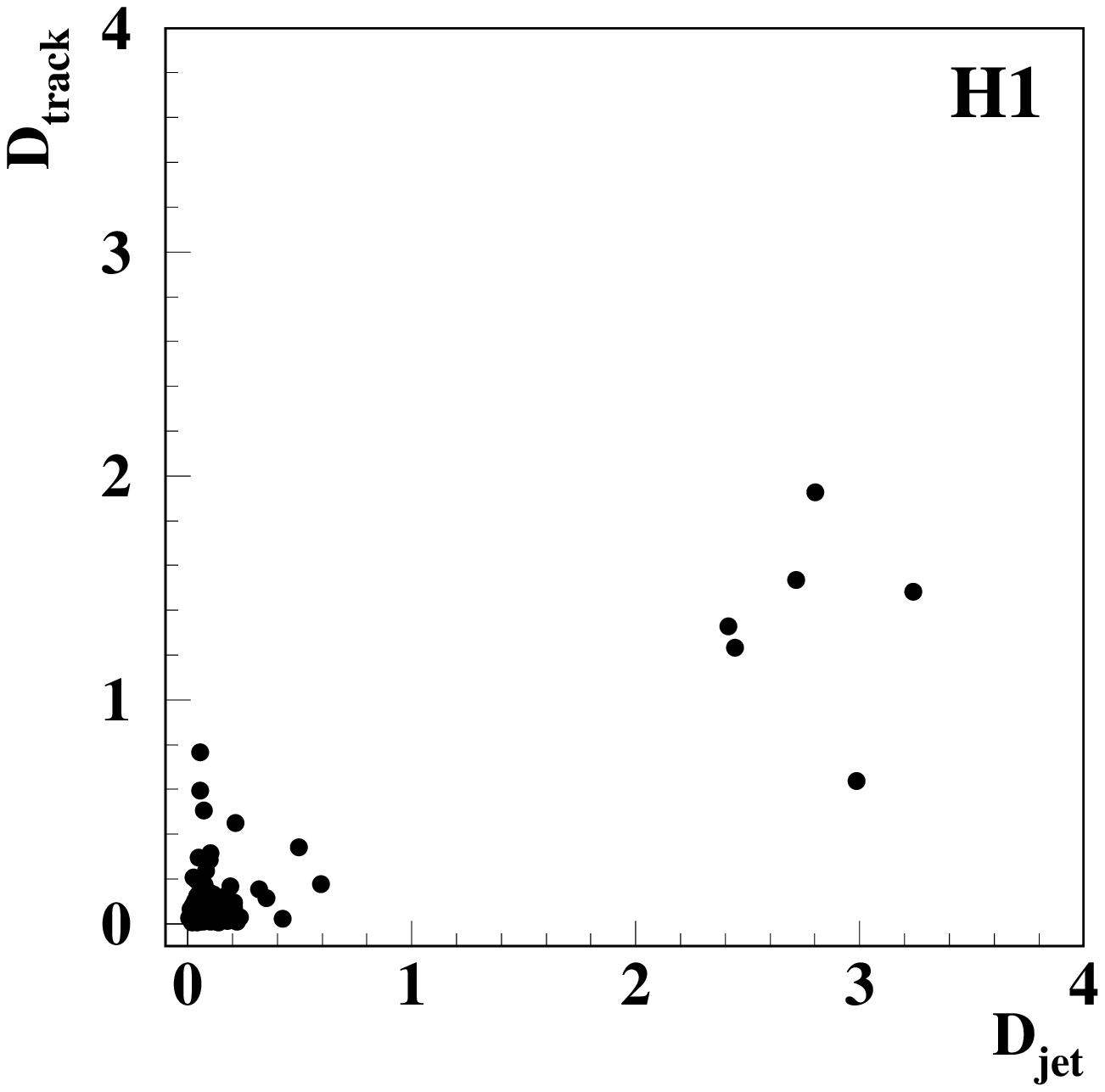}
\includegraphics[width=0.65\textwidth]{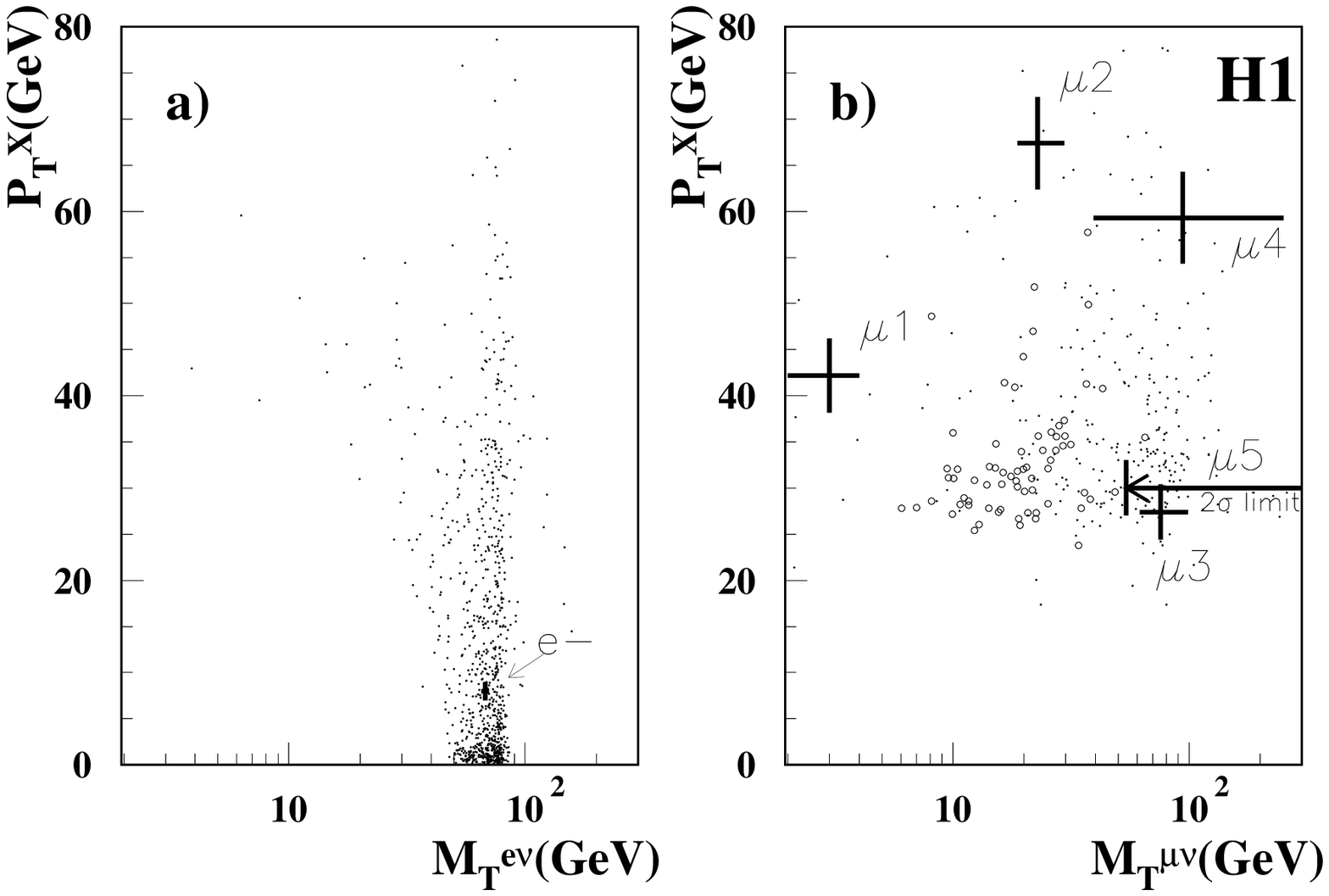}
\caption{ Early observations of events with isolated leptons and missing transverse momentum by H1. Left: correlation between the distances $D_{jet}$ and $D_{track}$
to the closest hadronic jet and track, for
all high-$P_T$ tracks in the inclusive event sample.
Right: distribution of  the selected events
 in  $P_T^{X}$ and $M_T^{l\nu}$ (see text): a) electron channel; b) muon channel.
}
\label{fig:isol}
\end{figure}
 \par
The analysis of a data sample collected in the period 1994-1997 in $e^+p$ collisions  corresponding to an integrated luminosity of 37~pb$^{-1}$ was performed for the phase space with large imbalance in the calorimetric transverse momentum $P_T^{calo}>25$~GeV~\cite{Adloff:1998aw}. The isolation of the charged tracks with large transverse momenta $P_T^{track}>10$~GeV  was calculated in each event with respect to the closest track ($D_{track}$) and with respect to the closest hadronic jet ($D_{jet}$). Figure \ref{fig:isol} shows the correlation between $D_{track}$
and $D_{jet}$ for the high-$P_{T}$  tracks in the inclusive event sample.
In most cases the high-$P_{T}$ tracks are not isolated.
This is expected, since the bulk of
events are  CC interactions with the  high-$P_{T}$ track
located within or close to the  hadronic shower.
However, six high-$P_{T}$ tracks are found  in a region well
separated from all other charged tracks and from hadron jets.
They belong to six events each with one single isolated
high-$P_{T}$ charged particle. All six isolated high-$P_{T}$ particles fulfill 
the lepton identification
criteria. One is an electron candidate and five are muon
candidates. 
\par
 The total yield of events
 with one isolated high-$P_T$ lepton which is
 expected from SM
processes is $2.4 \pm 0.5$ in the $e^{\pm}$ channel and $0.8 \pm 0.2$
in the $\mu^{\pm}$ channel. The main contribution is due to $W$ production,
estimated in leading order to be
$1.7 \pm 0.5$ and $0.5 \pm 0.1$ events respectively.
\par
The kinematics of the observed events is compared to the \sm prediction in figure~\ref{fig:isol}. Shown is the distribution of the events in the hadronic transverse momentum $P_T^X$ and the transverse mass~\footnote{ The transverse mass is calculated from the massless four vectors obtained by projecting the missing and the lepton momenta on the plane transverse to the beams direction:  $M_T^{\ell\nu} = \sqrt{(P_T^{miss} +P_T^{\ell})^2 -
(\vec{P}_T^{miss}+ \vec{P}_T^{\ell})^2}$} plane.  
The electron event and one of the muon events are found
  in a region of phase space likely to be
populated by $W$ production.
 Another muon event can, within its large measurement errors, also
be accommodated in a $W$ interpretation.
  The kinematic properties of the remaining three muon events,
together with the overall rate excess in the muon
channel, disfavour an  interpretation of these events
within the SM processes considered.

\subsubsection{Events with $\ell+P_T^{\mathrm miss}$ in HERA I data}

The search for events with isolated leptons and missing transverse momentum has been performed on the full HERA I dataset.
The H1 analysis~\cite{Andreev:2003pm} has been extended at lower transverse momentum $P_T^{\mathrm calo}>12$~GeV in order to increase the acceptance for the real $W$ boson production. The leptons with transverse momenta above 10~GeV are identified  in the polar angle domain $5^\circ<\theta<140^\circ$ using also the detection capabilities in the forward region. The background rejection is inforced~\cite{mireille:thesis} in order to cope with the huge cross sections of the photoproduction, NC and CC processes.
\par
The ZEUS analysis of the full HERA I data~\cite{Chekanov:2003yt} specifically searches for events with isolated leptons and missing transverse momentum having also a hadronic system at large transverse momentum. The analysis is done in the framework of the seach for anomalous top production. Events with significant calorimetric imbalance are selected with $P_T^{\mathrm calo}>20 GeV$. In each such event, the leptons are identified as isolated tracks with transverse momenta above 5~GeV in the polar angle range $17^\circ<\theta<115^\circ$. The tracks are required to be associated with specific calorimetric patterns for electrons or muons.
\begin{figure}[htbp]
\begin{center}
    \includegraphics[width=0.45\textwidth]{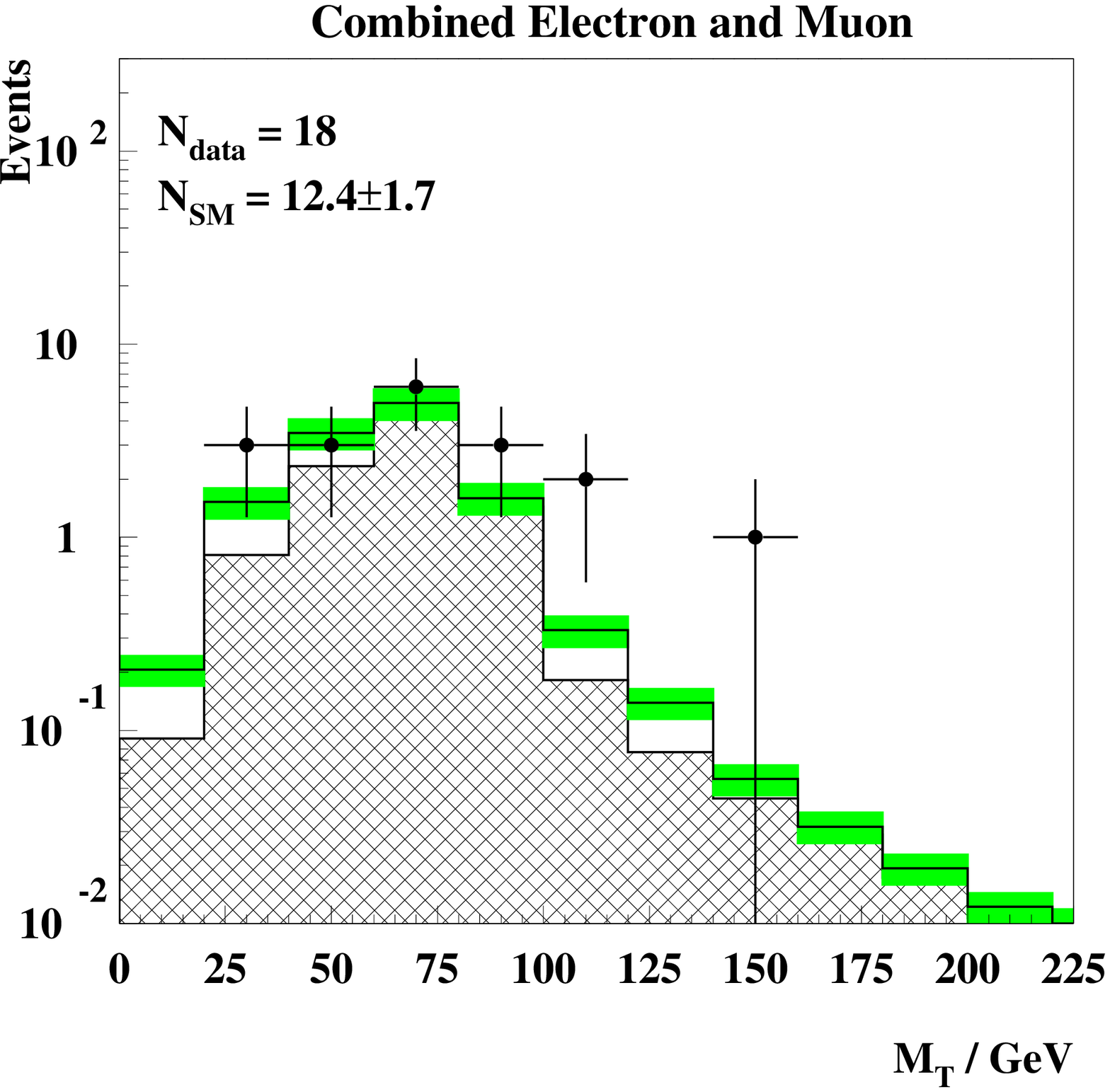}
    \includegraphics[width=0.45\textwidth]{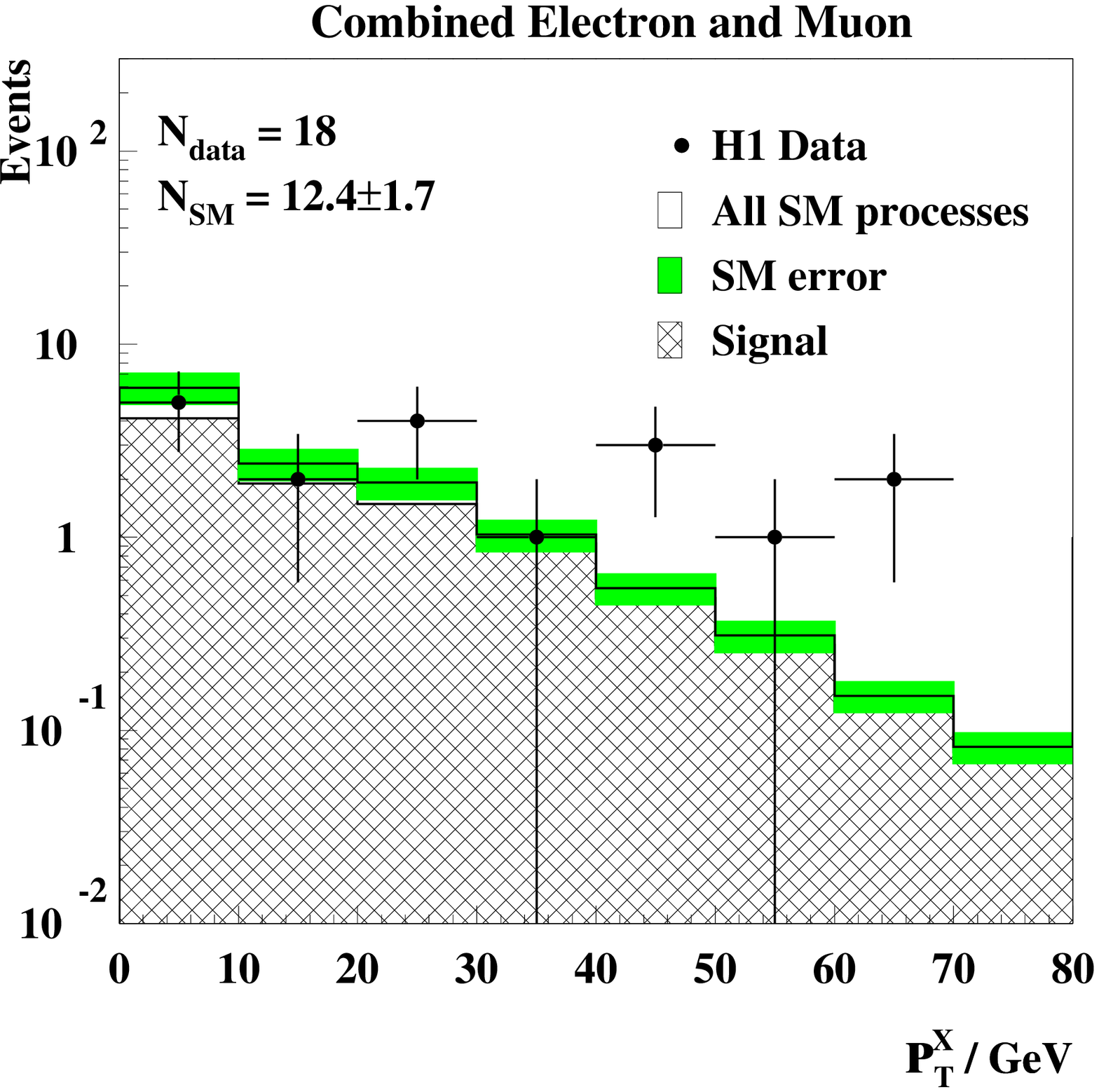}
    \includegraphics[width=0.45\textwidth]{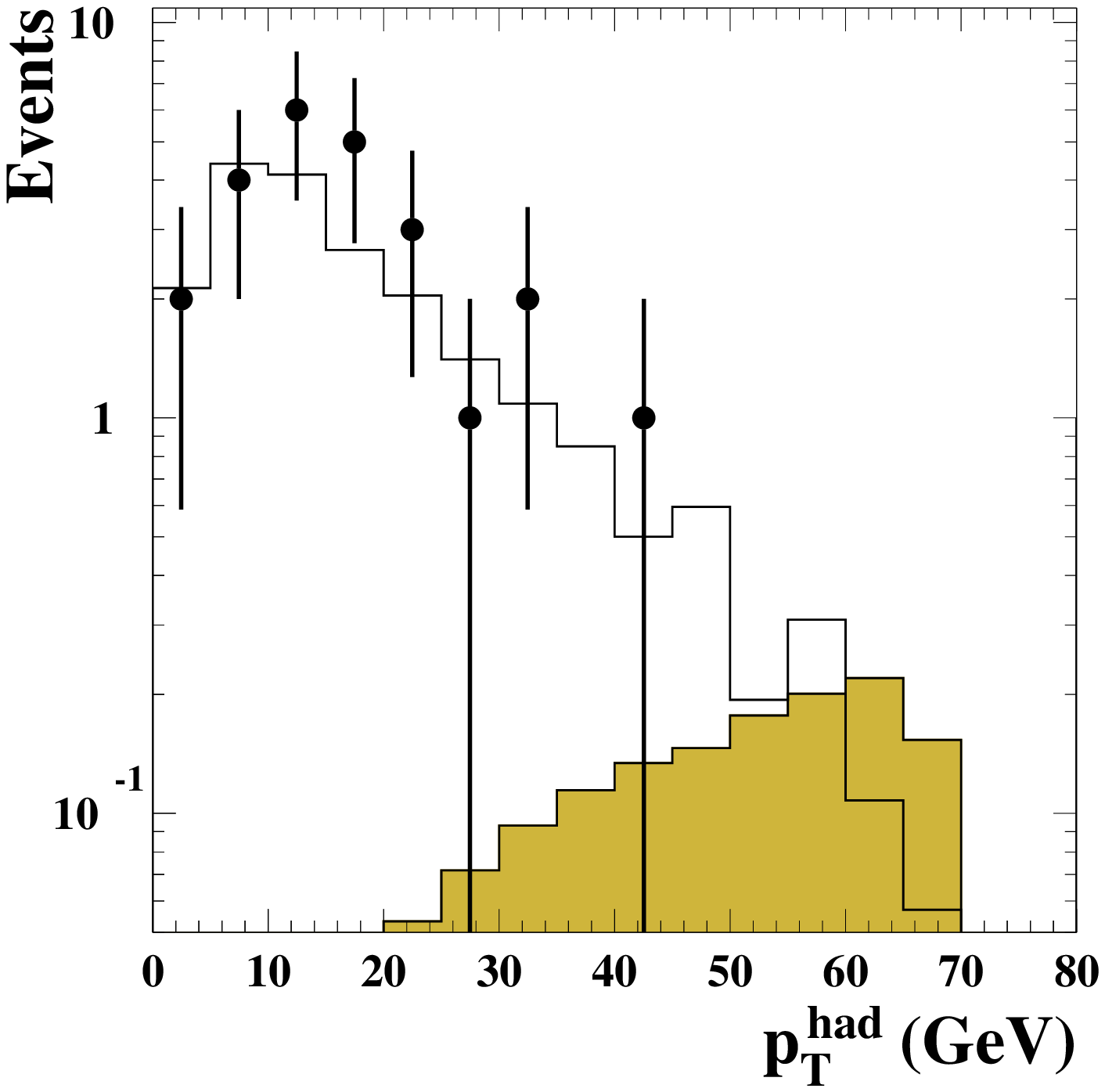}
    \includegraphics[width=0.45\textwidth]{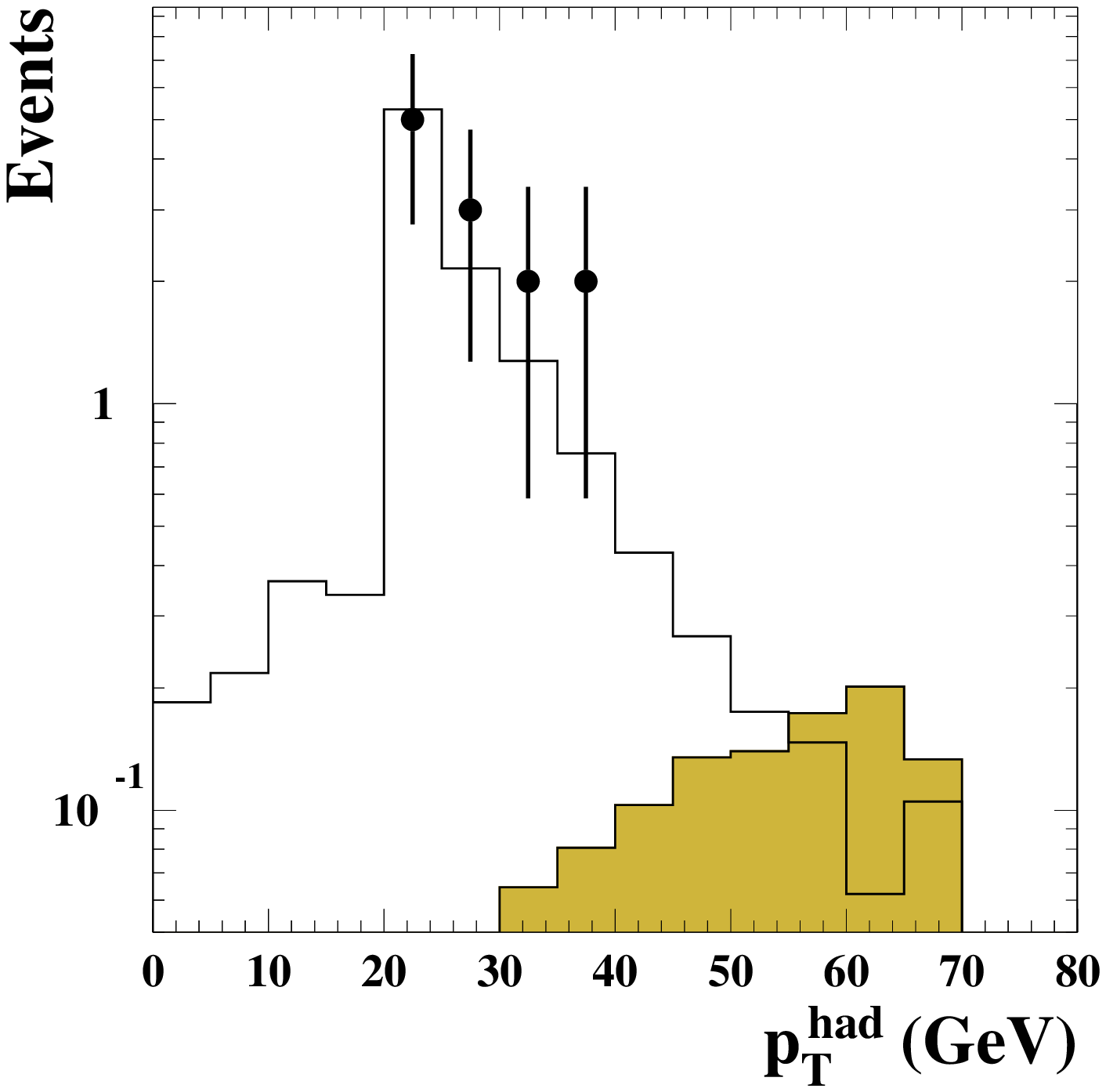}
    \includegraphics[width=0.8\textwidth]{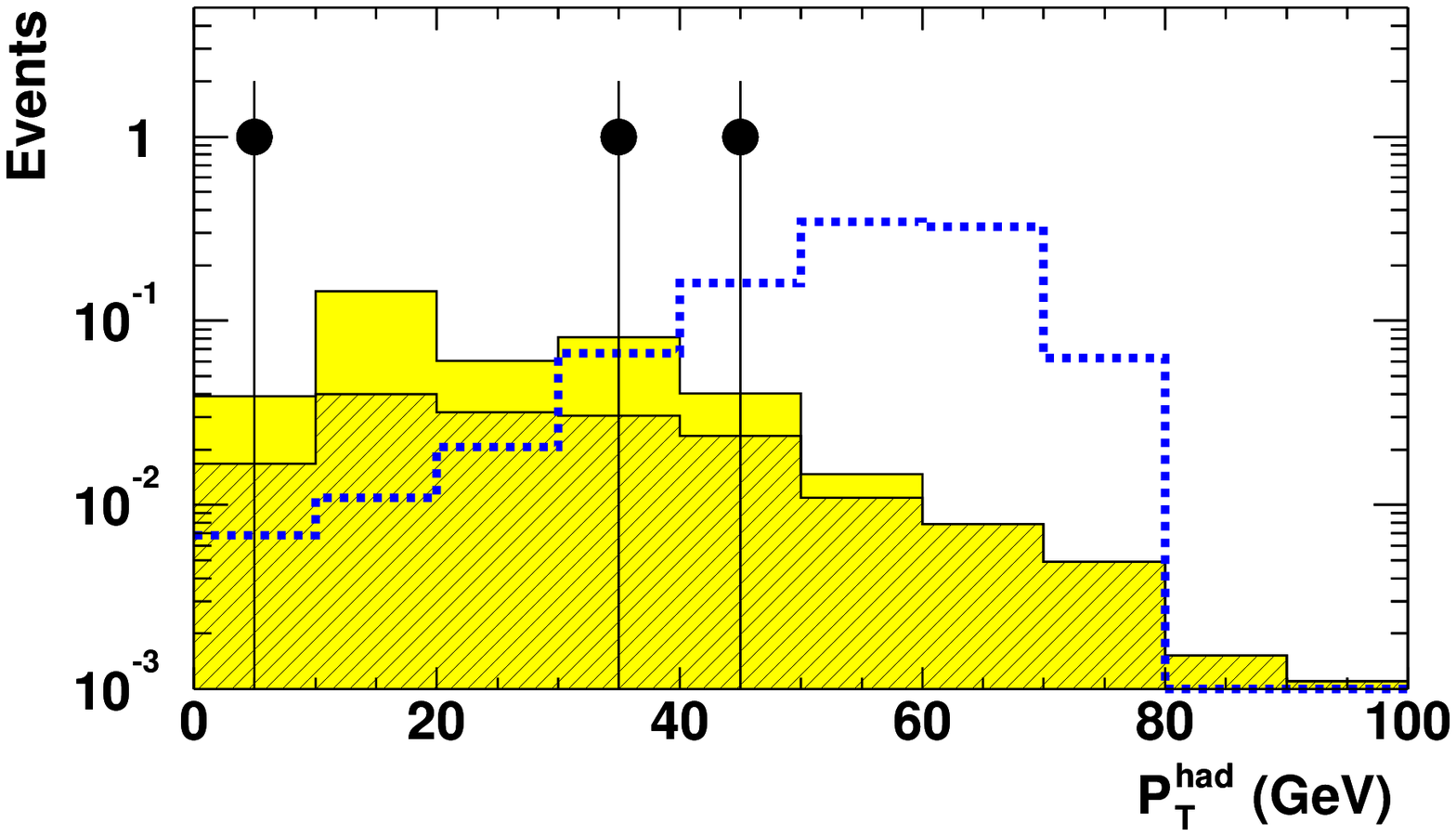}
\end{center}
\begin{picture} (0.,0.)
\setlength{\unitlength}{1.0cm}
\put (7.0,20.05){\bf\Large H1 $e+\mu$}
\put (5.0,13.05){\bf\Large ZEUS $e$}
\put (12.0,13.05){\bf\Large ZEUS $\mu$}
\put (11.0,5.55){\bf\Large ZEUS $\tau$}
\end{picture}
  \caption{The kinematical distribution of events with isolated leptons and missing $P_T$ for H1 and ZEUS analysis.
(the shaded ($e$ and $\mu$ plots) and the dashed ($\tau$) histograms in ZEUS plots represent the distributions of
the single-top MC normalised to an integral of one event).}
    \label{fig:ptx}
\end{figure}

\begin{figure}[hhh]
  \begin{center}
    \includegraphics[width=0.8\textwidth]{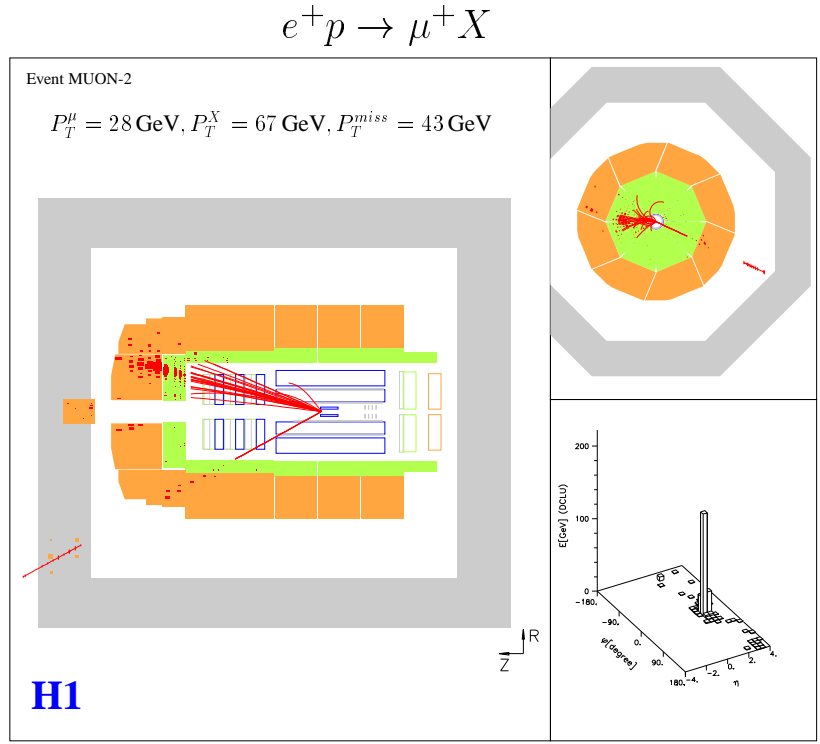}
\caption{ Display of an event with an isolated muon, missing transverse momentum and a prominent hadronic system. The acoplanarity between the lepton and the hadronic system is the signature of an undetected particle.}
\label{fig:isol_evd}
\end{center}
\end{figure}

\begin{table}
 \renewcommand{\arraystretch}{1.6}
\begin{center}
\begin{tabular}{|c|c|c|c|c|} \hline
  \multicolumn{2}{|c|}{ } & Electron & Muon & Tau \\
\multicolumn{2}{|c|}{1994-2000 $e^\pm p$} & obs./exp. & obs./exp. & obs./exp.  \\
\multicolumn{2}{|c|}{ } & {\footnotesize($W^\pm$ contribution)} & {\footnotesize ($W^\pm$ contribution) }&  {\footnotesize ($W^\pm$ contribution) }\\ \hline
                
                & Full sample &  11 / $11.54~\pm1.50~(71\%)$   & 8 / $2.94~\pm0.50~(86\%)$  &  \\ \cline{2-5}
{\large \bf H1} & $p_T^\mathrm{had}~>~25\gev$ &  5 / $1.76~\pm0.30~(82\%)$   & 6 / $1.68~\pm0.30~(88\%)$  &  \\ \cline{2-5}
{\footnotesize $118.4$~pb$^{-1}$}&  $p_T^\mathrm{had}~>~40\gev$ &  3 / $0.66~\pm0.13~(80\%)$   & 3 / $0.64~\pm0.14~(92\%)$  &  \\ \hline
\hline
    & Full sample &  24 / $20.6~^{+1.7}_{-4.6}~(17\%)$   & 12 / 11.9~$^{+0.6}_{-0.7}$ (16\%)  & 3 / 0.40~$^{+0.12}_{-0.13}$~(49\%) \\  \cline{2-5}
{\large \bf ZEUS } & $p_T^\mathrm{had}~>~25\gev$ &  2 / $2.90~^{+0.59}_{-0.32}~(45\%)$   & 5 / 2.75~$^{+0.21}_{-0.21}$ (50\%)  & 2 / 0.20~$^{+0.05}_{-0.05}$~(49\%) \\  \cline{2-5}
{\footnotesize $130.2$~pb$^{-1}$} & $p_T^\mathrm{had}~>~40\gev$ &  0 / $0.94~^{+0.11}_{-0.10}~(61\%)$   & 0 / $0.95~^{+0.14}_{-0.10}~(61\%)$  & 1 / 0.07~$^{+0.02}_{-0.02}$~(71\%) \\ \hline
\end{tabular}
\end{center}
\caption{Summary of the results of searches for events with isolated
leptons, missing transverse momentum and large $p_T^\mathrm{had}$ at HERA. The number
of observed events is compared to the SM prediction. The $W^\pm$
component is given in parentheses in percent. The statistical and systematic uncertainties added
in quadrature are also indicated.}
\label{tab:heraisolep}
\end{table}

\begin{table}[htb]
  \renewcommand{\arraystretch}{1.3}
  \begin{center}
    \begin{tabular}{|c||c|c||c|c|} \hline
      &  \multicolumn{4}{c|}{Cross Section  / pb}\\ \hline
      & Measured & SM NLO & SM LO  & SM LO \\
      & & & Diener {\it et al.} & Baur {\it et al.} \\ \hline
      \hline
      $P_T^X<25$~GeV    &  0.146 $\pm$ 0.081 $\pm$ 0.022  &  0.194 $\pm$ 0.029 & 0.147 $\pm$ 0.044 & 0.197 $\pm$ 0.059 \\ \hline
      $P_T^X>25$~GeV    &  0.164 $\pm$ 0.054 $\pm$ 0.023  &  0.043 $\pm$ 0.007 & 0.041 $\pm$ 0.012 & 0.049 $\pm$ 0.015 \\ \hline
    \end{tabular}
  \end{center}
\label{tab:h1_xsec}
  \caption{The measured cross section for events with an isolated high energy 
    electron or muon with missing transverse momentum (H1 collaboration). The cross sections are 
    calculated in the kinematic region: $5^\circ < \theta_l < 140^\circ$; 
    $P_T^l > 10$~GeV; $P_T^{\rm miss} > 12$~GeV and $D_{jet} > 1.0$. Also shown    are the signal expectations from the Standard Model where the dominant 
    contribution $ep \rightarrow eWX$ is calculated at next to leading order (SM NLO) \cite{Diener:2002if,Diener:2003df}
    and at leading order (SM LO) \cite{Diener:2002if} and \cite{Baur:1992pp}.}
\end{table}

\par
A search for events with isolated tau's at HERA has been recently published by ZEUS collaboration~\cite{zeus_tau}. In this case the isolated track is required to correspond to a calorimetric deposit compatible with a narrow jet with small track multiplicity produced by a hadronic tau decay. The discrimination of the tau signal from the quark induced jets 
is done by using a multivariate analysis based on six variables related to the shape of the calorimetric deposit. The analysis of the H1 data in the $\tau$ channel is currently in progress.
\par
The results of the search for isolated leptons with missing transverse momentum at HERA is summarized in table~\ref{tab:heraisolep} and figure~\ref{fig:ptx}. The data is compared with the \sm expectation for the full HERA I period. The phase space region at large hadronic transverse momentum is also shown. The H1 analysis has better non-$W$ background rejection at low and high transverse momentum. 
\par
With a more powerfull background rejection and an extended phase space at lower transverse momentum~\cite{mireille:thesis}, the H1 analysis provide clear evidence for the single $W$ boson production at HERA. The cross section measured at low and high hadronic transverse momentum is shown in table~\ref{tab:h1_xsec}. It is compared to the LO and NLO predictions and found in good agreement at low $P_T^X$ and slightly in excess (about $2\sigma$) at high $P_T^X$.
\par
Both experiments observe events with isolated leptons and missing transverse energy. Good agreement is found with the expectation from the \sm in the phase space at low hadronic transverse momentum $P_T^X<25$~GeV. An excess is observed by the H1 analysis at large trasverse momentum $P_T^X>25$~GeV. One such event is from the H1 analysis is shown in figure~\ref{fig:isol_evd}. The probability of the \sm expectation to fluctuate above the observed number of events is 10\% for the full sample and 0.15\% at $P_T^X>25$~GeV. The ZEUS analysis does not find such a prominent excess in the electron and muon channels, but observes tau events with large $P_T^X$. In total at HERA I, in the region $P_T^X>25$~GeV, 20 events with isolated leptons are observed for a total expectation of roughly $9.3\pm1.8$. The HERA II data, with an increase of a factor of ten in the integrated luminosity, will help in the clarification of this observation\footnote{The quantification of the excess based on the statistical $CL_b$ method applied to the $P_T^X$ spectrum can be found in~\cite{jochen:thesis}. Some possible scenarios of the excess and a discussion of the compatibility between H1 and ZEUS observations are studied in~\cite{dannheim:thesis,carli}.}.

\subsection{Events with several charged leptons}

\begin{figure}[hhh]
  \begin{center}
    \includegraphics[width=0.65\textwidth]{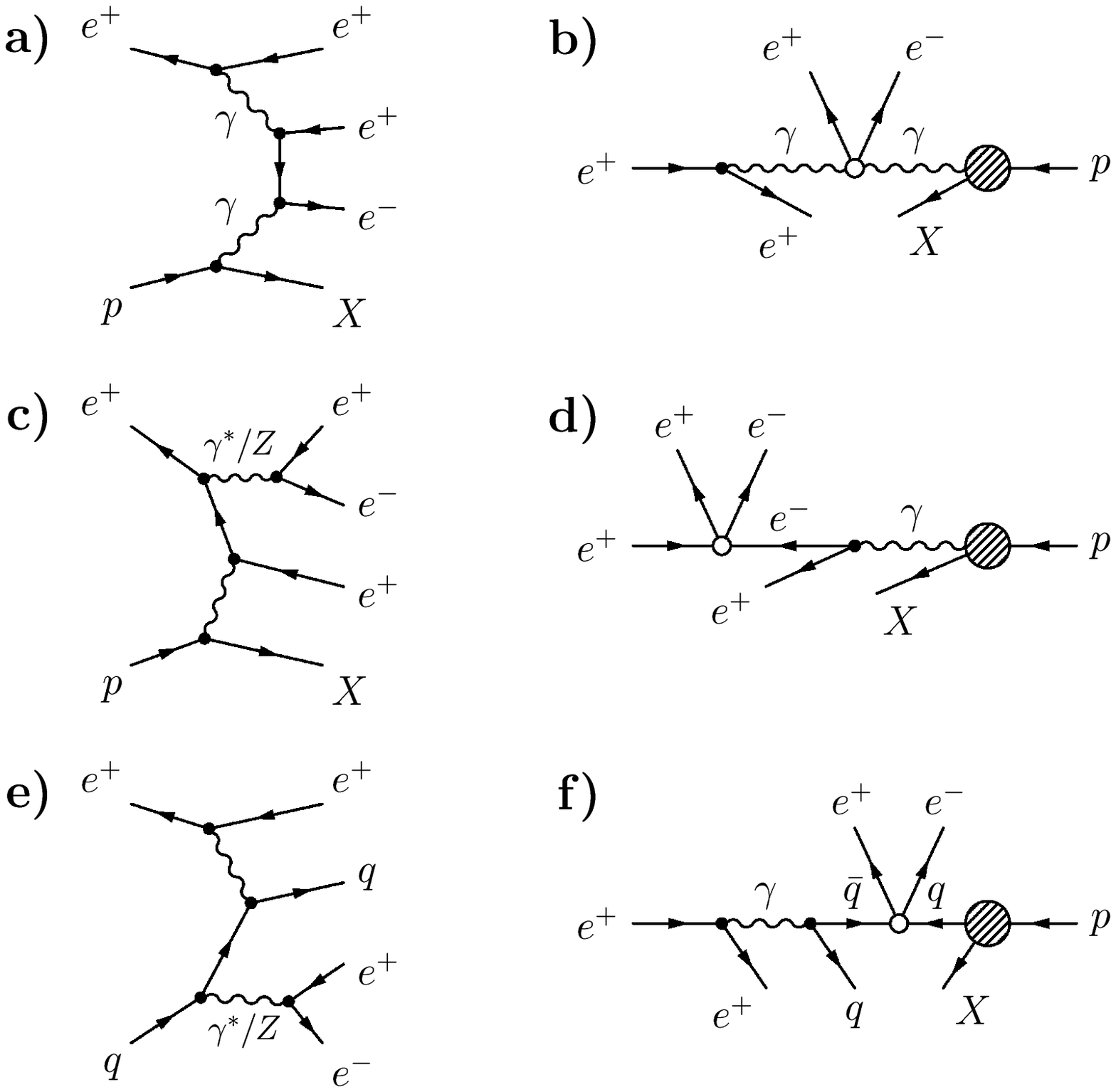}
    \includegraphics[width=0.45\textwidth]{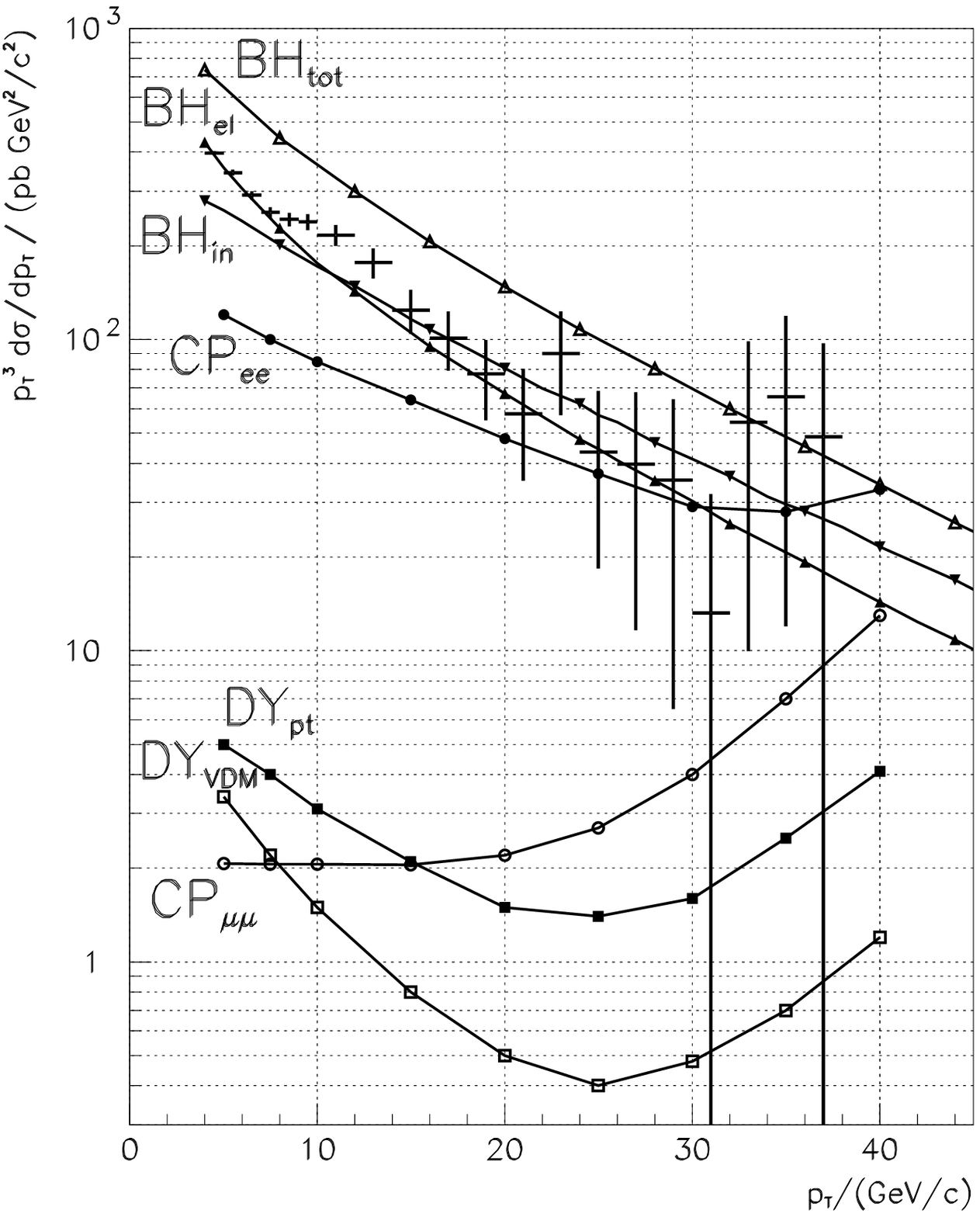}
    \caption{ 
 Main processes involved in lepton pair production (left and middle) and their cross sections(right). Example of Feynman diagrams (left) for a): photon--photon interaction;  c) and e): $\gamma^*/Z$ boson conversion. The kinematical poles for basic $2\rightarrow 2$  lepton pair production processes are sketched (middle) for b): photon--photon collisions $\gamma\gamma \rightarrow e^+e^-$ ;  d):
Cabibbo-Parisi $e^+e^- \rightarrow e^+e^-$ ; f): Drell-Yan  $q\bar{q} \rightarrow e^+e^-$.
}
    \label{fig:mel-diag}
  \end{center}
\end{figure}

The main SM processes involved in multi--electron production at HERA
are summarized in figure~\ref{fig:mel-diag}.
\par
The dominant contribution is
the interaction of two photons radiated from the incident electron and proton, also called Bethe-Heitler (BH) contribution and shown in diagram~\ref{fig:mel-diag}a. The kinematical pole corresponds to a two photons collision sketched in figure~\ref{fig:mel-diag}b. The hadronic final state (X) can be  a proton  (elastic process),  a proton resonance (quasi--elastic process) or a high mass system in case of high virtuality of the photon from the proton side (inelastic process).

\par
The produced electron pair can also originate from a $\gamma^*/Z$ boson, radiated  either from the electron line (diagram~\ref{fig:mel-diag}c ) or from the quark line (diagram~\ref{fig:mel-diag}e ).   This process, also called internal conversion, has two kinematical poles.
First, the Cabibbo--Parisi (CP) process, sketched in figure~\ref{fig:mel-diag}d), involves an $e^+e^-$
interaction where one of the electrons is issued from a photon radiated
from the proton\footnote{Pairs of leptons for which the kinematics do not correspond to the $\gamma\gamma$ collisions have first been observed in $e^+e^-$ collisions at ADONE and interpreted as an underlying $e^+e^-$ annihilation by Cabibbo and Parisi in a private communication to the experimentalists~\cite{Bacci:1972xx,Barbiellini:1974zp}}.
Its contribution is one order of magnitude lower than the photon-photon contribution, except at high transverse
momentum where it contributes more significantly due to the s-channel Z boson contribution.
The second pole of internal conversions is the Drell--Yan (DY) process
(figure~\ref{fig:mel-diag}f), involving a quark--antiquark interaction where
the anti-quark (close to its mass shell) is issued from a photon radiated from the incident electron~\footnote{The contribution from Drell-Yan process has been proposed as an interesting probe of proton or photon structure at HERA~\cite{Jaffe:1971we,Bawa:1993qr}.}.
Its contribution is small compared to photon-photon and Cabibbo-Parisi processes~\cite{Arteaga-Romero:1991wn}. 
\par The contribution of different processes can be seen in figure~\ref{fig:mel-diag} (right)~\cite{Hoffmann:1999jg} where the differential cross section multiplied by the cube of the electron transverse momentum is shown as a function of electron $P_T$. The increase of the CP and DY contributions in the region of $P_T\simeq M_Z/2$ is due to the contribution of the Z propagator in $s$-channel for those processes. 
The multi--electron production at HERA has been first computed in the photon-photon mode in~\cite{Vermaseren:1983cz} and implemented in the LPAIR generator~\cite{Baranov:1991yq}. The full matrix element calculation, except for the Drell-Yan pole, is done in~\cite{Abe:2000cv} and implemented in the GRAPE generator which is used for signal simulation. The background is composed from NC events with a second fake electron from the hadronic final state and from Compton scattering $ep\ra e\gamma X$ where the photon is wrongly identified as the second electron in the event.

\subsubsection{Measurement of multi-electron events}
Both H1 and ZEUS experiments at HERA have measured multi-lepton production at high transverse momenta~\cite{Aktas:2003jg,Aktas:2003sz,zeus_multilep_prelim}. 
\begin{table*}[htb]
 \renewcommand{\arraystretch}{1.6}
\begin{center} 
\begin{tabular}{|l|c|c||c|c|}\hline
   Selection   & DATA  &  SM &  GRAPE &  NC-DIS + Compton \\ \hline \hline
\multicolumn{1}{|l}{\bf H1 115 pb$^{\mathrm \bf -1}$}& \multicolumn{4}{l|}{$20^\circ < \theta^{e1,2} < 150^\circ$,
$P_T^{e1}>10$~GeV, $P_T^{e2}>5$~GeV }\\
\hline
  ``2e''              &   108 &  $117.1\pm   8.6$ &  $ 91.4\pm   6.9$ &  $ 25.7\pm   5.2$ \\
  ``3e''              &    17 &  $ 20.3\pm   2.1$ &  $ 20.2\pm   2.1$ &  $  0.1\pm   0.1$ \\
\hline \hline
 ``2e''   $M_{12}>100$~GeV &     $3$ &  $ 0.30\pm  0.04$ &  $ 0.21\pm  0.03$ &
$ 0.09\pm  0.02$ \\
 ``3e''   $M_{12}>100$~GeV &     $3$ &  $ 0.23\pm  0.04$ &  $ 0.23\pm  0.03$ &
$ <0.02$ {\footnotesize ( 95\% C.L.)} \\
 
   \hline
\hline
\multicolumn{1}{|l}{\bf ZEUS 130 pb$^{\mathrm \bf -1}$}&\multicolumn{4}{l|}{$17^\circ < \theta^{e1,2} < 167^\circ$, $P_T^{e1}>10$~GeV, $E^{e2}>10$~GeV }\\
\hline
  ``2e''           &   191 &  $213.9\pm  3.9$ &  $ 182.2\pm  1.2$ &  $ 31.7\pm   3.7$ \\
  ``3e''           &    26 &  $ 34.7\pm  0.5$ &  $ 34.7\pm   0.5$ &  $-$ \\
\hline \hline
 
``2e'' $M_{12}>100$~GeV &   2 &  $ 0.77\pm  0.08$ &  $ 0.47\pm  0.05$ &  $ 0.30\pm  0.07$ \\
 ``3e'' $M_{12}>100$~GeV &   0 &  $ 0.37\pm  0.04$ &  $ 0.37\pm  0.04$ &  $-$ \\
   \hline
 \end{tabular}
 
\caption{ Observed and predicted multi-electron event rates for masses $M_{12}>100$~GeV as function of the number of identified electrons. The prediction errors for H1 analysis include model uncertainties and experimental systematical errors added in quadrature. For ZEUS analysis, the predicted rates are shown with the statistical errors of the Monte Carlo only.}
 
\label{tab:h1zeusme}
\end{center}
\end{table*}

\par
The electron identification is based on calorimetric information inforced by tracking conditions for efficient background rejection. Electrons are measured in a large acceptance range $5^\circ < \theta_e < 175^\circ$, where $\theta_e$ is the electron polar angle measured with respect to the incoming proton direction.  The electron energy measured from calorimetric information has to be above 5 GeV. This energy threshold is inforced in H1 (ZEUS) analysis to 10 GeV for electrons candidates with $\theta_e<20^\circ$ ($\theta_e<164^\circ$). The electron candidates have to be isolated from other calorimetric deposits. In the central region defined by $20^\circ < \theta_e < 150^\circ$ for the H1 analysis and  $17^\circ < \theta_e < 164^\circ$ for the ZEUS analysis, an isolated charged track measured in the central tracking system has to be associated to the calorimetric deposit. The identified electrons  are indexed in decreasing transverse momentum $P_T$: $P_T^{e_i} > P_T^{e_{i+1}}$.
\begin{figure}[hhh]
  \begin{center}
    \includegraphics[width=0.36\textwidth,height=7cm]{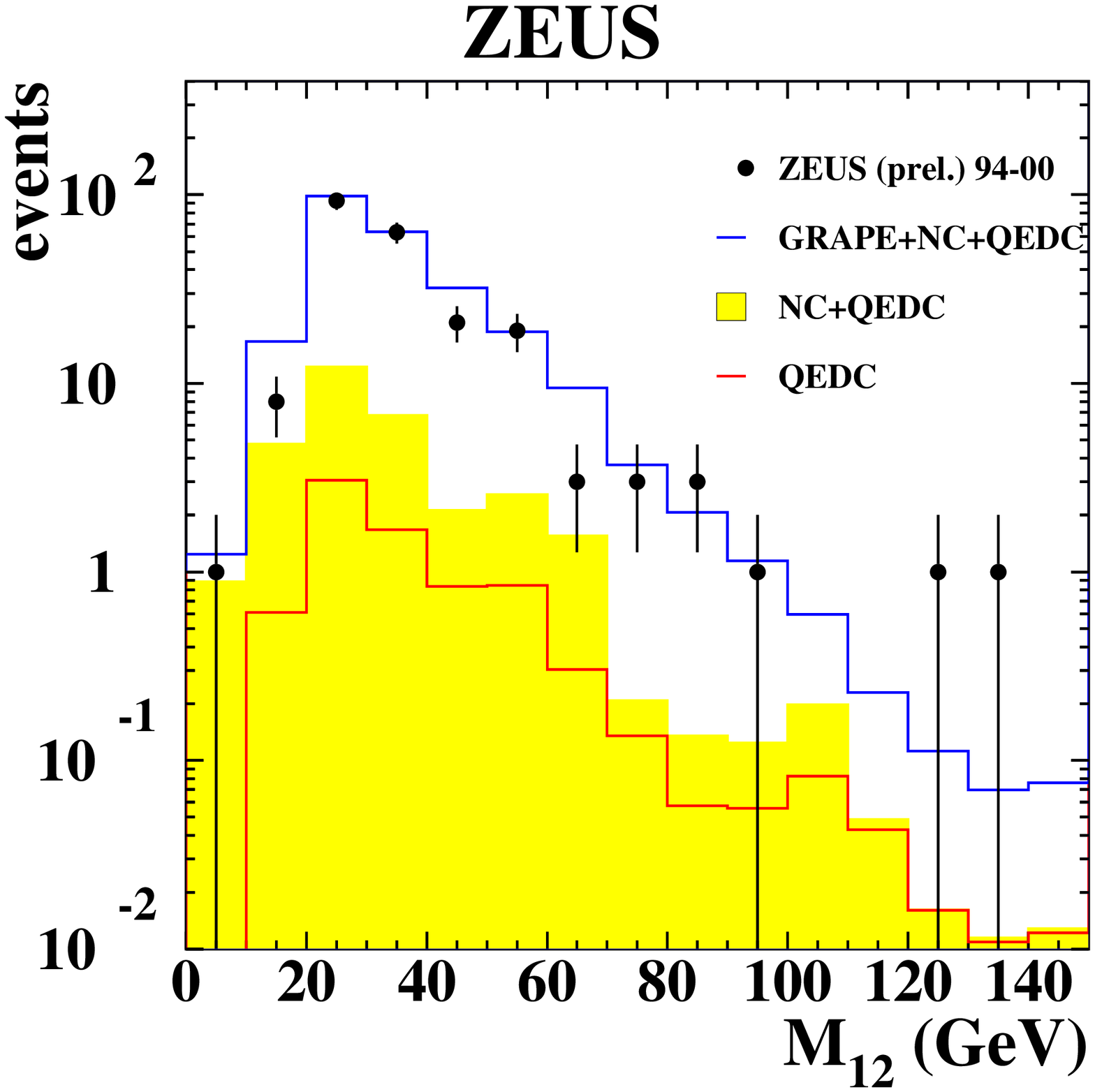}
    \includegraphics[width=0.32\textwidth,height=6.5cm]{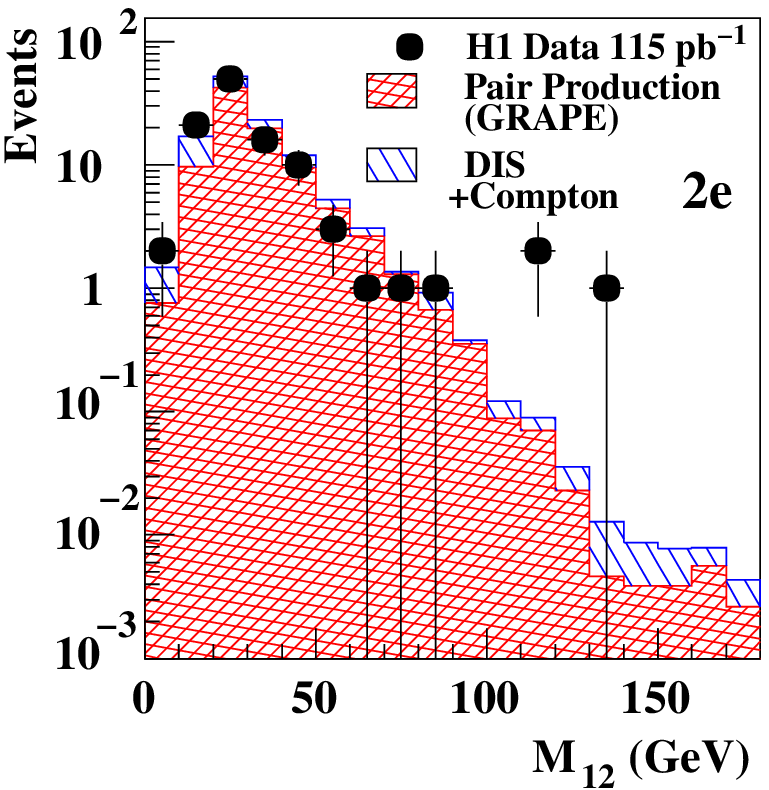}
    \includegraphics[width=0.32\textwidth,height=6.5cm]{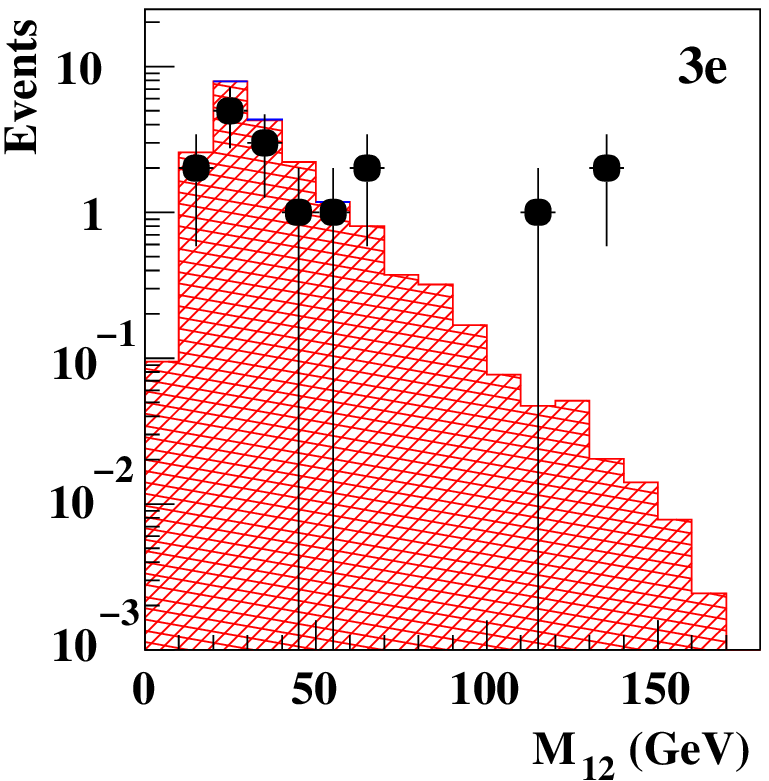}
\caption{ Distribution of the invariant mass $M_{12}$ of the two
     highest $P_T$ electrons for the the ZEUS analysis (left). All observed events with two electrons or more are shown. 
     Distribution of the invariant mass $M_{12}$ of the two
     highest $P_T$ electrons for the H1 analysis (center and right).                    
    Events are classified as di-electrons (center) and tri-electrons (right).}
\label{fig:mel_mass}
\end{center}
\end{figure}
\par
The selection of multi-electron events is based on the requirement of two central electrons with high energy or transverse momentum. Both the H1 and ZEUS analyses require the first central electron to have the transverse momentum above 10~GeV. The second central electron is required to have $P_T^{e2}>5$~GeV ($E^{e2}>10$~GeV) in H1 (ZEUS) analysis. Any other electron identified is also counted and the selected events are classified by the number of identified electrons in the event. 
\par
The results of the H1 and ZEUS analyses are presented in  table~\ref{tab:h1zeusme}. The H1 analysis, based on an event sample corresponding to 115 pb$^{-1}$, measured 125 multi-electron events, while ZEUS, with an integrated luminosity of 130 pb$^{-1}$, detected 217 such events. The di-electron sample is dominated by the signal  with a 15-20\% contribution from the background. In the tri-electron sample, the background contribution is negligible.
Both H1 and ZEUS observations are in good agreement with the predicted rates. The main difference between the H1 and ZEUS acceptances for the signal is due to a different angular range for the central electrons. 
\par
\begin{figure}[hhh]
  \begin{center}
    \includegraphics[width=1.1\textwidth]{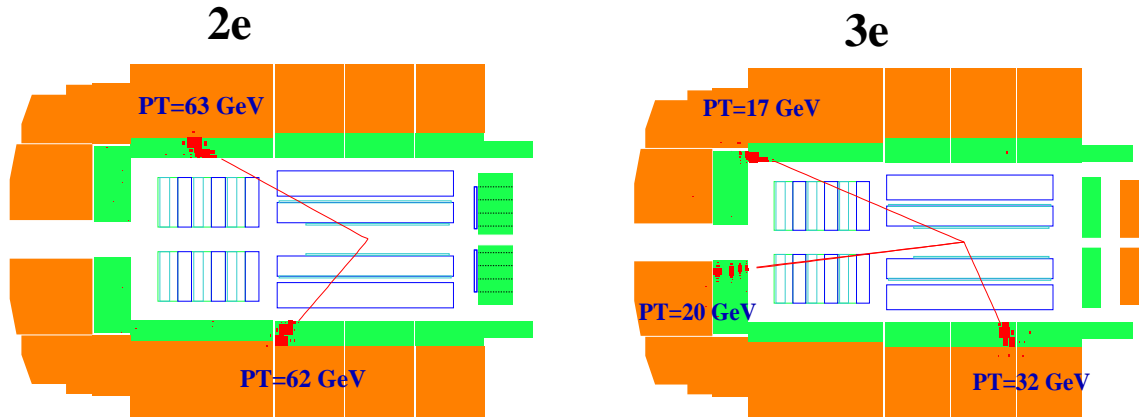}
\caption{ Event displays of two of the H1 multi-electron candidates: a di-electron (left) with large electron transverse momentum and a tri-electron (right) at lower $P_T$ with forward-backward topology of the highest $P_T$ electron pair that form $M_{12}$.}
\label{fig:mel_evd}
\end{center}
\end{figure}

The distributions of the invariant mass of the two highest $P_T$ electrons are shown in figure~\ref{fig:mel_mass}. Data is in good overall agreement with the Standard Model prediction. A few events with masses $M_{12}>100$~GeV are observed in a region where the standard model prediction is low. H1 measured 3 di-electron events for 0.30 expected. ZEUS observed 2 di-electron events for 0.77 expected. In the tri-electron sample, H1 observed 3 events with $M_{12}>100$~GeV for an expectation of 0.23 while ZEUS do not observe events in that mass region for an expectation of 0.37. For the high mass di-electron events, the transverse momenta of the two electrons is also important (above 50~GeV). The topology of the observed tri-electron high-mass events is different: the transverse momenta of the two highest $P_T$ electrons is lower (around 30 GeV) and the high mass value is associated with a larger polar opening angle between the two electrons (``forward-backward'' topology). One di-electron and one tri-electron at high mass are shown in figure~\ref{fig:mel_evd}.
\par
Alltogether, at high mass $M_{12}>100$~GeV, H1 and ZEUS observe 8 multi--electron events for an expectation of $1.67\pm0.20$.
A possible interpretation of the observed multi-electron events in terms of doubly charged higgs production at HERA is presented in the next chapter. 

\subsubsection{Measurement of multi-muon events}
Muons are identified using central tracker, calorimeter and  muon chambers signals.
Search for multi-muon events has been performed by  H1~\cite{Aktas:2003sz} and ZEUS~\cite{zeus_multilep_prelim}.
\par
\begin{figure}[hhh]
  \begin{center}
    \includegraphics[width=0.34\textwidth]{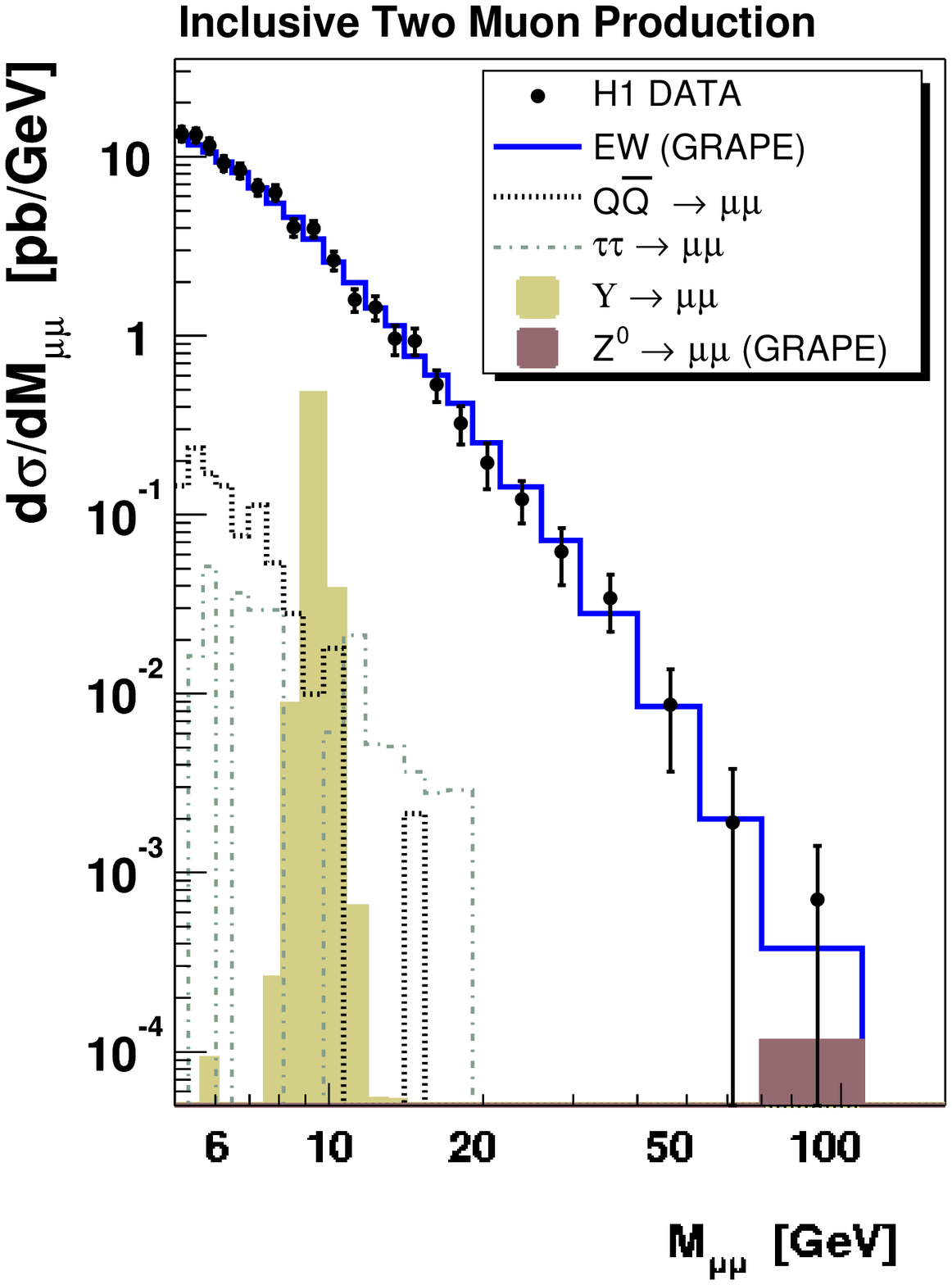}
    \includegraphics[width=0.66\textwidth]{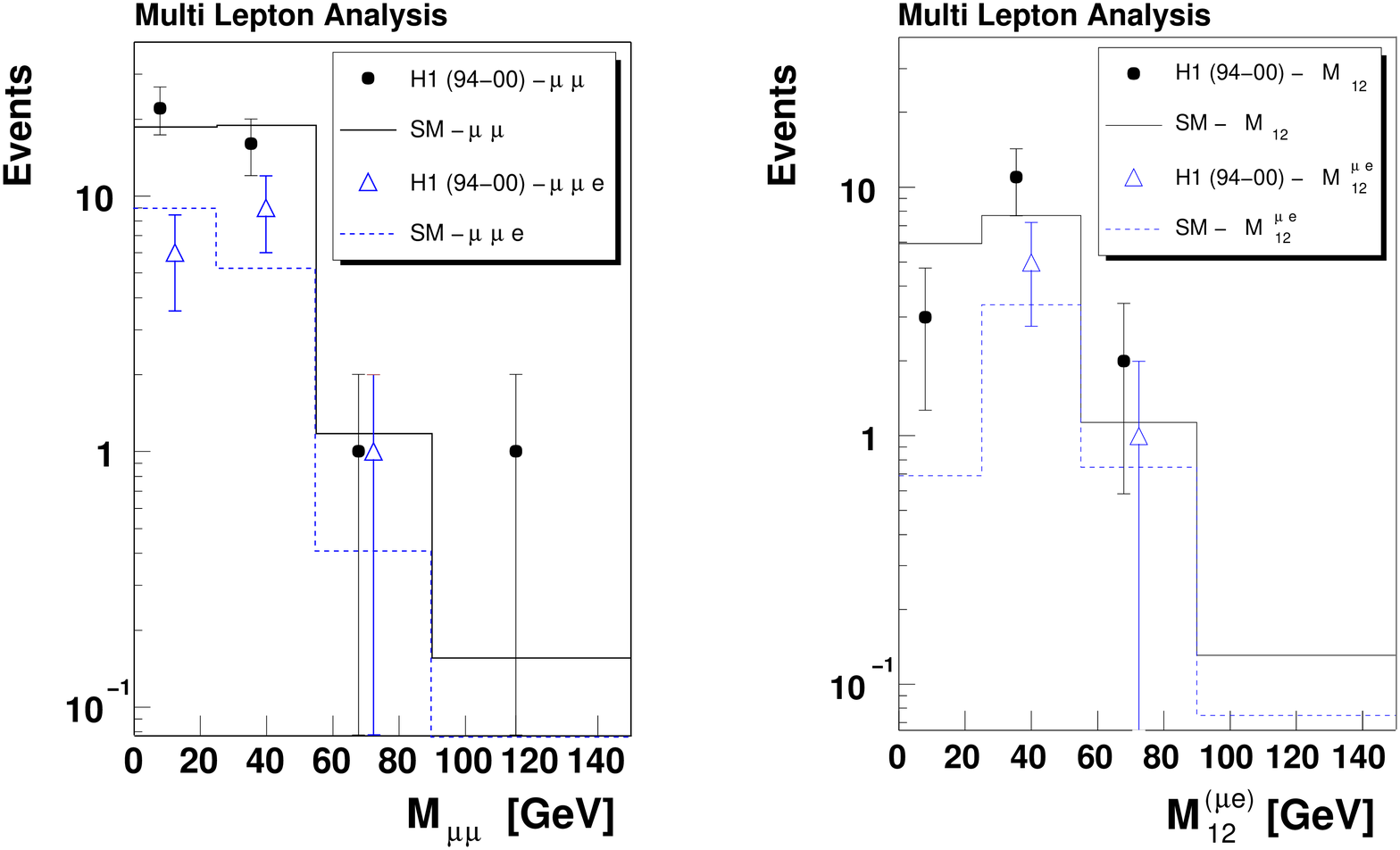}
\caption{ Muon pair production cross section as a function of the invariant di-muon mass (left). Distributions of the invariant mass $M_{\mu\mu}$ for
           compared with the Standard Model predictions (right).
           Distribution of $\mu\mu e$ events as function of mass
           $M_{12}$ of the two highest $P_t$ leptons, and of the mass
$M_{12}^{\mu e}$ for those 
           events where  the leptons with the highest $P_t$ are a muon and the electron.}
\label{fig:mel_mus}
\end{center}
\end{figure}

The H1 collaboration measures the di-muon production for  muons with $P_T>2$~GeV 
in the angular range $20^\circ < \theta < 160^\circ$.  Figure~\ref{fig:mel_mus} (left)  presents the visible cross section measured by H1~\cite{Aktas:2003sz} as a function of the invariant mass of the muon pair compared to the Standard Model prediction. Backgrounds and also other sources of muon pair production like heavy hadron decays are negligible. Very good agreement with Standard Model prediction is observed up to the highest masses and  over a decrease of four orders of magnitude. The integrated cross section in the visible phase space has been measured to  $46.5\pm4.7$~pb which is in good agreement with the prediction of 46.2~pb.
\par
A search for multi-lepton events ($\ell\mu\mu$) at high $P_T$ has been performed by requiring at least two muons  in the region  $20^{\circ}<\theta < 150^{\circ}$ with  transverse momenta $P_t^{\mu_1}>10$~GeV and $P_t^{\mu_2}>5$~GeV. Additional muons must be detected in the central region  of the  detector, $20^{\circ}< \theta_{\mu} < 160^{\circ}$,
with a minimum transverse momentum of $1.75$~GeV.
Additional electrons are searched for in the polar angle  range  $5^{\circ} < \theta_{e} <  175^{\circ}$
and are required to have a minimum energy of $5$~GeV. With this selection  56 di-muon events are found, among which 
 16 events have an extra identified electron ($\mu \mu e$ events).
In figure~\ref{fig:mel_mus} (center) the di-muon mass distributions of events
classified as $\mu \mu$ events or as $\mu \mu e$ events  are compared with the theoretical expectations.
Both mass distributions are in agreement with the Standard Model calculations.
The distribution in $M_{12}$, the invariant mass of the two leptons
with the largest $P_t$,  is shown for the $\mu \mu e$ sample in figure~\ref{fig:mel_mus} (right).
This mass combination is selected in order to 
compare with the H1 multi-electron analysis, where the scattered electron cannot
be identified uniquely.
For approximately half of these events, the two leptons with the highest $P_T$ are the
electron and  a muon.
For these events, the mass distribution  $M_{12}^{\mu e}$ is also shown
in figure~\ref{fig:mel_mus} (right).
\par
For masses $M_{12} > 100$~GeV ($> 80$ GeV)  one $\mu\mu$ event is found, while $0.08 \pm 0.01$ \mbox{($0.29 \pm 0.03$)} are expected. This inelastic event with two well identified muons
has a mass of $M_{\mu\mu} = 102\pm11$~GeV.
No event classified as $\mu \mu e$ with $M_{12} > 100$~GeV is observed. The prediction is $0.05 \pm 0.01 $.
These results at high di-lepton masses are in agreement with the Standard
Model predictions. In view of the present limited statistics, they cannot
be used to draw firm conclusions concerning the high mass excess observed
in the multi-electron analysis.
\par
In the ZEUS analysis two muons with $P_T^{\mu1,2}>5$~GeV in the angular range $20^\circ < \theta < 160^\circ$ are required. 
With an analyzed data sample corresponding to an integrated luminosity of 105 pb$^{-1}$, ZEUS detects 200 events for an expectation of 213$\pm11_{stat}$. No event with two muons at high mass $M_{\mu\mu}>100$~GeV is observed.
\par
It should be noted that the detection of the muon as a ``third'' lepton is presently done for a polar angle $\theta>20^\circ$, while the electrons are detected in a more forward region $\theta>5^\circ$. It is important to extend in  future the search for muons at a lower polar angle because high mass states tend to be boosted in the forward direction (at low polar angles) at HERA.
\par
In the case of a multi--electron final state, there exists the possibility that the ``scattered'' electron is a part of the highest $P_T$ electron pair and that one of the ``produced'' electrons is lost down the beam pipe. This is expected to be the case for half of the di-electron events at high masses ($M>100$~GeV), as predicted by the GRAPE generator~\cite{Abe:2000cv} . The full cross check of the production of multi-lepton events with muons should include the $e\mu$ final states in the future.

%% file: bsm.tex
\chapter{Isolated Leptons as Signals Beyond the Standard Model}

\section{Welcome to BSM-land!}
What is usually called ``physics beyond the Standard Model'' (BSM) is a part of the unknown. The theories not yet positively confirmed or falsified by the experimental data (including the mechanism of symmetry breaking of the Standard Model) have an equal probablity to be true. 
\par
At colliders, high $P_T$ lepton production is a signal easy to measure. The processes producing leptons in the final state span a large range of \sm tests, as  has been illustrated in the previous chapter. The events with high energy leptons also provide gold plated channels for the observation of the new physics beyond the Standard Model.
\par
There are many patterns in which the theories extending the \sm are classified. The most extensive ones define problems of the \sm and classify the models according to the solutions to those problems. There are theories that would like to save the \sm from family replication, from fine tunning, from leptons and quarks non-symmetry, from neutrinos being too light or from the top quark being too heavy. It might well be that some assumption or the theory as a whole is part of the NSM (next Standard Model). Only new experimental data can solve the puzzle.
\par
Here we adopt a sequential scheme to classify the models of the new physics, according to the proximity of the new phenomena to the final state leptons. 
The idea is that, if some signal above the background show up in the data, the source of the deviation can be localized at various levels during the evolution from the initial state to the final state. This is illustrated in figure~\ref{fig:bsm_scheme}. The particle production is represented as a tree where the new (BSM) mechanism modifies one of the vertices during the hard interaction process:
\begin{figure}[hhh]
  \begin{center}
    \includegraphics[width=0.8\textwidth]{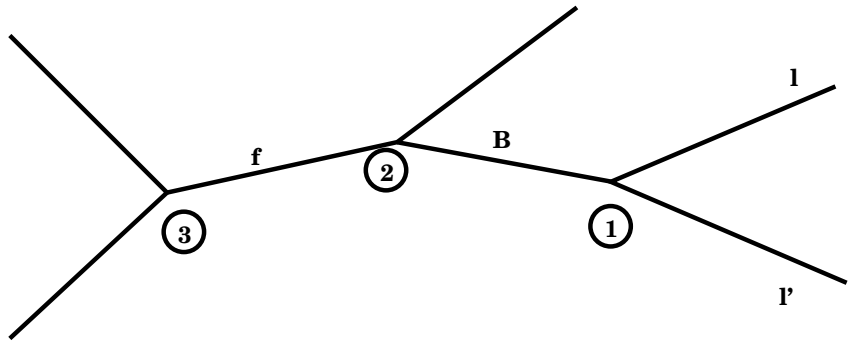}
    \caption{ 
Lepton production in a fermion-boson-leptons cascade. The anomaly beyond the \sm may occur in lepton (1), boson (2) or fermion (3) production couplings (see text).
}
    \label{fig:bsm_scheme}
  \end{center}
\end{figure}

\begin{enumerate}
\item The lepton couplings are modified. This is the case if the leptons are produced by a new scalar or vector boson ($H^{++},W',Z'$). 
\item The vector bosons of the \sm are anomalously produced by tri-boson anomalous couplings or in the decays of new fermions, for instance the excited fermions. 
\item A special case in the \sm is the top quark, the only fermion that decays into real $W$ bosons. The source of anomalies in the lepton production may also be traced back to an anomalous top production.

\end{enumerate}
Examples related to these patterns for physics beyond the \sm are discussed in this chapter.

\section{Anomalous lepton production}

Anomalous production of leptons at high transverse momenta can be predicted by  theories that incorporate new symmetries and  new bosons decaying to \sm leptons. The most constrained BSM theory  is the \sm itself for the scalar sector, not yet measured experimentally. However couplings are predicted to be proportional to the mass and little impact from the direct Higgs decay to leptons is expected. 
\par
New scalar and vector bosons are proposed in  models that extend the Standard Model, like the Left-Right Symmetric, supersymmetric, GUT  models or combinations of those~\cite{Kuze:2002vb}. Examples of searches for such new bosons are described below.

\subsection{New charged and neutral vector bosons}

The search for extra gauge boson production ($W',Z'$) in $p\bar{p}$ collisions  at the  Tevatron  is done using the leptonic decay channel. The hadronic decay channel has little sensitivity due to the large QCD background. 
\par
The search for a neutral $Z'$ boson is based on the detection of two charged leptons (electrons or muons). The main \sm contribution is due to the Drell-Yan process, as mentioned in the previous chapter. The searched signal is a peak in the di-lepton invariant mass spectrum. No such peak has been detected up to now and the data is in good agreement with the expectation from the \sm up to the highest observed masses, as mentioned in the previous chapter (figure~\ref{fig:d0_dilep}). An asymmetry in the angular distribution for high mass lepton pairs can also be a sign of a new boson and can be used for instance to investigate the indirect effects of a compositeness model~\cite{Affolder:2001ha}. 
Limits on the production of $Z'$  are obtained by combining the electron and muon channels. Neutral boson masses are excluded at 95\% confidence level for masses below an upper limit in the range between 545 and 730~GeV, depending on the models. 
\par
A search for an extra charged gauge boson $W'$ is performed in events with  leptons and missing energy. In the analysis done by CDF on run I data~\cite{Affolder:2001gr}  assuming the strength of the Standard Model  coupling, the W' mass limit is set to be $M(W') > 755$~GeV at the 95\% confidence level. 
\par
The Tevatron sensitivity to extra gauge bosons decaying to SM leptons should finally reach about 1~TeV in run 2, while $4\div  5$ TeV should be accessible at the LHC~\cite{eperez:lp03}.

\subsection{Doubly charged higgs $H^{++}$ }
\begin{figure}[hhh]
  \begin{center}
    \includegraphics[width=0.23\textwidth]{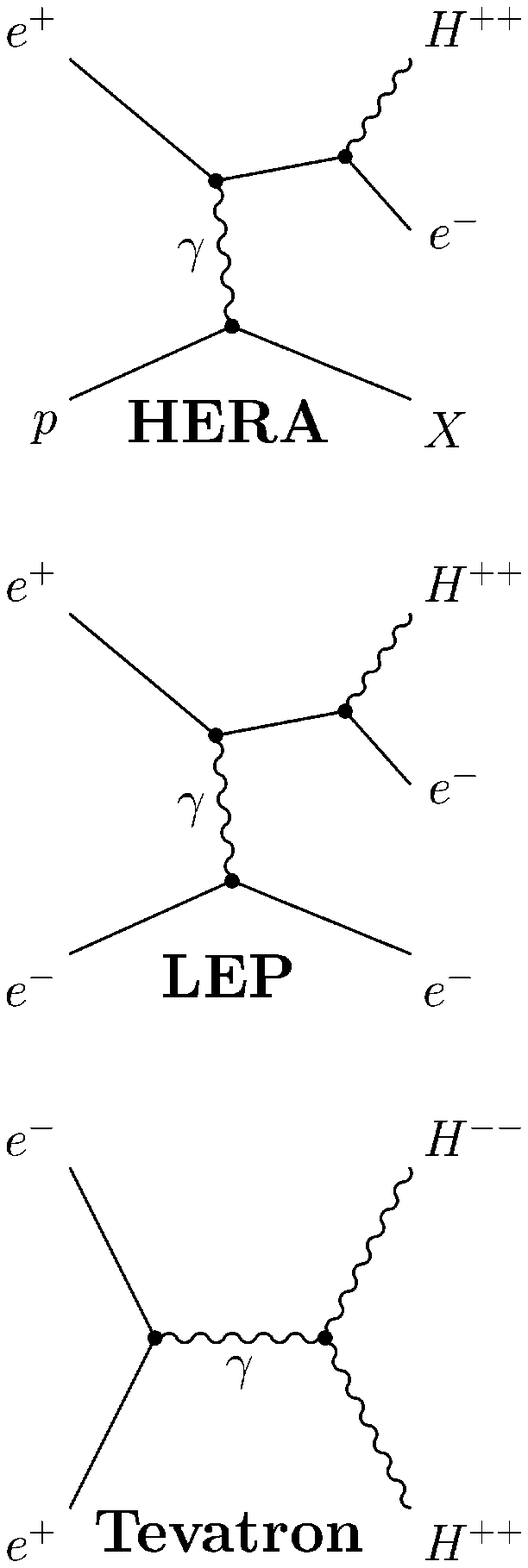}
    \includegraphics[width=0.81\textwidth]{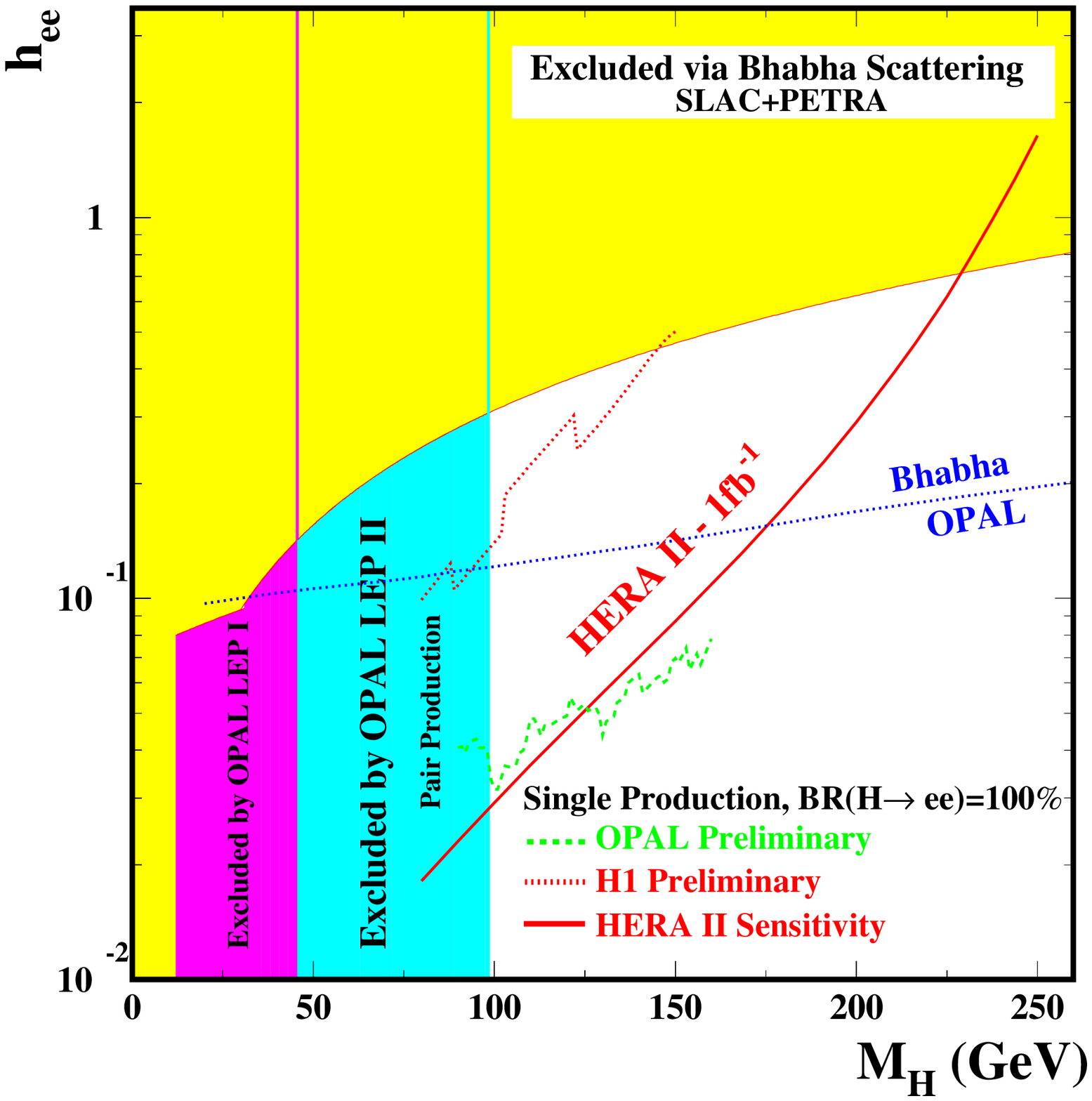}
    \caption{ 
Mechanisms involving  doubly charged higgs at colliders (left) and world status of the limits on the coupling to electrons $h_{ee}$ versus Higgs mass. The previsions for next running period at HERA II is also indicated.
}
    \label{fig:h++}
  \end{center}
\end{figure}

The Left-Right (LR) symmetric models~\cite{Senjanovic:1975rk,Mohapatra:1980qe,Mohapatra:1980ia} provide some interesting features like the see-saw mechanism that explains the smallness of the neutrino mass together with the non-observation of right handed currents. The LR \sm extensions of the \sm predict higgs boson triplets containing doubly charged Higgs $H^{++}$. Due to the charge, they can only couple to leptons. This is an example of a BSM process that affects only the leptonic sector~\cite{Kuze:2002vb}.  Since the Higgs triplet is not related to the symmetry breaking mechanism, the coupling is not necessarily proportional to the mass and can therefore be large also in the case of electrons and muons. The mechanisms of $H^{++}$ production at colliders are shown in the diagrams of figure~\ref{fig:h++}.

\par
Doubly charged higgs particles can be pair produced at the Tevatron through a Drell-Yan mechanism $e^+e^-\rightarrow H^{++}H^{--}$. The final state consists of four charged leptons and has little background. The production cross section is independent of the $H^{++}$ couplings.  At the Tevatron the search for resonances in a sample of events with four electrons or two electrons and two muons has started for Run II data~\cite{eperez:lp03}. A sensitivity up to $M_{h^{\pm\pm}}\simeq 180$~GeV is forseen for the full run II data taking. The pair production mechanism has also been studied at LEP~\cite{Abbiendi:2001cr}. No deviation with respect to the \sm has been found and  a lower limit on the $H^{++}$ mass of about 100~GeV has been derived at 95\% confidence level, independent of the couplings. 
\par
At HERA, single $H^{\pm\pm}$ can be produced in a hard $e\gamma^{(*)}$ scattering. The final state consists of two or three charged leptons which may make the $H^{++}$ production an attractive interpretation of the high mass multi-electron events observed at HERA and discussed in the previous chapter. But this is not the case since among the six events with two or three electrons at high mass only one remains consistent with the $H^{++}$ interpretation~\cite{h1hpp}.
\par
The single $H^{++}$ production is also possible at LEP~\cite{Abbiendi:2003pr} through a mechanism very similar to HERA. The final state consists of two or three leptons, searched for in an flavour independent analysis. The main \sm backgrounds are the fermion pair production and the production of events with four leptons. The requirement that same-sign leptons form the higgs candidate at high mass reduces the background typically by a factor of 10-20 in the final stage of the analysis. The final sample consists of 55 events, in rough agreement with the \sm expectation of $44\pm2$. The events are uniformely distributed in invariant mass from 80 GeV up to roughly 150 GeV. In contrast, in the H1 analysis, the \sm  expectation falls steeply with the invariant mass. However, the larger luminosity of LEP ensures a better sensitivity for $H^{++}$ searches. 
\par At LEP, the doubly charged Higgs boson can modify the Bhabha scattering cross section due to the exchange in the $t$-channel. The Bhabha cross section has been measured and found in very good agreement with the \sm prediction. Indirect limits on the doubly charged higgs model parameters (mass and couplings) can be obtained.
\par
The status of $H^{++}$ searches is shown in figure~\ref{fig:h++}. The exclusion limit at 95\% confidence level on the couplings to electrons $h_{ee}$ is shown as a function of the $H^{++}$ mass. 
\par
\subsection{Bileptons }
New bosons can also be studied in a more general framework of bileptons~\cite{Cuypers:1998ia}, bosons that couple to leptons but not to quarks. A special case are the bileptons for which the leptonic number $L=2$ which explicitely prevents the boson coupling to quarks.
\par
 Simply charged bileptons $L^\pm$, coupling to a lepton-neutrino pair, can be produced by pairs at LEP or Tevatron and yield final state topologies with two leptons and missing transverse momentum,  similar to the $W$--pairs in the case in which both W bosons decay in the leptonic channel\footnote{The search for charged Higgs pairs can also be interpreted in terms of searches for bilepton pairs if the hadronic decays are ignored. In the ALEPH analysis~\cite{Heister:2002ev}, a lepton-flavour blind search for tau decays a limit for the production cross section around 0.8 pb$^{-1}$, assuming 100\% branching ratio to $\tau\bar{\nu_\tau}$. However the rejection of the $W$--pairs relies on the charged higgs model. Supposing the effect of the signal modelling is small, this analysis can also be used to extract a limit on the bilepton pair production at LEP.}. At HERA, single  $L^\pm$ can be produced in $e\gamma$ collisions ($e^+\gamma\ra \nu L^+$) and yield final state topology similar to the $W$ production process discussed in the previous chapter. The hadronic system is expected to have low transverse momentum and an important fraction of the production cross section should belong to the elastic process, for which the proton is left unbroken and no hadrons are observed in the main detector. This makes unlikely the interpretation of the outstanding $\ell+P_T^{miss}$ events as bilepton $L^\pm$ production.  The single charged bileptons can also be simply produced at LEP in $e\gamma$ collisions, with a similar final state as the single $W$ production in case of the $W$ leptonic decay. This production channel should provide a sensitivity to bileptons with mass up to the available $\sqrt{s}$ at LEP. 
\par
The doubly charged bileptons can also be produced by pairs at LEP and Tevatron and singly at LEP and HERA. The production of doubly charged bileptons can yield final states with multi-leptons, particularly interesting in view of the observed multi--lepton events at HERA. The case of the doubly charged higgs presented above gives an idea about the relative sensitivities. However, it would be interesting to perform  a  complete analysis of the ``hadro-phobic'' boson production at LEP, HERA and Tevatron in the framework defined in~\cite{Cuypers:1998ia}.

\section{Anomalous EW boson production}
The lepton production through the \sm  weak bosons can present anomalies if those bosons result from new particle decays. Two examples are given here.
\subsection{Excited fermions}
An anomalous production of weak bosons can result from the radiative decay of higher mass excited states of the known fermions.
Excited fermions represent a natural scenario for the compositeness
extensions of the Standard Model.
The production and decay of excited fermions are described in terms of the phenomenological Lagrangian proposed in \cite{Hagiwara:1985wt}.
\begin{equation}
L_{ff^*} =
\frac{e}{\Lambda} \sum_{V}^{\gamma,Z,W} \bar{F^*} \sigma^{\mu\nu}(c_{V}-d_{V}\gamma^5)F
\partial_{\mu}V_{\nu} + h.c.
\end{equation}
where $c_{V}$ and $d_{V}$ are coupling constants at the fermion $F \leftrightarrow F^*$ vertex labelled
for each vector boson V, and $\Lambda$ is the compositeness scale. The precise measurement of
muon $g-2$ and the absence of an electron or muon electric dipole moments  implies that
 $c_{V}=d_{V}$ for compositeness scales less than 10--100 TeV. This fact leads to an interaction
Lagrangian where the coupling constants of excited fermions to Standard Model gauge bosons are
 described in terms of three constants corresponding to the three gauge groups: $f$ for
 $SU(2)_L$, $f'$ for $SU(1)_Y$ and $f_s$ for $SU(3)_C$. For specific assumptions relating $f$,
 $f'$ and $f_s$ the branching ratios can be predicted and the cross sections are described by
 a single parameter (e.g. $f/\Lambda$).
\par
%%%%%%%%%%%%%%%%%%%%%%%%%%%%%%%%%%%%%%%%%%%%%%%%%%%%%%%%%%%
\begin{figure}[hhh]
  \begin{center}
    \includegraphics[width=0.28\textwidth]{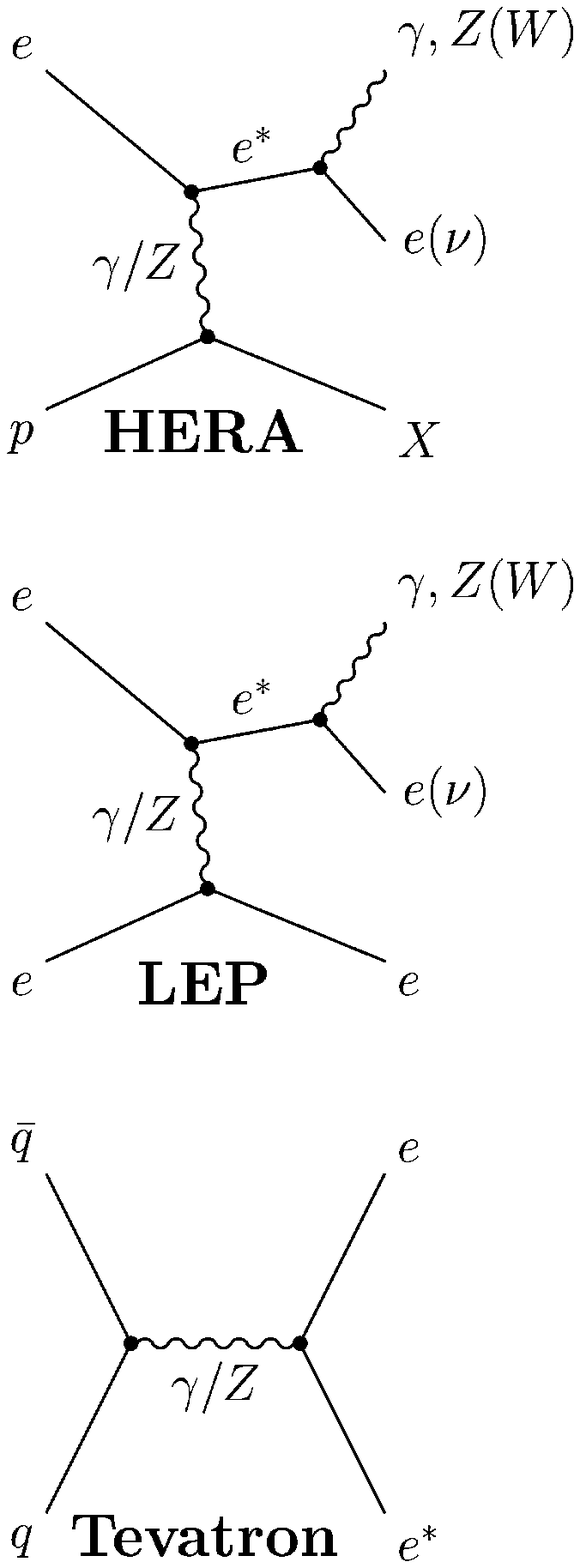}
    \includegraphics[width=0.80\textwidth]{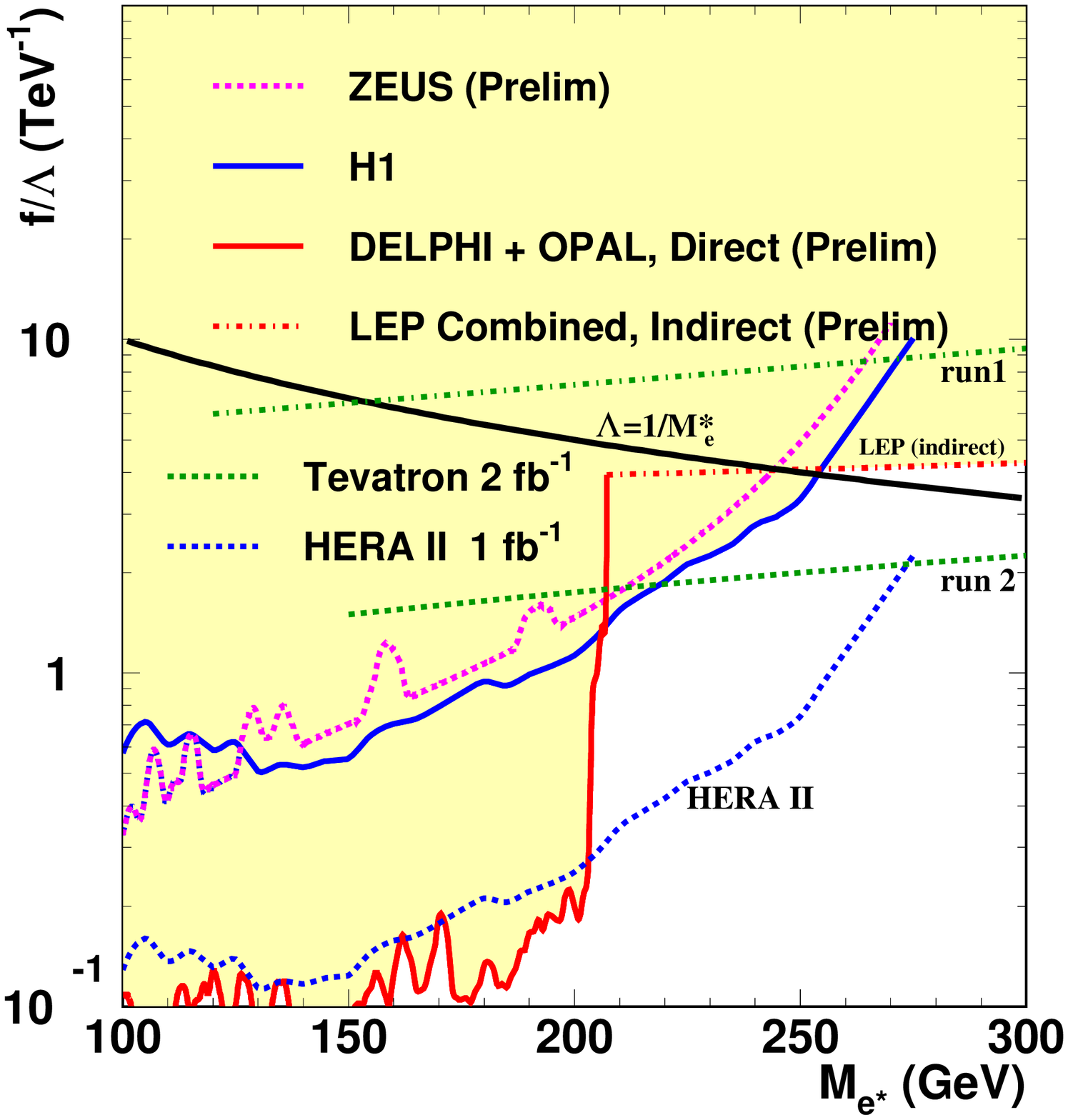}
    \caption{ 
Mechanisms of excited electron production  at colliders (left) and the 95\% confidence limits of the coupling normalized to the compositeness scale as a function of the excited electron mass (right, figure from~\cite{eperez:lp03}). Previsions for next running periods at HERA and Tevatron are also indicated.
}
    \label{fig:estar}
  \end{center}
\end{figure}
The search for excited fermions is performed at all colliders by  searching for fermion-boson resonances. As an example, the mechanisms of excited electron production are shown in figure~\ref{fig:estar} (left). Although the branching to a photon is favoured in the majority of parameter configurations, the coupling to the photon can vanish in some cases. For instance the $\nu^*\nu\gamma$ coupling tends to zero for $f=f'$. In this case  the decay $\nu^*\ra \nu\gamma$ is forbidden and the search for $\nu^*$ must rely on final states with weak bosons~\cite{Adloff:2001me}. Similarly, the decay $e^*\ra e\gamma$ is forbidden for $f=-f'$ and the excited electron decays only in $eZ$ channel~\cite{Adloff:2002dy}.
\par 
No signal for radiative decays of a high mass fermion has been found. The limits on coupling normalized on the compositeness scale ($f/\Lambda$) are calculated. The present status for the excited electron $e^*$ is shown in figure~\ref{fig:estar}. LEP excludes, with high sensitivity to the  coupling, excited electrons with masses up to $\sqrt{s}$. Beyond that, HERA experiments, for which the $e^*$ production proceeds with photons from the proton, provide sensitivity for masses up to 250 GeV but for couplings ten times larger than at LEP. The production at Tevatron has less sesitivity for low couplings but it may directly  probe excited fermion masses well beyond HERA or LEP. For the next running period, HERA will have the priviledged discovery window at low couplings and for masses in the range between 200 and 300~GeV.

\subsection{Gauge bosons anomalously coupled}

The Standard Model of particle physics is a non-Abelian theory and therefore predicts that the gauge bosons interact with each other, allowing coupling vertices such as $WWZ$ and $WW\gamma$. The trilinear couplings can be measured experimentally. For instance, the flattening of the $e^+e^- \rightarrow W^+W^-$ cross section with $\sqrt{s}$ (figure~\ref{fig:lep_bos})   is the experimental proof of the existence of the tri-linear coupling~\cite{Spiesberger:2000ks,Godbole:2002mt} that damps the linear increase of the cross section predicted by the W pair production mechanism provided by a neutrino exchange in the $t$-channel. The value of the couplings may be modified by radiative corrections involving particles predicted by extensions to the Standard Model such as Supersymmetry. Models based on substructure of the gauge bosons would also lead to non Standard Model or "anomalous" couplings. Making experimental measurements of the couplings is therefore an important test of the Standard Model and its extensions.
\par
The most general Lorentz invariant Lagrangian which describes the triple gauge boson interaction has
fourteen independent complex couplings, seven describing the
WW$\gamma$ vertex and seven describing the WWZ vertex~\cite{Hagiwara:1987vm}.  Assuming
electromagnetic gauge invariance as well as C and P conservation, the
number of independent TGCs reduces to five.  A common set is \{$\gz,
\kz, \kg, \lz$, $\lg$\} where $\gz = \kz = \kg = 1$ and $\lz = \lg =
0$ in the \sm.  The parameters proposed  used by
the LEP experiments are $\gz$, $\lg$ and $\kg$ with the gauge constraints:
\begin{eqnarray}
\kz & = & \gz - (\kg - 1) \twsq \,, \\
\lz & = & \lg \,,
\end{eqnarray}
where $\theta_W$ is the weak mixing angle.  The
couplings are considered as real, with the imaginary parts fixed to
zero. 
 \par
Note that the photonic couplings $\lg$ and $\kg$ are related to the
magnetic and electric properties of the W-boson. One can write the
lowest order terms for a multipole expansion describing the W-$\gamma$
interaction as a function of $\lg$ and $\kg$. For the magnetic dipole
moment $\mu_{W}$ and the electric quadrupole moment $q_{W}$ one
obtains $e(1+\kappa_{\gamma}+\lambda_{\gamma})/2\MW$ and
$-e(\kappa_{\gamma}-\lambda_{\gamma})/\MW^2$, respectively.
\begin{table}[htbp]
\begin{center}
\renewcommand{\arraystretch}{1.3}
\begin{tabular}{|l|c|c|c|c|} 
\hline
Parameter (95\% C.L.) & \sm  & LEP   &  Tevatron (D0 - run I) & HERA (40$\pbi$)      \\
\hline
$\kg$     & 1.0 &  [$0.835,~~1.052$]  &   [$0.02,~~2.01]$ & [$-3.7,~~2.5$]\\ 
\hline
$\lg$      & 0.0 & [$-0.067,~~0.028$]  &  [$-0.33,~~0.31$]  & [$-3.2,~~3.2$]\\ 
\hline
\end{tabular}
\caption[]{ The 95\% C.L. intervals
  obtained combining the results from the four LEP experiments and  from the D0 and ZEUS analyses.  In
  each case the parameter listed is varied while the other two are
  fixed to their Standard Model values.  }
 \label{tab:tgc}
\end{center}
\end{table}

\par
The measurement of triple gauge couplings has been performed at LEP. For instance, in the analysis~\cite{Heister:2001qt,Barklow:2001ge} done by ALEPH collaboration, the $WW\gamma$ vertex is investigated in events with single photons, single $W$ bosons or $W$ pairs. No deviations from the Standard Model are found. 
The triple gauge  couplings are tested at Tevatron in the di-boson production, for instance in the D0 analysis of the run 1 data~\cite{Abachi:1997hw}. The best sensitivity is obtained in the $W\gamma$ channel where limits on the deviations of the couplings from their \sm values are obtained. At HERA, the single production of W bosons can in principle also be used to set limits on the anomalous triple gauge couplings. 
HERA sensitivity to an anomalous coupling has been investigated by the ZEUS collaboration~\cite{Breitweg:1999ie}.  
\par
The 95\% confidence intervals from the three colliders are presented in table~\ref{tab:tgc}.
Deviations from the \sm with $\lambda\ne 0$ or $\Delta \kappa = \kappa-1$ may lead to a production of $\ell+P_T^{miss}$ events with atypical kinematical properties at HERA~\cite{Baur:1989gh}. However, HERA sensitivity is significantly smaller than that of LEP or the Tevatron to this type of anomalus couplings and therefore the interpretation of the events with isolated leptons and missing transverse energy as  W production through an anomalous triple gauge coupling is largely disfavoured.

\section{Anomalous top production}
Due to the large top quark mass close to the electroweak scale, phenomena related to the top quark may have the best sensitivity to new physics phenomena. The top quark is only produced at Tevatron, mostly in pairs. LEP and HERA colliders provide sufficient \CoM energy for a single top production in $ee$ and $ep$ collision, respectively. However, the top production cross section within the \sm is very low of about 1~fb at HERA~\cite{Stelzer:1997ns,Moretti:1998dz} and $10^{-4}$~fb at LEP~\cite{Hagiwara:1994mg}.
\par
\begin{figure}[hhh]
  \begin{center}
    \includegraphics[width=0.23\textwidth]{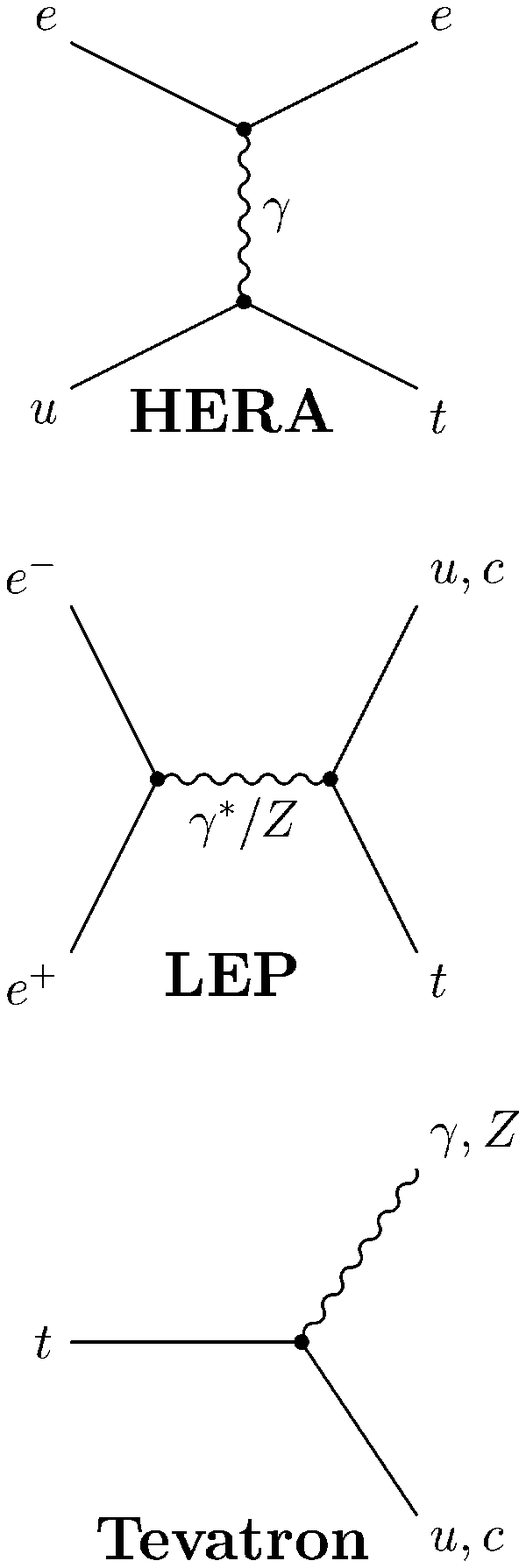}
    \includegraphics[width=0.81\textwidth]{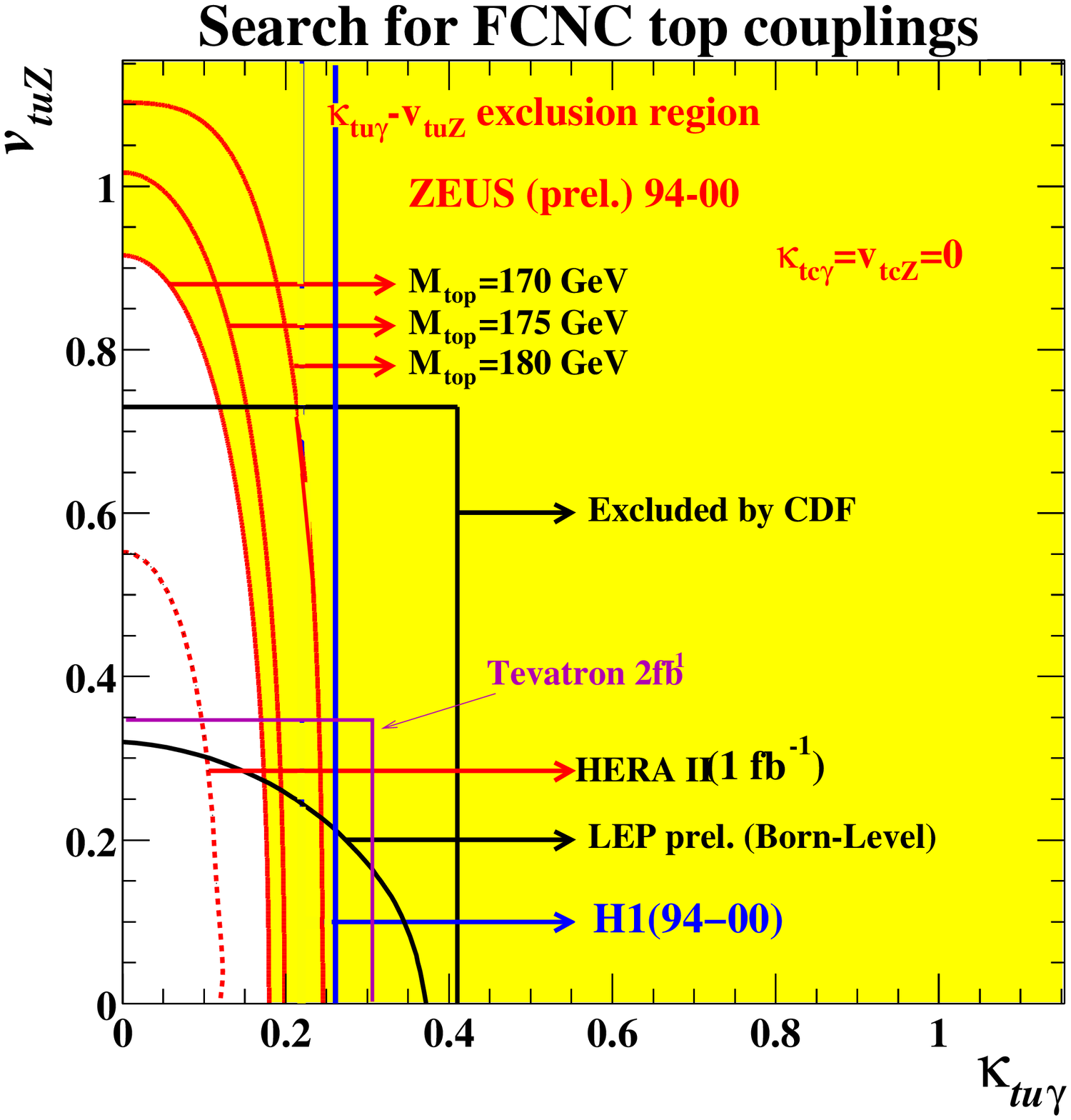}
    \caption{ 
Mechanisms involving FCNC top couplings at colliders (left) and world status of the limits on the $tu\gamma$ and $tuZ$ couplings. Previsions for next running periods at HERA and Tevatron are also indicated.
}
    \label{fig:anotop}
  \end{center}
\end{figure}

The top quark could be produced at HERA and LEP if  flavour changing neutral currents were allowed in the top sector~\cite{Han:1998yr}. The production mechanisms would be based on the non-vanishing  couplings of the top quarks to the light quarks of same charge (up and charm) and to the neutral bosons ($\gamma,Z$).  In this case,  the production of single top quarks at LEP and HERA would proceed through the mechanisms sketched in the diagrams of figure~\ref{fig:anotop}. In addition, anomalous top quark decays to neutral bosons should be observed in the top sample produced at the Tevatron. The process is usually described\footnote{For a complete langrangian with FCNC in the top sector see for example~\cite{Han:1998yr,Aktas:2003yd}} using the anomalous couplings $\kappa_{\gamma qt}$ (magnetic coupling to the photon) and $v_{Zqt}$ (vector coupling to the Z) with $q=u,c$.
\par
 The search for anomalous decays ($t\rightarrow q\gamma$ and $t\rightarrow qZ$)  of the top quark has been performed at Tevatron~\cite{Abe:1998fz}. Top pairs are searched for in events for in which one top quark decays via the dominating \sm mode $t\rightarrow bW$ (the branching ratio is $100\%$ in the \sm) and the second via FCNC decays. Two candidate events are found in this search: one event in the mode $t\rightarrow qZ$ for which 1.2 events are expected, the second event in the mode $t\rightarrow q\gamma$ for which the background is not calculated. Conservative limits are derived by assuming that the observed events belong to the signal. The 95\% confidence level limits on the branching fractions are:
\begin{displaymath}
B^{t\rightarrow q\gamma} <3.2\% \;\;\;\;\;\;\;\;\;\;
B^{t\rightarrow Z\gamma} <33\% 
\end{displaymath}
The branching fractions for FCNC decays are proportional to the square of the anomalous couplings~\cite{Han:1995pk}. Therefore upper limits on the anomalous FCNC couplings can be calculated. The limits related to the photon and $Z$ are independent. The flavour of the final state is not tagged and the limits apply to both $tu\gamma$ or $tc\gamma$ couplings. 
\begin{table}
 \renewcommand{\arraystretch}{1.6}
\begin{center}
\begin{tabular}{|c|c|c|c|} \hline
Single top search   & Electron & Muon & Hadrons \\
1994-2000 $e^\pm p$ & obs./exp. & obs./exp. & obs./exp.  \\
   & {\footnotesize(top efficiency)} & {\footnotesize (top efficiency) }&  {\footnotesize (top efficiency) }\\ \hline
              
{\large \bf H1}{\footnotesize $118.4$~pb$^{-1}$} &  $ 3 / 0.65~\pm0.10~(36\%)$   & $ 2 / 0.66~\pm0.12~(38\%)$  & $18/20.2\pm3.6 (30\%)$ \\
\hline
{\large \bf ZEUS } {\footnotesize $130.2$~pb$^{-1}$} &    $0/0.94~^{+0.11}_{-0.10}~(35\%)$  & $0/0.95~^{+0.14}_{-0.10}~(35\%)$   & $14 / 17.6^{+1.8}_{-1.2}$~(25\%) \\  \hline
\end{tabular}
\end{center}
\caption{Summary of the results of searches for anomalous top production in electron, muon and hadronic channels at HERA.
The number
of observed events is compared to the SM prediction. The efficiency to detect top quark decays in the specified channels is given in parentheses in percent.}
\label{tab:anotop}
\end{table}

\par
At LEP, the FCNC couplings may lead to associated top-charm or top-up production in a $e^+e^-$ annihilation via a $\gamma^*/Z$. The FCNC single top production depends on the couplings to $u$ and $c$ quarks, that is four types of couplings: $tu\gamma$, $tc\gamma$,  $tuZ$ and $tcZ$. The main dependence is proportional with the square of the couplings, with some small intererence effect. The events are searched with a top candidate recoiling against a hadronic jet~\cite{Achard:2002vv,Heister:2002xv,Abbiendi:2001wk}.  Limits on the single top production cross section are obtained in the range $0.2\div 0.7$~pb at various \CoM energies. 
\par
If single top quarks are produced at HERA, the experimental signature of the leptonic decays channel $t\ra bW \ra b \ell \nu$ corresponds to events with isolated leptons, missing transverse energy and a prominent jet. This is particularly interesting in view of the excess of events of this type observed by  H1  and discussed in the previous chapter.  For the hadronic channel, the signature is three or more jets with high transverse momenta. The search for anomalous top production at HERA is summarized in table~\ref{tab:anotop} (H1~\cite{Aktas:2003yd} and ZEUS~\cite{Chekanov:2003yt}). In the H1 analysis, a part of the events with isolated leptons, missing energy and a prominent hadronic jet are also found as top candidates in the electron and muon channels: 5 events for an expectation of $1.3\pm0.2$. In the hadronic channel the sensitivity is much reduced due to the high background originating from QCD multi-jet production. 
\par
The limits at 95\% confidence level on the anomalous couplings $\kappa_{\gamma ut}$ and $v_{Zut}$ obtained at LEP, Tevatron and HERA are shown in figure~\ref{fig:anotop}. The sensitivity to the Z couplings has been studied by the ZEUS collaboration and found to have a non-negligible impact at large values of $v_{Zut}$, a region however already excluded by LEP and Tevatron.
\par
In order to extract the top quark production cross section, a multivariate likelihood analysis is performed by H1 in
addition to the cut--based analyses. The top signal contribution in each channel is determined in a maximum--likelihood fit to the likelihood discriminator distributions.
The results from the hadronic channel do not rule out a single top interpretation of the candidates observed in the electron and muon channels. For the combination of the electron, muon and hadronic channels a cross section for single top production of $\sigma=0.29^{+0.15}_{-0.14}$~pb at $\sqrt{s} = 319$~GeV~is obtained. This corresponds to a magnetic coupling $\kappa_{tu\gamma}$ of $0.20^{+0.05}_{-0.06}$. This result is not in contradiction with limits obtained by other experiments.
The addition of a contribution from a model of anomalous single top production yields a better description of the data than is obtained with the Standard Model alone. The new data at HERA II will allow the improvement the sensitivity to the anomalous magnetic coupling  $\kappa_{tu\gamma}$  by roughly a factor of two. For the FCNC coupling to the $Z$ boson, a gain in  sensitivity beyond the present LEP limit is expected from the Tevatron run 2 data.

\section{Supersymmetry}
The scheme presented in the beggining of this chapter (figure~\ref{fig:bsm_scheme}) is of course not a theory, but rather a pretext to discuss the path to new physics interpretation, once some deviation from the \sm is found in the experimental data.  A popular theory that extends the Standard Model is the supersymmetry (SUSY). SUSY  provides the unification of internal symmetries with the Lorentz invariance and associates supersymmetric particles ($s$particles) to the known \sm particles. The supersymmetric models propose solutions to most of the \sm problems (hierarchy, fine--tunning, unification) and  predict spectacular final states to be obtained in particle collisions. Despite extensive studies at colliders and elsewhere, no trace of SUSY has been detected yet.
\par
As an example of the capabilities of the SUSY theories to predict spectacular topologies, we shortly present here several hypothesis, some of them not yet fully explored with the present data, that may explain the production of events with high $P_T$ leptons at HERA. The production of single $s$particles is  possible at colliders if the multiplicative quantum number  $R_p$ is violated (for a particle,  the $R$--parity is $R_p=(-{\bf 1})^{3B+L+2S}$ where $B$ is the baryon number, $L$ the lepton number and $S$ the spin). For instance, the violation of the $R$--parity  opens the possibility of squark production in the $s$--channel at HERA due to the electron-quark-squark Yukawa coupling. A special case is the stop ($\tilde{t}$) which in most SUSY scenarios is the lightest SUSY particle. 
\par
The produced squarks can decay to $eq$ or $\nu q$ trough $R_p$--violating coupling producing events with the same topology as the deep inelastic scattering process. However, gauge decays into a quark and a chargino or a neutralino may lead to complex final state topologies, including configurations with high energy leptons. 
\par
There are a few possibilities to produce events with isolated leptons and missing transverse momentum  in squark decays at HERA\footnote{In most scenarios the charge of the prominent isolated lepton in the event correspond to the lepton beam charge}:
\begin{itemize}
\item If the squark decay via gauge decays $\tilde{q}\ra q\chi^+$, the semi-leptonic decay of the chargino into a neutralino $\chi^+\ra \chi^- \ell \nu$ issue an isolated leptons and missing energy together with a large $P_T$ jet from the primary squark decay. If the neutralino is stable, it can travel undetected the apparatus and induce extra missing momentum.  The same topology is obtained in the case of the decay chain  $\tilde{q}\ra q\chi^+\ra q \nu \tilde{l}$. If the slepton is stable, then it can fake a muon signature in the detector. However, a stable neutralino or a stable slepton require a SUSY parameter space that is already  severely constrained from LEP analyses.

\item If in the same configuration as above the neutralino can decay via a $R_p$ violating coupling: $\chi^0\ra eq\bar{q}$ or  $\chi^0\ra \nu q\bar{q}$. This will give rise to events with either several jets or several charged leptons in the final state. Searches of squark production at HERA by the H1 collaboration have not detected any deviation from the \sm prediction~\cite{Adloff:2001at,johannes} and constrained the squark masses up to 290~GeV. The H1 events with isolated leptons and missing transverse energy do not correspond to the predicted topology in the SUSY parameter space, mostly because no multi--jet shape of the hadronic system is observed. 

\item If we consider the resonant production of stop quarks at HERA, a SUSY parameter space can be considered where the stop is heavier than the sbottom. If in addition the charginos are heavier than the stop, the following channel may have an important contribution to the stop quark decay $\tilde{t}\ra \tilde{b} W$ with subsequent $R_p$ violating decay $\tilde{b}\ra d \nu$. This decays chain produce  events with isolated leptons and missing energy accompanied by a prominent hadronic system if $W\ra\ell\nu$ or events with three jets and missing $P_T$ in the case of hadronic $W$ decays. In addition, the $R_p$ violating decays $\tilde{t}\ra eq$ are also present and provide a cross--check of the stop production at HERA in this scenario. 
 
\item In the case of two non--zero $R$--parity violating couplings, the slepton produced in chargino decay can undergo a leptonic decay via a $\lambda$ coupling: $\chi^+ \ra \tilde{\ell}^+ \nu \ra \ell^+ \nu \nu $. A large number of final states should be observed at HERA in $e^+p$ collisions~\cite{zerwas,Diaconu:1998rf} ($j$ denotes a hadronic jet): 
$$
e^+p \ra \tilde{t} \ra \tilde{\chi}_1^+ b \ra 
\ell^+ j \nu \nu\;,\;\; 
\ell^+\ell^+ \ell^- j\;,\;\;
\ell^+ jjj\;,\;\;
jjj\nu 
$$
$$
e^+p \ra \tilde{c} \ra \tilde{\chi}_1^0 c \ra 
\ell^+ \ell^- j \nu\;,\;\; 
\ell^\pm jjj\;,\;\;
jjj\nu 
$$

Events with an isolated lepton, missing $P_T$ and prominent hadronic system are predicted to appear in conjunction with tri-lepton events and a jet. The compatibility of this hypothesis with the observation of both the $\ell+P_T^{miss}$ events and the multi-lepton events at high mass has to be investigated\footnote{ No significant missing $P_T$ has been detected in those events. Only one among the six H1 multi-electron events at high mass have a clear hadronic system.}.
\end{itemize}
SUSY processes may also produce events with several energetic charged leptons.
For instance, the production of scalar neutrino~\cite{Kon:2002hy} via a $R_P$ violating coupling in the leptonic sector ($\lambda$ ), $e^+p\ra \mu^+\tilde{\nu_\tau} X \ra \mu^+ \mu^- e^+ X$. The signal should produce a peak in the invariant $e^+\mu^-$ mass distribution. If tau leptons are produced in the final state (trough a different $\lambda$ coupling, $e^+p\ra \tau^+\tilde{\nu_\tau} X \ra \tau^+ \tau^- e^+ X$),  events with narrow jets and missing transverse momentum  should be observed. 
\par
The supersymmetry can be at the origin of the outstanding events with leptons observed at HERA, altough the most popular scenarios are not likely. A rich phenomenology, with numerous decays channels - typical for the $R_p$ violating SUSY - provides a cross check of this hypothesis and invites for further detailed analyses.
 
\section{The model independent new physics search}
\begin{figure}[hhh]
\center
\includegraphics[width=0.6\textheight]{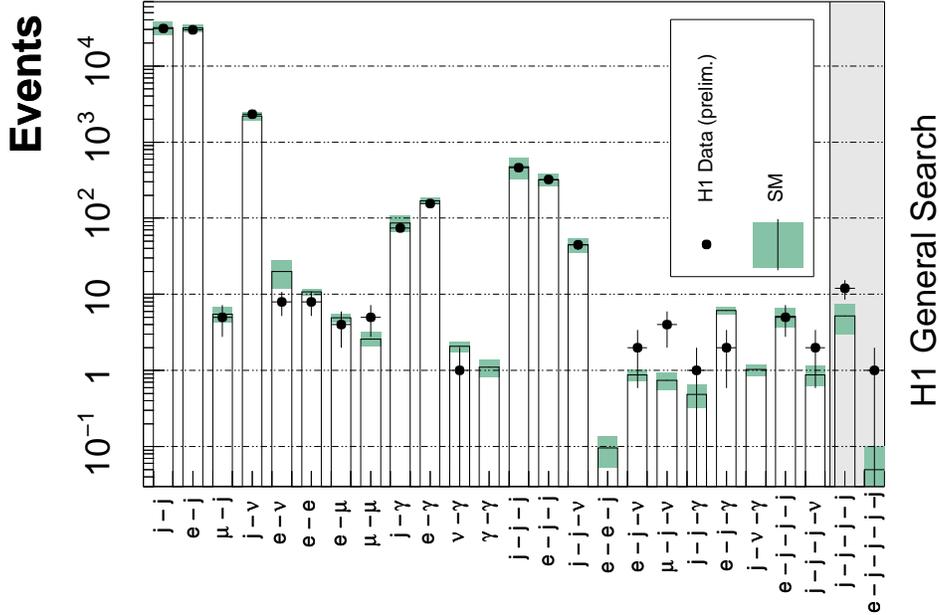}
\caption{The reasult of the H1 general search. 
The data ((points) and \sm  expectation (histogram)  for all event classes
with a SM expectation
greater than $0.1$ events are shown.
}
\label{fig:h1general}
\end{figure}
The data can also be investigated in an model independent approach in order to search for deviations from the \sm predictions. The general search for new phenomena has been pioneered by the D0 collaboration~\cite{Abbott:2000gx}. The idea is to define a common phase space for all types of final state identified particles. The events are then classified according to the particle content. Kinematical quantities are defined, like the mass  and the scalar transverse momentum, and a non-biased search algorithm is applied in order to look for local deviations that may appear due to for instance to a hypothetical multi--channel decay of a heavy particle.
\par
The analysis has also been performed by the H1 collaboration~\cite{h1_generalsearch}. `Objects'' are defined from particle identification: electron ($e$), muon ($\mu$),
photon ($\gamma$), jet ($j$) and neutrino ($\nu$) (or
non-interacting particles). All final states are analysed having at least two objects
with a transverse momentum ($P_T$) above $20$~GeV and in the
polar angle range $10^\circ < \theta < 140^\circ$. All selected events are then classified into exclusive event classes
(e.g.  $ej$, $jj$, $j\nu$)
according to the number and types of objects detected in the final state.
The event samples selected in the different classes are shown in figure~\ref{fig:h1general}. 
A very good agreement is found for nearly  all those topologies which is a great achievement, taking into account the complexity of the final states that are studied. 
The data fluctuate over the \sm in the $\mu j\nu$, as expected from the previous observations. The statistical analysis of the mass and $P_T$ spectra also identifies the observed fluctuation in the multi--electron channel, no visible on the global rate of $ee$ channel shown in figure~\ref{fig:h1general}. No other significant deviation with respect to the \sm has been found in addition. 
\par
The general search at high $P_T$ gives a global view of the physics rates as a function of the final state topology and require a good understanding of the \sm and of the detector.
This type of analysis is necessary for the searches program at present and future colliders, as it provides an extra security belt against the unexpected phenomena that may occur in a new pattern, different from what is predicted from the existing models.

%% file: conclusion.tex
\chapter{Conclusion}

The detection of isolated leptons at large transverse momenta at colliders  provides an unique tool  for the study of the processes involving the basic structure of the matter. 
The present colliders LEP ($e^+e^-$ collisions), HERA ($ep$ collisions) and Tevatron ($p\bar{p}$) collisions are complementary and span a wide range of hard scaterring processes. 
\par
Within the \sm framework, the lepton production proceeds either through boson splitting or via boson--boson scattering.
The investigation of the final states with isolated leptons allows to test the electroweak sector of the \sm. The production of weak bosons, single or by pair, at LEP and the Tevatron is found in agreement with the precise calculations based on the Standard Model. The boson production at HERA is a rather rare process and the production of the $W$ boson has been experimentally demonstrated. The observation of a few spectacular events with isolated leptons, missing transverse momentum and a prominent hadronic jet is puzzling and needs more integrated luminosity to be clarified. The measurement of the production of events with several charged leptons is particularly interesting at HERA and the Tevatron due to the particular conditions of production: QED processes coupled to the intimate structure of the protons entering the reaction. The study  of the spectacular multi-electron events with high masses observed in the  HERA I data will continue at HERA II with large statistics. 
\par
The search for new physics is particularly interesting in the leptonic channels due to usually low and theoretically well understood rates of  the \sm processes. The search for new phenomena suppose that ``something'' may start to deviate from the \sm laws at lepton couplings, vector boson couplings or top quark couplings. The examples given in this paper shed light on the complementarity between the three colliders for the search of the new physics beyond the Standard Model. 
\par 
The forthcoming years will bring an increased data sample at HERA and the Tevatron,  comparable with the excellent LEP performances. The precision of the measurements will increase, new tests of the \sm will be possible and hopefully a discovery in the BSM--land. Otherwise the Large Hadron Collider will have to  demonstrate that there is something beyond the celebrated Standard Model.

\vspace{2cm}
{\large \bf Acknowledgments}\\
I am grateful to Emmanuelle Perez, Mireille Schneider, Claude Vallee, Jochen Dingfelder, Emmanuel Sauvan, Joachim Meyer for many useful discussions and fruitful collaboration. Many thanks to Joachim Meyer and Jos Vermaseren for the careful reading of the manuscript. I thank Elie Aslanides, Jean-Jacques Aubert, Phillipe Bloch, Jean-Fran\c cois Grivaz, Sylvain Tisserant and Jos Vermaseren for having taken part to the jury. It is a pleasure to thank also all my colleagues from the H1 collaboration. I acknowledge DESY hospitality during the completion of this work.